\documentstyle[11pt,aaspp4]{article}
\begin{document}

\title{The Diffuse Interstellar Clouds toward 23 Orionis\footnotemark}

\footnotetext{Based on observations with the NASA/ESA {\it Hubble Space Telescope}, obtained at the Space Telescope Science Institute, which is operated by the Association of Universities for Research in Astronomy, Inc., under NASA contract NAS5-26555.}

\author{Daniel E. Welty}
\affil{University of Chicago, Astronomy and Astrophysics Center, 5640 S. Ellis Ave., Chicago, IL 60637}
\affil{\it welty@oddjob.uchicago.edu}

\author{L. M. Hobbs}
\affil{University of Chicago, Yerkes Observatory, Williams Bay, WI 53191}
\affil{\it hobbs@yerkes.uchicago.edu}

\author{James T. Lauroesch}
\affil{Northwestern University, Department of Physics \& Astronomy, 2131 Sheridan Rd., Evanston, IL 60208}
\affil{\it jtl@elvis.astro.nwu.edu}

\author{Donald C. Morton}
\affil{Herzberg Institute of Astrophysics, National Research Council, 5071 W. Saanich Rd., Victoria, BC, V8X 4M6 CANADA}
\affil{\it don.morton@hia.nrc.ca}

\author{Lyman Spitzer\altaffilmark{2}}
\affil{Princeton University Observatory, Peyton Hall, Princeton, NJ 08544}
\altaffiltext{2}{deceased}

\and

\author{Donald G. York\altaffilmark{3}}
\affil{University of Chicago, Astronomy and Astrophysics Center, 5640 S. Ellis Ave., Chicago, IL 60637}
\affil{\it don@oddjob.uchicago.edu}
\altaffiltext{3}{also, Enrico Fermi Institute}

\begin{abstract}

Spectra obtained with the {\it Hubble Space Telescope} GHRS are combined with high-resolution optical spectra and UV spectra from {\it Copernicus} to study the abundances and physical conditions in the diffuse interstellar clouds seen along the line of sight to the star 23 Ori.
Multiple absorption components are present for each of several distinct types of gas, which are characterized by different relative abundance and depletion patterns and physical conditions.

Strong low-velocity (SLV) absorption, due to cool, moderately dense neutral gas and representing about 92\% of the total $N$(\ion{H}{1}), is seen for various neutral and singly ionized species at +20 km s$^{-1}$ $\la$ $v_{\odot}$ $\la$ +26 km s$^{-1}$.
Most typically severely depleted species are less depleted by factors of 2-4, compared to the ``cold, dense cloud'' pattern found, for example, in the main components toward $\zeta$ Oph.
For the two strongest SLV components, $T$ $\sim$ 100 K and the thermal pressure log($n_{\rm H} T$) $\sim$ 3.1 cm$^{-3}$K; we thus have $n_{\rm H}$ $\sim$ 10--15 cm$^{-3}$ and a total thickness of 12--16 pc.
The adopted average SLV electron density, $n_e$ = 0.15$\pm$0.05 cm$^{-3}$,
implies a relatively large $n_e$/$n_{\rm H}$ $\sim$ 0.01, and thus some ionization of hydrogen in these predominantly neutral components. 

Weaker low-velocity (WLV) absorption, probably largely due to warmer neutral gas, is seen primarily for various singly ionized species at 0 km s$^{-1}$ $\la$ $v_{\odot}$ $\la$ +30 km s$^{-1}$.
The depletions in the WLV gas are typically less severe by a factor of 2--3 than in the SLV gas, and are somewhat similar to the ``warm cloud'' pattern seen in lines of sight with low reddening, low mean density, and/or low molecular fraction.
If $T$ $\sim$ 3000 K for the WLV components, then we have log($n_{\rm H} T$) $\sim$ 4.7--4.8 cm$^{-3}$K, $n_{\rm H}$ $\sim$ 15--20 cm$^{-3}$,
$n_e$ $\sim$ 0.2 cm$^{-3}$, $n_e$/$n_{\rm H}$ $\sim$ 0.01, and a total thickness of 0.7--0.9 pc.

Absorption from a number of singly and doubly ionized species, perhaps due to a radiative shock, is seen at $-$108 km s$^{-1}$ $\la$ $v_{\odot}$ $\la$ $-$83 km s$^{-1}$.
While the depletions in these ionized components are uncertain due to unobserved ionization stages, aluminum (typically severely depleted) is probably depleted there by only a factor $\sim$ 3, even at cloud velocities in excess of 100 km s$^{-1}$.
The individual high-velocity components typically have $T$ $\sim$ 8000$\pm$2000 K, $n_e$ = $n_{\rm H}$ $\sim$ 0.4--0.5 cm$^{-3}$, thermal pressure log(2$n_e T$) $\sim$ 3.7--4.0 cm$^{-3}$K, and thicknesses of order 0.1 pc.

Weak absorption components from ionized (\ion{H}{2}) gas are seen in \ion{C}{2}, \ion{Mg}{2}, and \ion{Si}{3} at intermediate velocities ($-$43 km s$^{-1}$ $\la$ $v_{\odot}$ $\la$ $-$4 km s$^{-1}$). 
Broad, weak absorption from the higher ions \ion{S}{3}, \ion{C}{4}, \ion{Si}{4}, and \ion{N}{5} is centered at $-$5 km s$^{-1}$ $\la$ $v_{\odot}$ $\la$ +6 km s$^{-1}$.
No obvious absorption is discerned from a circumstellar \ion{H}{2} region around 23 Ori itself.

The large range in $n_e$ (from 0.04 cm$^{-3}$ to 0.95 cm$^{-3}$) derived independently from nine pairs of neutral and singly ionized species in the SLV gas suggests that additional processes besides simple photoionization and radiative recombination affect the ionization balance.
Charge exchange with protons may reduce the abundances of \ion{S}{1}, \ion{Mn}{1}, and \ion{Fe}{1}; dissociative recombination of CH$^+$ may help to enhance \ion{C}{1}.
The large apparent fractional ionization in the SLV and WLV gas may be due to an enhanced flux of X-rays in the Orion region, to mixing of neutral and ionized gas at the boundary of the Orion-Eridanus bubble, or perhaps (in part) to charge exchange between singly ionized atomic species and large molecules (in which case the true $n_e$ would be somewhat smaller).
Comparisons of the SLV depletions and $n_{\rm H}$ with those found for the strong ``component B'' ($v_{\odot}$ $\sim$ $-$14 km s$^{-1}$) blend toward $\zeta$ Oph hint at a possible relationship between depletion and {\it local} density for relatively cold interstellar clouds.
Calcium appears to be more severely depleted in warm, low density gas than has generally been assumed.

An appendix summarizes the most reliable oscillator strengths currently available for a number of the interstellar absorption lines analyzed in this work.

\end{abstract} 


\section{Introduction}
\label{sec-intro}

Determination of the physical properties of individual interstellar clouds from observations of absorption lines requires both high spectral resolution and access to the satellite UV.
High spectral resolution is needed to distinguish the individual components contributing to the generally complex absorption-line profiles along most lines of sight.
High-resolution (FWHM = 0.3--1.2 km s$^{-1}$) surveys of interstellar \ion{Na}{1} and \ion{Ca}{2} absorption indicate, for example, that the median widths (FWHM) of individual interstellar components seen in \ion{Na}{1} and the median separation of adjacent components, in both species, are all $\la$ 1.2 km s$^{-1}$ (Welty, Hobbs, \& Kulkarni 1994; Welty, Morton, \& Hobbs 1996).
Furthermore, those adjacent components may be characterized by very different abundances and physical characteristics.
Many ground-state transitions for the dominant ionization states of the most abundant elements --- and thus many of the crucial diagnostic lines for determining such quantities as the local electron density $n_e$, the local hydrogen density $n_{\rm H}$, the temperature $T$, and the ambient radiation field --- are found in the UV.
The moderate resolution (FWHM $\sim$ 15--25 km s$^{-1}$) UV spectra obtained by the {\it Copernicus} and {\it IUE} satellites provided considerable information on total line-of-sight abundances and general physical conditions in the interstellar medium (e.g., Spitzer \& Jenkins 1975; Cowie \& Songaila 1986; de Boer, Jura, \& Shull 1987; Jenkins 1987), but could not, in general, provide detailed information for individual interstellar clouds.
Even detailed studies which applied profile-fitting techniques to high signal-to-noise ratio (S/N) {\it Copernicus} spectra of multiple transitions (e.g., Ferlet et al. 1980; Martin \& York 1982) did not discern all the components present along those lines of sight.
The increased resolution (FWHM $\sim$ 3.5 km s$^{-1}$) and S/N (up to $\sim$ 1000) achievable in spectra obtained with the {\it Hubble Space Telescope} Goddard High Resolution Spectrograph (GHRS) allow individual interstellar clouds to be studied in more detail, where the clouds are warm and/or isolated in velocity [e.g., the halo clouds studied by Spitzer \& Fitzpatrick (1993, 1995) and by Fitzpatrick \& Spitzer (1997)].
For the colder clouds and more complex lines of sight commonly found in the Galactic disk, however, even the GHRS is not able to discern and characterize most of the detailed component structure present.
For example, the two component groups studied by Savage, Cardelli, \& Sofia (1992) in GHRS spectra of $\zeta$ Oph are found to be composed of at least 4 and 8 components, respectively, in high-resolution \ion{Na}{1} and \ion{Ca}{2} spectra (Welty et al. 1994, 1996; Barlow et al. 1995).
If this underlying complexity is not incorporated into analyses of the UV spectra, the derived abundances and physical properties must be viewed as averages over several potentially quite disparate parcels of gas, and column densities obtained from moderately saturated lines may be systematically somewhat underestimated.

The Orion region has provided a fruitful laboratory for the investigation of various interstellar processes, even apart from studies of the background giant molecular cloud.
High-resolution surveys of optical absorption lines of \ion{Na}{1} and \ion{Ca}{2} have revealed very complex interstellar component structure in the neutral gas at relatively low LSR velocities toward many of the stars in Orion (Hobbs 1969; Marschall \& Hobbs 1973; Welty et al. 1994, 1996).
UV spectra of a number of stars in the region have revealed in addition absorption from ionized gas at intermediate ($-$80 km s$^{-1}$ $\la$ $v_{\rm LSR}$ $\la$ $-$20 km s$^{-1}$) and high velocities ($-$120 km s$^{-1}$ $\la$ $v_{\rm LSR}$ $\la$ $-$80 km s$^{-1}$) (Cohn \& York 1977; Cowie, Songaila, \& York 1979; Trapero et al. 1996).
While some of the components arise in relatively nearby gas within about 70 pc (Frisch, Sembach, \& York 1990), much of the complex structure is likely to be due to interactions between the young, luminous Orion stars (and earlier supernovae) and the ambient ISM.
Maps of the \ion{H}{1} 21 cm emission in the Orion-Eridanus region exhibit numerous filamentary structures, whose systematic behavior with velocity strongly suggests that some of the filaments trace the boundary of an expanding shell (Heiles 1976; Brown, Hartmann, \& Burton 1995).
Similar filamentary structures seen in H$\alpha$ and [\ion{N}{2}] emission suggest that the inner edge of the shell is ionized, presumably by UV radiation from the Orion OB1 association stars (Sivan 1974; Reynolds \& Ogden 1979).
Enhanced soft X-ray emission at $\onequarter$ and/or $\threequarters$ keV, in the region bounded by the H$\alpha$ filaments, suggests that the shell encloses a superbubble of hot gas [log($T$) $\sim$ 6.2 K] produced by supernovae and stellar winds over the past several million years (Burrows et al. 1993; Snowden et al. 1995).
Comparisons of the 21 cm data, the soft X-ray maps, the {\it IRAS} 100 $\mu$m maps, and optical interstellar absorption-line data toward stars in the region have yielded information on the relative locations and distances of some of the neutral and ionized material (Guo et al. 1995).

This study of the line of sight to 23 Ori is the first in a series which will combine high-resolution optical spectra with GHRS spectra to discern and investigate the properties of individual interstellar clouds in diverse Galactic environments.
The high-resolution optical spectra reveal much of the underlying component structure that is unresolved in the UV spectra, enabling the latter to be interpreted more accurately.
In Section~\ref{sec-obs}, we describe the new high-resolution optical and UV spectra obtained toward 23 Ori and introduce our analysis of the observed absorption-line profiles.
In Sections~\ref{sec-neutral} and~\ref{sec-ionized}, we describe in detail the profile analysis and discuss the abundances and physical properties derived for several ``natural'' groups of neutral and ionized components along the line of sight.
In Section~\ref{sec-disc}, we discuss those component groups in the context of prior studies of the Orion region, we explore possible explanations for the wide range in electron density derived from different ratios of trace and dominant species in the strongest neutral components, and we compare the elemental abundances and depletions with those found for other lines of sight.
In an appendix, we briefly summarize some recent results on atomic data relevant to this work.

\section{Observations and Data Reduction}
\label{sec-obs}

\subsection{23 Ori Line of Sight}
\label{sec-los}

The B1 V star 23 Ori [V = 5.00; $E(B-V)$ = 0.11; $v$sin$i$ = 295 km s$^{-1}$] is located at ($l$, $b$) = (199$\fdg$2, $-$17$\fdg$9), about 3 degrees south of $\gamma$ Ori (the northwest shoulder of Orion).
The parallax 3.39$\pm$0.87 mas measured by {\it Hipparcos} implies a distance of 295$^{+102}_{-60}$ pc (ESA 1997), consistent with, but perhaps somewhat smaller than the value 430$\pm$100 pc inferred from the spectral type and apparent magnitude (using the absolute magnitudes of Blaauw 1963).
Using {\it Hipparcos} parallaxes for a number of Orion stars, Brown, Walter, \& Blaauw (1998) have suggested that Orion OB association subgroup 1a (of which 23 Ori is a member) lies at an average  distance of about 330 pc, roughly 100 pc closer than subgroups 1b and 1c.
Morrell \& Levato (1991) determined a stellar radial velocity $v_{\rm rad}$ = 37$\pm$6 km s$^{-1}$, somewhat larger than the previous estimate of 18 km s$^{-1}$ due to Campbell \& Moore (1928), and concluded that $v_{\rm rad}$ does not vary.
The line of sight to 23 Ori has a total neutral hydrogen column density log[$N$(\ion{H}{1} + 2H$_2$)] = 20.74 cm$^{-2}$ (Bohlin, Savage, \& Drake 1978), with a relatively modest molecular hydrogen column density log[$N$(H$_2$)] $\sim$ 18.3 cm$^{-2}$ (Savage et al. 1977; Frisch \& Jura 1980; this paper).
23 Ori may lie behind a detached wisp of H$\alpha$ emission probably associated with Barnard's Loop; it does lie behind the high-velocity ionized gas dubbed ``Orion's Cloak'' by Cowie et al. (1979).
While 23 Ori has been observed in connection with several surveys of optical and UV interstellar absorption-lines (e.g., Hobbs 1978b; Chaffee \& White 1982; Jenkins, Savage, \& Spitzer 1986; Crane, Lambert, \& Sheffer 1995), the line of sight has not previously been studied in detail.

\subsection{Optical Spectra}
\label{sec-obsopt}

High resolution, relatively high S/N observations of the optical absorption lines of \ion{Na}{1}, \ion{Ca}{1}, \ion{Ca}{2}, \ion{K}{1}, and several molecular species have been obtained using the 0.9 m coud\'{e} feed telescope and coud\'{e} spectrograph at Kitt Peak National Observatory, the 3.9 m Anglo-Australian Telescope and Ultra-High-Resolution Facility (UHRF) at the Anglo-Australian Observatory, and the 2.7 m telescope and coud\'{e} spectrograph at the McDonald Observatory (Table~\ref{tab:opt}).
The references noted in Table~\ref{tab:opt} give details of the various instrumental setups and data-reduction procedures used to obtain and process the optical spectra.

\subsection{UV Spectra}
\label{sec-obsuv}

We observed 23 Ori with the {\it HST} GHRS during Cycle 1 (on 1992 April 14--15 and 1992 September 23), as part of GO programs 2251 and 3993 aimed at determining the properties of individual interstellar clouds.
Additional observations were obtained in Cycle 5 (on 1996 January 13), as part of GO program 5896 studying high-velocity ionized gas.
All observations reported here were obtained through the small science aperture, in order both to achieve the highest possible spectral resolution and to avoid complications due to the extended wings of the large science aperture instrumental function (for the pre-COSTAR spectra).
Thirteen wavelength regions between 1740 and 2805 \AA, each covering 9--15 \AA, were observed using echelle-B at resolutions of 3.3--3.7 km s$^{-1}$ (Table~\ref{tab:ghrs}).
Owing to the (temporary) failure of side 1, and the consequent unavailability of echelle-A for obtaining high-resolution spectra at shorter wavelengths, ten wavelength regions between 1130 and 1680 \AA, each covering approximately 37--38 \AA, were observed with grating G160M at intermediate resolutions of 13--20 km s$^{-1}$ (Table~\ref{tab:ghrs}).
The two ECH-A exposures, near 1260 and 1334 \AA, were obtained in Cycle 5.
In order to obtain proper sampling and to reduce the effects of photocathode sensitivity variations, all exposures were divided into a number of slightly shifted subexposures.
For the shorter wavelength G160M spectra, FP-SPLIT = FOUR was used to obtain four subexposures, successively shifted by typically 0.7 \AA;
FP-SPLIT = DSFOUR was used for the ECH-A observations. 
For the longer wavelength ECH-B spectra, the WSCAN procedure was used to obtain three or four subexposures, each offset by 0.7--1.0 \AA.
(See, e.g., Duncan \& Ebbets 1989 for discussions of these observing strategies.)
Additional G160M spectra of 23 Ori, obtained from the {\it HST} archive (GO program 2584), were used either to measure absorption lines not covered in our spectra or to increase the S/Ns for weak lines.

Among the subexposures for each spectral region obtained in Cycle 1, the generally small relative offsets in the nominal wavelength scales supplied by STScI were determined by fitting the profiles of interstellar and/or stellar lines in the spectra.
These relative offsets were typically less than 0.5 km s$^{-1}$ for the ECH-B data, and usually less than about 10 km s$^{-1}$ for the G160M data.
The individual subexposures were shifted according to those relative offsets, sinc interpolated (cf. Spitzer \& Fitzpatrick 1993) onto a roughly twice-oversampled standard wavelength grid, and summed.
Obvious deviant pixels in the individual subexposures (due to detector pattern features, bad pixels, radiation hits, etc.) were given zero weight in the sums; no other explicit detector pattern noise identification and removal was performed, however.
Adjacent pixels were then summed to yield approximately optimal sampling (i.e., two points per resolution element) in the final spectra.
For the Cycle 5 ECH-A spectra, the STSDAS routines POFFSETS, DOPOFF, and SPECALIGN were used to determine the detector pattern and to align and sum the subexposures.

Portions of the spectra near interstellar absorption lines of interest (usually including 150--250 points) were normalized by fitting the continuum regions with Legendre polynomials of orders typically 7--15.
For the Cycle 1 spectra, standard values for the on-order scattered light background (Cardelli, Ebbets, \& Savage 1993) were subtracted from the normalized ECH-B spectra.
The applicability of those standard values was confirmed via the strongly saturated \ion{Mg}{2} lines at 2796 and 2803 \AA, where the resulting relative intensity in the saturated line cores was within 0.1\% of zero.
The default (constant) backgrounds subtracted from the ECH-A spectra obtained in Cycle 5, however, left some residual flux (0.5--2.5\%) in the otherwise clearly saturated cores of several lines; that residual flux was subtracted from those spectra.
Except for the shortest wavelength G160M exposure centered near 1133 \AA, the continuum signal-to-noise ratios achieved for the G160M data are typically between 90 and 120, yielding 2$\sigma$ equivalent width upper limits or errors of 1-2 m\AA~ for weak, unresolved absorption lines.
The corresponding continuum S/Ns for the echelle data range from 30--40 for the initial set (I) of wavelength regions, from 60--100 for the second and third sets (II and III) of regions, and is about 120 for the region near the \ion{C}{2}] line at 2325 \AA.
The 2$\sigma$ equivalent width errors or upper limits for the echelle data range from 0.6 m\AA~ near the \ion{C}{2}] line to 1--2 m\AA~ for other weak lines.
The quoted equivalent width errors include contributions from both photon noise and continuum placement (Jenkins et al. 1973; Sembach \& Savage 1992), added in quadrature.

The equivalent widths of absorption lines due to various atomic and ionic species found in the optical and UV spectra are listed in Table~\ref{tab:ew}. 
For most of the lines, the equivalent widths correspond to the strong low-velocity absorption seen at 0 km s$^{-1}$ $\la$ $v$ $\la$ +30 km s$^{-1}$; absorption from intermediate-velocity gas at $-$45 km s$^{-1}$ $\la$ $v$ $\la$ 0 km s$^{-1}$, blended with the low-velocity absorption, is also included where it is detected\footnotemark.
\footnotetext{In this study, we will use heliocentric velocities; the zero point for velocities with respect to the Local Standard of Rest is at v$_{\odot}$ $\sim$ +16.4 km s$^{-1}$.}
Equivalent widths and limits for the high-velocity (HV) components found at $-$108 km s$^{-1}$ $\la$ $v_{\odot}$ $\la$ $-$83 km s$^{-1}$ are noted separately in the last column (updating values given by Trapero et al. 1996).
Lines observed at high resolution (FWHM $\la$ 4 km s$^{-1}$) are noted by an ``H'' following the wavelength; dominant contributors to blended lines are marked with a ``B'' following the equivalent width.
Minor contributors to blended lines are included in the table, but are referred to the dominant contributors for the total equivalent widths.
The letters ``P'' and ``S'' denote primary and secondary transitions used in determining the column densities for each species.
Equivalent widths and column densities for molecular absorption lines (including {\it Copernicus} data for H$_2$) are given in Table~\ref{tab:mol}.

Offsets in velocity, relative to the zero point set by the high-resolution optical spectra of \ion{Ca}{2}, are listed in the next-to-last column of Table~\ref{tab:ew}. 
These offsets, determined in the analysis of the line profiles as described below, are consistent with those derived from the various wavelength calibration and ``spybal'' spectra obtained along with the stellar spectra.
The offsets between the initial set of ECH-B spectra and the optical spectra are small, ranging from $-$0.1 to $-$1.7 km s$^{-1}$, but the offsets for the second set of ECH-B spectra are significantly larger, ranging from $-$5.4 to $-$8.7 km s$^{-1}$.
The zero point offset for the \ion{Al}{3} spectra could not be determined by comparisons with the optical lines, since \ion{Al}{3} is likely to be distributed significantly differently from those species.
Since the offsets for each observing session generally showed systematic trends with time, however, we  estimated an offset of $-$4.1 km s$^{-1}$ for the \ion{Al}{3} spectrum from the average of the offsets found for the immediately preceding spybal spectrum ($-$2.7 km s$^{-1}$) and the immediately following \ion{Fe}{2} $\lambda$2249 spectrum ($-$5.4 km s$^{-1}$).
The relative velocity offsets determined from fits to the HV components are very similar to those derived independently from fits to the stronger low-velocity absorption.
The average velocity offsets found between the G160M spectra and the optical and ECH-B data ranged from $-$40 km s$^{-1}$ for the shorter wavelength settings to $-$25 km s$^{-1}$ for the longer wavelength settings. 
Systematic variations of several km s$^{-1}$, with more negative values at shorter wavelengths, were obtained for the offsets within each G160M spectrum.

\subsection{Profile Analysis}
\label{sec-prof}

The normalized absorption-line profiles of various neutral and singly ionized species seen in the high-resolution optical spectra are shown in Figure~\ref{fig:lowv}.
Selected UV echelle and lower resolution G160M profiles, covering a larger velocity range, are shown in Figures~\ref{fig:uv} and~\ref{fig:highv}.
While the lower resolution (FWHM $\ga$ 1.5 km s$^{-1}$) optical and GHRS echelle profiles seem to show a single strong component at $v$ $\sim$ +23 km s$^{-1}$, the higher resolution (FWHM $\sim$ 0.3--1.3 km s$^{-1}$) \ion{Ca}{2}, \ion{Ca}{1}, and \ion{K}{1} spectra reveal (at least) two components dominating that strong observed absorption.
A number of weaker (i.e., lower total hydrogen column density) components can be seen in the profiles of the stronger lines of various singly ionized species at 0 km s$^{-1}$ $\la$ $v$ $\la$ +30 km s$^{-1}$. 
Additional weak intermediate-velocity (IV) components at $-$20 km s$^{-1}$ $\la$ $v$ $\la$ $-$4 km s$^{-1}$ are apparent in the strong \ion{Mg}{2} $\lambda$2796 and $\lambda$2803 lines; still weaker IV components at $-$43 km s$^{-1}$ $\la$ $v$ $\la$ $-$32 km s$^{-1}$ can be seen in the very strong \ion{C}{2} $\lambda$1334 and \ion{Si}{3} $\lambda$1206 lines.
The high-velocity (HV) gas comprising Orion's Cloak appears at $-$108 km s$^{-1}$ $\la$ $v$ $\la$ $-$83 km s$^{-1}$ in the strongest lines of \ion{C}{2}, \ion{N}{2}, \ion{Mg}{2}, \ion{Al}{2}, \ion{Al}{3}, \ion{Si}{2}, and \ion{Si}{3}; four components can be discerned in the echelle profiles of \ion{Mg}{2}, \ion{Si}{2}, and \ion{C}{2}.
Weak, broad absorption due to more highly ionized species (\ion{S}{3}, \ion{C}{4}, \ion{Si}{4}, \ion{N}{5}) is centered between $-$5 km s$^{-1}$ and +6 km s$^{-1}$.
A schematic diagram of the observed absorption is given in Figure~\ref{fig:schem}.

Since the primary goal of this investigation is the determination of abundances and physical conditions for individual interstellar clouds, we have used the method of profile fitting in an attempt to discern and characterize the individual components giving rise to the observed absorption profiles (e.g., Welty, Hobbs, \& York 1991).
In this analysis, we attempt to obtain column densities ($N$), Doppler line width parameters [$b$ = $(2kT/m + 2v_t^2)^{1/2}$] and heliocentric velocities ($v$) for as many components as are needed (at minimum) to reproduce the observed line profiles.
Comparisons among the various high-resolution optical line profiles reveal both the underlying component structure (number of components and their relative velocities, constraints on $b$-values and relative abundances) and also some general characteristics of the individual components which together can be used to understand and model the lower resolution UV line profiles.
In the analysis of the line profiles detailed below (and in the subsequent interpretation of the derived component column densities), we will make frequent use of the equation of photoionization equilibrium, 
$\Gamma(X^0)~n(X^0)~=~\alpha(X^0,T)~n_e~n(X^+)$, where $\Gamma$ is the photoionization rate and $\alpha$ is the recombination rate. 
For example, where abundances for several trace neutral species are available for a given component, ratios of that equation for two different elements yield estimates for the relative total elemental abundances for the two elements (York 1980; Snow 1984; Welty et al. 1991).
We note, however, that other processes besides simple photoionization and radiative recombination may affect the ionization balance for some elements.

Where we had multiple lines from the same species, it was generally possible to find a consistent set of component parameters for fitting all the lines, within the uncertainties in the line profiles and/or equivalent widths.
This consistency suggests both that the derived component model adequately describes this line of sight and that the adopted atomic data are accurate.
For atomic transitions, we use the wavelengths, oscillator strengths, and damping constants tabulated by Morton (1991), except where more recent work has provided more accurate values (as noted in Table~\ref{tab:ew} and discussed in the Appendix).
References for the corresponding molecular data are given in Table~\ref{tab:mol}.

The detailed component structures for the predominantly neutral gas at low velocities and for the predominantly ionized gas at intermediate and high velocities, derived primarily from the high-resolution optical and UV profiles, are described in Sections~\ref{sec-ncomp} and~\ref{sec-compion} below. 
The individual components are marked above the spectra in Figures~\ref{fig:lowv}, ~\ref{fig:uv}, and~\ref{fig:highv} and are numbered sequentially, with the letters H, I, W, and S appended to associate them with the high-velocity, intermediate-velocity, weak low-velocity, and strong low-velocity component groups defined above and below.
The absorption from the higher ions \ion{Al}{3}, \ion{S}{3}, \ion{C}{4}, \ion{Si}{4}, \ion{N}{5} seen at relatively low velocities is discussed in Sections~\ref{sec-compion} and~\ref{sec-abhigh}.

\subsection{$N$(H I) and Reference Abundances}
\label{sec-href}

We have adopted the total line of sight \ion{H}{1} column density log[$N$(\ion{H}{1})] = 20.74 cm$^{-2}$ determined by Bohlin et al. (1978) from {\it Copernicus} spectra, instead of the lower value 20.54 cm$^{-2}$ obtained by Diplas \& Savage (1994) from {\it IUE} spectra.
Profile fits to archival GHRS G160M spectra encompassing the \ion{H}{1} Lyman $\alpha$ line, using the component structure derived for \ion{Zn}{2}, yield log[$N$(\ion{H}{1})] $\sim$ 20.7 cm$^{-2}$.
In addition, the lower value obtained by Diplas \& Savage would imply essentially solar (or slightly higher) relative abundances for the typically very mildly depleted elements N, S, and Zn, as well as higher than usual abundances for C, O, and P.

For calculating depletions, we use the meteoritic (C1 carbonaceous chondritic) reference abundances of Anders \& Grevesse (1989) and Grevesse \& Noels (1993) for all elements except C, N, and O (which are solar photospheric values), as summarized in Table~\ref{tab:refab}.
For most elements, the meteoritic and solar photospheric abundances agree to within $\pm$0.06 dex; the exceptions are P ($A_m$$-$$A_p$=0.12), Cl (-0.23), Mn (0.14), Ga (0.25), Ge (0.22), and Sn (0.14) (though the listed uncertainties for photospheric Cl and Sn are both 0.3 dex).
For comparison with the depletions found toward 23 Ori, the Galactic ``warm'' and ``cold cloud'' interstellar depletion patterns listed in Table~\ref{tab:refab} have been adapted and updated from Jenkins (1987); the ``halo cloud'' pattern has been adapted from Savage \& Sembach (1996) and Fitzpatrick (1996b).
Slight differences from similar compilations in our previous papers (Lauroesch et al. 1996; Welty et al. 1997, 1999a) are due to the meteoritic reference abundances adopted here, to some newly-determined $f$-values (see the Appendix), and to recent results for
the abundances of several elements (C: Sofia et al. 1997; N: Meyer et al. 1997;
O: Meyer et al. 1998; Kr: Cardelli \& Meyer 1997).
The warm cloud depletion for Ca is estimated in this paper ($\S$~\ref{sec-deplwlv}).
Depletions for the heavy elements Cu, Ga, Ge, and Sn are based on data in Hobbs
et al. (1993), Cardelli (1994), and Lambert et al. (1998).

\section{Predominantly Neutral Gas at Low Velocities}
\label{sec-neutral}

\subsection{Neutral Gas Component Structure}
\label{sec-ncomp}

For the absorption at relatively low velocities ($v$ $\ga$ 0 km s$^{-1}$, or $|v_{\rm LSR}|$ $\la$ 20 km s$^{-1}$), \ion{O}{1}, \ion{N}{1}, and various singly ionized species show strong absorption.
The main absorption component(s) near $v$ $\sim$ 23 km s$^{-1}$ show the strongest lines both of various trace neutral species and of the corresponding dominant first ions. 
The absorption lines from the trace neutral species are remarkably strong, given the relatively modest total $N$(\ion{H}{1}); for example, some are stronger than the corresponding lines observed toward $\zeta$ Oph, which has log[$N$(H$_{\rm tot}$)] $\sim$ 21.15 cm$^{-2}$.
\ion{C}{1}, \ion{Na}{1}, and perhaps \ion{Mg}{1} are the only trace neutral species detected in the somewhat weaker absorption at 0 km s$^{-1}$ $\la$ $v$ $\la$ 20 km s$^{-1}$.
As expected, the component structure found for trace neutral species seen in the UV is very similar to that of \ion{Na}{1}, \ion{K}{1}, and \ion{Ca}{1} (Welty et al. 1991).
The absorption-line profiles for many of the dominant first ions of typically depleted elements (e.g., \ion{Fe}{2}, \ion{Si}{2}, \ion{Cr}{2}, \ion{Ni}{2}) appear somewhat similar to the \ion{Ca}{2} profile, both toward 23 Ori and for a number of other lines of sight for which we have GHRS data.
The latter similarity is somewhat remarkable, as \ion{Ca}{2} is probably a trace ionization stage for many of those components, but is likely attributable to the offsetting effects of the Ca ionization balance and the Ca depletion for the different types of neutral (\ion{H}{1}) clouds (Welty et al. 1996).
The component structure found for \ion{K}{1}, \ion{Ca}{1}, and \ion{Na}{1} was therefore used to model the UV profiles of other trace neutral species, while the \ion{Ca}{2} component structure was used to model the UV profiles of other (generally dominant) first ions of comparable strength\footnotemark.
\footnotetext{In fitting the UV lines, the GHRS instrumental profile was assumed to be adequately described by a Gaussian, with the FWHM appropriate for each wavelength region taken from Duncan \& Ebbets (1990); see column 4 and footnote b of Table~\ref{tab:ghrs}.}
The low-velocity component structures derived for species observed at high resolution are given in Table~\ref{tab:h1cmp}.

\subsubsection{Optical Spectra of Na I, K I, Ca I, and Ca II}
\label{sec-optcomp}

We begin the profile analysis with the high-resolution \ion{K}{1} and \ion{Ca}{2} profiles, where the detailed component structure is most clearly seen (Figure~\ref{fig:lowv}).
Independent fits to the highest resolution \ion{Ca}{2} and \ion{K}{1} profiles indicate that the strong absorption near +23 km s$^{-1}$ arises primarily from two components, separated by about 2.10--2.15 km s$^{-1}$ and of comparable strength (within a factor of $\sim$ 2); several weaker outlying components are present as well.
We will refer to these components (16S--19S) as the ``strong low-velocity'' (SLV) components.
Six relatively weak blueward components (10W--15W) at 0 km s$^{-1}$ $\la$ $v$ $\la$ +18 km s$^{-1}$, and two others (20W-21W) at $v$ $\sim$ +28 and +30 km s$^{-1}$, which we will call ``weak low velocity'' (WLV) components, are required to fit the profiles of \ion{Ca}{2}, \ion{Na}{1}, and some of the UV lines.
The fits to the coud\'{e} feed and higher resolution UHRF \ion{Ca}{2} spectra are quite consistent for the SLV components at +20 km s$^{-1}$ $\la$ $v$ $\la$ +26 km s$^{-1}$; the UHRF spectrum is somewhat noisy, and not as useful for confirming details in the WLV absorption, however.

Because the SLV \ion{Na}{1} components are strongly saturated, we fit the \ion{Na}{1} and \ion{K}{1} profiles together, for various ``reasonable'' values of the ratio $N$(\ion{Na}{1})/$N$(\ion{K}{1}) for the two strongest components (17S and 18S).
That ratio is typically within a factor of 2--3 of 70 in the nearby ISM (Welty et al. 1991).
The \ion{Na}{1} and \ion{K}{1} column densities and $b$-values for components 16S and 19S, and the overall velocity shift between the \ion{Na}{1} and \ion{K}{1} profiles, were determined via iterative fits for each set of those initial conditions.
The resulting fits were evaluated on the basis of goodness of fit to the observed \ion{Na}{1} profile and of the reasonableness of $N$(\ion{Na}{1})/$N$(\ion{K}{1}) for those two outlying SLV components.
$N$(\ion{Na}{1})/$N$(\ion{K}{1}) $\sim$ 70 for the two strongest SLV components gave quite acceptable fits and very similar column density ratios for the two outlying components; $N$(\ion{Na}{1})/$N$(\ion{K}{1}) ratios smaller than about 35 or larger than about 100 for the two strongest components yielded noticeably poorer fits.
For the reasonable assumption that \ion{Na}{1} and \ion{K}{1} coexist, $b$(\ion{Na}{1})/$b$(\ion{K}{1}) will be in the range 1.0--1.3 (fully turbulent to fully thermal broadening). 
Since the best fits to the profiles were obtained for $b$(\ion{Na}{1}) $\sim$ $b$(\ion{K}{1}), and since the $b$-values are significantly larger than would be obtained for thermal broadening at $T$ $\sim$ 100 K (as estimated in Section~\ref{sec-phst}), the broadening is apparently dominated by turbulence. 
The adopted \ion{Na}{1} SLV component structure could be checked via future high-resolution observations of the \ion{Na}{1} $\lambda$3303 lines.
The \ion{Na}{1} WLV components were fitted using the relative velocities determined from fits to the \ion{Ca}{2} profile.

The slight disagreement in relative velocities for SLV components 16S and 19S seen in the initial independent \ion{Ca}{2} and \ion{Na}{1} fits, the apparently larger $b$-values for \ion{Ca}{2} for those components, and initial fits to some of the UV lines all suggested the presence of additional structure.
We therefore re-fit the \ion{Ca}{2} and \ion{Na}{1} profiles, with the SLV components fixed at the same relative velocities found for \ion{Na}{1} and \ion{K}{1} and with two additional WLV components suggested by the UV lines.
The resulting \ion{Ca}{2} $b$-values are still slightly larger than those found for the corresponding components in \ion{Na}{1} and \ion{K}{1}, suggesting that the \ion{Ca}{2} may occupy a somewhat larger volume (as found for a larger sample of clouds seen in both \ion{Na}{1} and \ion{Ca}{2} by Welty et al. 1996)\footnotemark.
\footnotetext{The component structure given in Table~\ref{tab:h1cmp} for \ion{Ca}{2} differs slightly from that given in our \ion{Ca}{2} survey (Welty et al. 1996), which was based on fits to the \ion{Ca}{2} spectra alone.}
The weak \ion{Ca}{1} absorption was fitted with two components having the same separation and $b$-values as the two strongest \ion{K}{1} components.

\subsubsection{Modelling the UV Spectra}
\label{sec-modcomp}

For the two strongest SLV components (17S and 18S), which are not resolved in the UV spectra, the column density ratio found for \ion{Ca}{1} (1.0e10/0.6e10 $\sim$ 1.7) was assumed to typify the relative abundances of the trace neutral lines of heavily depleted elements (\ion{Ca}{1}, \ion{Fe}{1}), while the column density ratio found for \ion{K}{1} (1.37e11/1.61e11 $\sim$ 0.85) was assumed to hold for the trace neutral lines of lightly depleted elements (\ion{C}{1}, \ion{Na}{1}, \ion{S}{1}, \ion{K}{1}).
Taking the ratio of the $N$(\ion{Ca}{1})/$N$(\ion{Ca}{2}) ratios, for the same two components, yields the ratio of electron densities (0.026/0.010 $\sim$ 2.6) in the two components (assuming similar $\Gamma$/$\alpha$). 
The $N$(\ion{Ca}{1})/$N$(\ion{Ca}{2}) ratios also suggest that \ion{Ca}{2} is the dominant ionization state of Ca in both components.
The column density ratio for \ion{Ca}{2} in the two strong components (3.9e11/6.3e11 $\sim$ 0.6) was therefore assumed to apply for the dominant ions of other heavily depleted elements. 
Dividing the \ion{K}{1} ratio by the electron density ratio (0.85/2.6 $\sim$ 0.33) then yields an estimate for the relative abundances of the dominant ions of lightly depleted elements.
For those two strongest components, the abundance ratios ($r$ = $N_{17S}/N_{18S}$) of trace and dominant ionization states of elements typically depleted less than Ca but more than K were assigned intermediate values (i.e., 0.85 $\la$ $r$ $\la$ 1.7 for trace species and 0.33 $\la$ $r$ $\la$ 0.6 for dominant species), while maintaining the electron density ratio of 2.6.
These predicted abundance ratios depend on the assumption that the $N$(\ion{X}{1})/$N$(\ion{X}{2}) ratios for different elements X all depend in the same way on $n_e$ in the two strongest SLV components --- as would be the case if photoionization and radiative recombination dominate the ionization balance.
We will find in Sections~\ref{sec-phsne} and~\ref{sec-ne}, however, that different neutral/first ion ratios yield different values for $n_e$ (for a given $T$ and radiation field) --- suggesting that additional processes may be affecting the ionization balance.
In particular, the $n_e$ inferred from the $N$(\ion{Ca}{1})/$N$(\ion{Ca}{2}) ratio is often significantly higher than the values derived from other such ratios.
The relative column densities for components 17S and 18S may therefore differ from the values assumed for the profile fits.
We note, however, that consistent abundance ratios were obtained from fits in which $N_{17S}$ and $N_{18S}$ were allowed to vary -- both for the lightly depleted \ion{Zn}{2} and for the more severely depleted \ion{Fe}{2}.
Since at least the two strongest SLV components appear to consist of relatively cold, but turbulent gas, the $b$-values for most species are very similar for each component.
For the apparently warmer WLV components, the $b$-values for various species were derived from the WLV \ion{Ca}{2} $b$-values assuming a temperature of 3000 K ($\S$~\ref{sec-condwlv}).

\subsubsection{UV Echelle Spectra}
\label{sec-uvcomp}

For lines due to most trace neutral species and for the weaker lines due to some first ions, the WLV components were not detected in the UV echelle profiles, so that it is very difficult to unambiguously determine both the individual component column densities for the SLV components and the overall velocity offset, given the typical $\sim$ 1 km s$^{-1}$ inherent uncertainties in the GHRS echelle velocities. 
In such cases, the optical SLV column densities were uniformly scaled to fit the UV profiles, after constraining the ratio of the two strongest components as just described --- in effect yielding only an overall SLV column density ratio and velocity offset.
While this uniform scaling gave an acceptable fit in all cases, the UV line profiles do permit some component-to-component variations.
Where the WLV components are present in the echelle spectra, however, they provide additional constraints on the overall velocity offset, and it was thus possible to estimate individual component column densities for \ion{Zn}{2} ($\lambda$2026, $\lambda$2062), \ion{Fe}{2} ($\lambda$2344, $\lambda$2374), \ion{Si}{2} ($\lambda$1808), and \ion{S}{2} ($\lambda$1259), as listed in Table~\ref{tab:h1cmp}. 
Separate simultaneous fits were made to the three \ion{Fe}{2} lines and to the two \ion{Zn}{2} lines observed with the echelle.
In the \ion{S}{2} $\lambda$1259 profile, the SLV components are saturated, and were thus poorly constrained in the profile fits.
For ``reasonable'' values of SLV $N$(\ion{S}{2}) and the velocity zero-point offset (constrained by \ion{Si}{2} $\lambda$1260 and \ion{C}{1} $\lambda$1260), however, $N$(\ion{S}{2}) is well determined for all of the WLV components except 15W and 20W, which are adjacent to the SLV components.
Since the WLV $N$(\ion{S}{2})/$N$(\ion{Zn}{2}) ratios suggest that sulfur is undepleted in the WLV components, we estimated the total $N$(\ion{S}{2}) by using the curve of growth defined by the other dominant, little-depleted species and by the derived \ion{Zn}{2} component structure ({$\S$~\ref{sec-cog}).
Integrated over the line of sight, sulfur appears to be essentially undepleted, within $\pm$0.2 dex.
The very strong \ion{C}{2} $\lambda$1334 and \ion{Mg}{2} $\lambda$2796 and $\lambda$2803 lines are saturated over the entire WLV--SLV velocity range, and exhibit relatively weak, but definite damping wings (Figures~\ref{fig:uv} and~\ref{fig:highv}).
We obtained total WLV + SLV column densities for \ion{C}{2} and \ion{Mg}{2} by using the component structures determined for \ion{Zn}{2} and \ion{Si}{2}, respectively, to fit the damped profiles --- uniformly scaling the WLV and SLV column densities until a ``smooth'' continuum was achieved (as often done in fitting damped lines of \ion{H}{1}).
For those fits, the velocity zero points were constrained by requiring alignment of the high-velocity components in the echelle spectra of \ion{C}{2}, \ion{Mg}{2}, and \ion{Si}{2}, which show very similar HV component structure.
The resulting WLV + SLV column densities for \ion{C}{2} and \ion{Mg}{2} are consistent with the limit determined from the weak \ion{C}{2}] $\lambda$2325 line and with the value obtained from the weak \ion{Mg}{2} $\lambda$1239 doublet, respectively, using the revised $f$-values for these weak transitions discussed in the Appendix.
In view of the complexity of the component structure and the relatively low S/N obtained for some of the profiles, the derived individual component parameters listed in Table~\ref{tab:h1cmp} must still be regarded as somewhat uncertain.

\subsubsection{G160M Spectra}
\label{sec-gcomp}

For the typically higher S/N but lower resolution G160M spectra, individual component parameters could not be reliably derived.
We therefore used the component velocity structure and relative column densities derived from the optical and UV echelle spectra to model the G160M spectra. 
For the WLV and SLV gas, the echelle profiles of \ion{Zn}{2}, \ion{Si}{2}, and \ion{Fe}{2} (representing a range of depletion behavior for species dominant in \ion{H}{1} regions) were especially useful for interpreting the lines of other dominant neutral and singly ionized species observed with G160M.
Where the WLV components could be seen (e.g., as a blueward wing in the absorption-line profiles in some of the longer wavelength G160M spectra), the total SLV and WLV column densities could be scaled separately, guided by fits to the echelle spectra of similarly depleted species, to fit the observed G160M profiles.
Where the WLV components could not be separately discerned (e.g., for weak lines and for the shorter wavelength, lower resolution G160M spectra), relative SLV and WLV column density scalings from species with similar overall depletion were used, essentially to fit the observed equivalent width.
In such cases, the uncertainties in $N$ correspond to the range in values required to yield $W_{\lambda}$ $\pm$ 1$\sigma$.
The derived WLV and SLV column densities are typically less reliable for species (such as \ion{Al}{2}) which have only a single strong line in the G160M spectra.

\subsection{Abundances and Depletions for SLV and WLV Gas}
\label{sec-abundlv}

In Table~\ref{tab:h1abund}, we list column densities or upper limits for the predominantly neutral (\ion{H}{1}) gas --- for the total line of sight and for the SLV and WLV component groups at 0 km s$^{-1}$ $\la$ $v$ $\la$ +30 km s$^{-1}$.
The code in the last column of the table indicates how the component group column densities were determined, as discussed in the previous section. 
Species whose WLV and SLV column densities were obtained via scaling from similarly depleted species (typically observed at higher resolution) are given in parentheses in Table~\ref{tab:h1abund}.
For several species (e.g., \ion{Mn}{1}, \ion{Kr}{1}), we have averaged spectra from several weak transitions in order to obtain either a detection or a more stringent limit. 
\ion{Mn}{1} is detected (tentatively) for the first time in the diffuse Galactic ISM toward 23 Ori.

In Figure~\ref{fig:relind}, we show several column density ratios for the individual WLV and SLV components to illustrate some of the differences between those two sets of neutral components.
At the right in some of the panels, we show the ratios for the meteoritic reference abundances and for the three representative Galactic interstellar cloud types listed in Table~\ref{tab:refab}.
The $N$(\ion{Na}{1})/$N$(\ion{Ca}{2}) ratio provides the clearest distinction between the WLV and SLV components, though the $N$(\ion{Si}{2})/$N$(\ion{Zn}{2}), $N$(\ion{Fe}{2})/$N$(\ion{Zn}{2}), and $N$(\ion{Fe}{2})/$N$(\ion{S}{2}) ratios reveal some differences in depletion ($\S$$\S$~\ref{sec-deplslv} and~\ref{sec-deplsw}).
As noted above, the ratio $N$(\ion{Ca}{1})/$N$(\ion{Ca}{2}) suggests that \ion{Ca}{2} is the dominant ionization state of Ca in the two strongest SLV components (17S and 18S) (Welty et al. 1996, 1999b).
The smaller $N$(\ion{Ca}{2})/$N$(\ion{Fe}{2}) ratios for the outlying SLV components (16S and 19S), however, may indicate that \ion{Ca}{2} is a trace species there.
Similarly, comparing the ratio $N$(\ion{Na}{1})/$N$(\ion{Ca}{2}) with $N$(\ion{Na}{1}) suggests that \ion{Ca}{2} is likely a trace species in the weaker SLV and WLV components (Welty et al. 1996).
The $N$(\ion{Na}{1})/$N$(\ion{Zn}{2}) ratio provides constraints on the electron densities in the individual WLV components (if a typical interstellar radiation field is assumed).
Finally, we note that the slightly sub-solar $N$(\ion{Zn}{2})/$N$(\ion{S}{2}) ratio found for the WLV components in the profile fits (and the essentially identical value obtained for the SLV components from the curve of growth) are consistent with the results of Fitzpatrick \& Spitzer (1997) for the predominantly halo clouds toward HD 215733.

While the \ion{Al}{3}, \ion{Si}{3}, and \ion{N}{2} profiles indicate that there is apparently some ionized (i.e., \ion{H}{2}) gas present at 0 km s$^{-1}$ $\la$ $v$ $\la$ +30 km s$^{-1}$ (Figure~\ref{fig:highv}), comparison of the column densities with those of \ion{Al}{2}, \ion{Si}{2}, and \ion{N}{1} at similar velocities implies that the SLV and WLV gas is predominantly neutral.
For example, the limit on $N$(\ion{N}{2}) ($\S$$\S$~\ref{sec-compion} and~\ref{sec-abhigh}) suggests that the low velocity gas is less than about 5\% ionized (though the limit on the fractional ionization is somewhat higher for the WLV gas and somewhat lower for the SLV gas).
The relative column densities of \ion{O}{1}, \ion{N}{1}, and \ion{C}{2}, very similar to those typically found in \ion{H}{1} gas, also suggest that the SLV and WLV gas is mostly neutral, since only \ion{C}{2} would be present in any ionized gas.
As we will show below ($\S$~\ref{sec-ionized}), the various IV and HV components at $v$ $\la$ 0 km s$^{-1}$ are apparently predominantly ionized.
It thus seems reasonable to ascribe the measured $N$(\ion{H}{1} + 2H$_2$) to the SLV and WLV gas.
If the contribution of any \ion{H}{2} gas to the observed \ion{Zn}{2} and \ion{S}{2} absorption is negligible, then Zn and S are nearly undepleted in the SLV and WLV gas (D$_{\rm Zn}$ = $-$0.12 dex; D$_{\rm S}$ $\sim$ 0.0 dex).
We may then use the \ion{Zn}{2} and \ion{S}{2} column densities found for the individual SLV and WLV components to infer neutral hydrogen column densities for those individual components (assuming the same small level of depletion in each).
Unsurprisingly, the SLV components contain the bulk of the neutral gas, with total log[$N$(H)] = 20.71 cm$^{-2}$.
The WLV components have a total log[$N$(H)] = 19.61 cm$^{-2}$.

The depletions of various elements listed in Table~\ref{tab:h1abund} and plotted in Figure~\ref{fig:depl} for the SLV and WLV components (integrated over four and seven components, respectively) are with respect to the neutral hydrogen column densities inferred from \ion{Zn}{2} and \ion{S}{2}.
In Figure~\ref{fig:depl}, the elements are ordered according to the condensation temperature, as tabulated by Wasson (1985).
The average depletions for the line of sight are very close to those found for the SLV components, which contain most of the total \ion{H}{1}.
Where we do not have independent values for the WLV and SLV depletions, we have therefore plotted only the average line-of-sight values (with an ``s'').
For comparison in Figure~\ref{fig:depl}, the Galactic ``warm'' and ``cold cloud'' depletion patterns (Table~\ref{tab:refab}) are plotted as dotted and solid lines, respectively.
The $\zeta$ Oph ``component A'' (warm, low density gas near $v_{\odot}$ $\sim$ $-$28 km s$^{-1}$ ) and ``component B'' (cold, dense gas near $v_{\odot}$ $\sim$ $-$14 km s$^{-1}$) depletions are derived from data in Savage, Cardelli, \& Sofia (1992), Hobbs et al. (1993), Cardelli et al. (1994), Sofia et al. (1994), Lambert et al. (1998), and from our own analyses of various optical and archival GHRS spectra.
We have adopted log[$N$(H$_{\rm tot}$)] = 19.95 cm$^{-2}$ for $\zeta$ Oph component A by assuming D(Zn) $\sim$ $-$0.2 dex --- as found for warm, low density gas by Sembach et al. (1995).

\subsubsection{SLV Depletions}
\label{sec-deplslv}

For most elements, the 23 Ori SLV depletions (Table~\ref{tab:h1abund} and Figure~\ref{fig:depl}) appear to be less severe, by factors of 2--4, than the typical cold, dense cloud depletions or the depletions found for the strong component B blend seen toward $\zeta$ Oph.
It is conceivable that such an intermediate depletion could arise from a mixture of several components, each characterized by either "typical" warm or cold cloud depletions.
As noted above, the depletions in the two strongest SLV components (17S, 18S) differ by factors $\la$ 2 for different elements, and the weaker SLV components (16S, 19S) may be warmer and less dense than the two strong SLV components.
Where we have echelle data, however, the line profiles do not seem to permit a mixture extreme enough to yield the overall factor of 2--4 (see Fig.~\ref{fig:relind}). 
While the $N$(\ion{Fe}{2})/$N$(\ion{Zn}{2}) ratios suggest that the depletions for component 16S may be somewhat less severe than those for components 17S--19S, the $N$(\ion{Na}{1})/$N$(\ion{Ca}{2}) ratio is somewhat higher for both the outlying SLV components (16S and 19S) than for the WLV components.
The possible presence of a small amount of ionized gas would affect the depletions by less than 10\% (0.04 dex).
Interestingly, the SLV depletions are much closer to the cold cloud pattern when viewed as [X/Zn] = (X/Zn)$_{obs}$ $-$ (X/Zn)$_{\odot}$, since Zn also appears to be somewhat less severely depleted than in the cold cloud pattern.

\subsubsection{SLV Molecules}
\label{sec-slvmol}

Absorption from the molecular species H$_2$, CH, and CH$^+$ is also present at velocities corresponding to the SLV components.
The H$_2$ column densities for rotational levels $J$=0 through $J$=5 listed in Table~\ref{tab:mol} have been derived by comparing the observed ({\it Copernicus}) H$_2$ line profiles with model profiles based on the adopted SLV component structure for \ion{Na}{1} --- using either all four SLV components or just the two strongest ones.
While Frisch \& Jura (1980) used different $b$-values for the odd and even rotational levels to obtain estimates for the H$_2$ column densities, our fits do not suggest any obvious systematic difference in component structure between the odd and even levels (though the resolution and S/Ns are rather poor).
The column densities in the lowest four rotational levels yield an ortho- to para-H$_2$ ratio of $\sim$3, similar to the values found by Spitzer, Cochran, \& Hirshfeld (1974) for lines of sight with somewhat lower $N$(H$_2$).
The CH absorption in the SLV components toward 23 Ori is remarkably strong --- a factor of at least 40 stronger than would be expected from the usually fairly tight correlation between $N$(CH) and $N$(H$_2$) (e.g., Danks, Federman, \& Lambert 1984)\footnotemark. 
\footnotetext{The even higher $N$(CH) reported by Crane et al. (1995) appears to be in error (P. Crane, private communication).}
The $N$(CH)/$N$(CO) ratio is also higher than in most other lines of sight.
Federman, Welty, \& Cardelli (1997) have proposed that the CH observed toward 23 Ori is due to the same (poorly understood) non-equilibrium process(es) responsible for the observed CH$^+$, rather than to the equilibrium gas-phase chemical reactions responsible for the diatomic molecules (e.g., CH, CN, C$_2$) observed in many other diffuse cloud lines of sight.
While the absorption from both CH and CH$^+$ may be acceptably fitted with a single component (Table~\ref{tab:mol}), the presence of two strong SLV components in the atomic species and the possible asymmetry in the CH$^+$ profile (Figure~\ref{fig:lowv}) suggest that two components may be present in CH and CH$^+$ as well (see also Crane et al. 1995).
The relatively broad components required even for two-component fits and the lambda doubling present for CH make more complex fits not well constrained, however.

\subsubsection{WLV Depletions}
\label{sec-deplwlv}

The 23 Ori WLV depletions (Table~\ref{tab:h1abund} and Figure~\ref{fig:depl}) are somewhat less well known for the species whose relative SLV and WLV column densities were constrained to be the same as for \ion{Zn}{2}, \ion{Si}{2}, or \ion{Fe}{2}, based on similarities in average line-of-sight depletion.
Overall, however, the WLV depletions seem to be less severe by an additional factor of 2--3 than the SLV depletions, and are similar to the typical warm cloud and $\zeta$ Oph component A values.
We obtained the Ca depletion for the WLV gas by using the $N$(\ion{C}{1})/$N$(\ion{C}{2}), $N$(\ion{Na}{1})/$N$(\ion{Na}{2}), and $N$(\ion{Mg}{1})/$N$(\ion{Mg}{2}) ratios to estimate $n_e$, then using $n_e$ to derive the column density of the presumably dominant \ion{Ca}{3}.
For temperatures from 100 K to 3000 K, $n_e$ ranges over roughly an order of magnitude ($\S$~\ref{sec-condwlv}), but D(Ca) $\sim$ $-$2.4 dex throughout --- roughly a factor 8 more severe than that of Fe in the WLV gas.
A similar analysis applied to the $\zeta$ Oph A components yields very similar values for the Ca depletion there, and Ca also seems more severely depleted than Fe in the component groups exhibiting ``warm cloud'' depletions toward SN 1987A (Welty et al. 1999a).
If the $N$(\ion{Ca}{2})/$N$(\ion{Ca}{3}) ratio depends on $n_e$ in the same way as do the other three ratios, then Ca may generally be more severely depleted than Fe in warm, diffuse gas, contrary to what is often assumed.
Figure 4 in Jenkins (1987), for example, shows D(Ca) $\sim$ D(Fe) $\sim$ $-$1.2 dex for log($<n_{\rm H}>$) = $-$1.5 cm$^{-3}$, where the Ca depletion is based largely on relatively low resolution optical and {\it IUE} spectra obtained by Phillips, Pettini, \& Gondhalekar (1984). 
Improved atomic data, allowance for the (very mild) depletion of Zn, and better data for several stars (especially $\alpha$ Vir; Hobbs 1978a) make the Ca depletion at that low $<n_{\rm H}>$ more severe by about a factor of 5.
We have therefore adopted a ``warm cloud'' Ca depletion of $-$2.0 dex in Table~\ref{tab:refab} and in Figure~\ref{fig:depl}.

\subsubsection{Empirical Curves of Growth}
\label{sec-cog}

Given the column densities for various neutral and singly ionized species in the SLV and WLV gas derived from fits to the observed line profiles and/or equivalent widths, we may construct an empirical curve of growth for the neutral gas along the line of sight (Figure~\ref{fig:cog}).
Curves of growth have often been used to estimate line-of-sight column densities from lower resolution spectra, usually by comparing the observed log(W$_\lambda$/$\lambda$) vs. log($f$$\lambda$) with theoretical curves derived for assumed ``representative'' $b$-values.
In the figure, we have used separate symbols for species that are trace ions in the \ion{H}{1} gas (squares), species that are dominant ions of mildly depleted elements (e.g., \ion{C}{2}, \ion{O}{1}, \ion{Zn}{2}) (triangles), and species that are dominant ions of more severely depleted elements (e.g., \ion{Si}{2}, \ion{Ti}{2}, \ion{Fe}{2}) (circles).
The solid lines are the curves that would be expected from the adopted component structures for \ion{Na}{1} (trace species), \ion{Zn}{2} (dominant, mildly depleted), and \ion{Fe}{2} (dominant, severely depleted); the diagonal line shows the relationship for unsaturated lines.
It is apparent that the three groups define distinct curves of growth, reflecting the differences in distribution and physical properties among the various SLV and WLV components.
The curve defined by the trace neutral species, which are concentrated in the two strongest SLV components, is the first to show signs of saturation in the absorption lines as log($Nf\lambda$) increases, and lies below the other two curves for 7.5 cm$^{-1}$ $\la$ log($Nf\lambda$) $\la$ 10.5 cm$^{-1}$.
The curve occupied by the dominant, depleted species, which are less severely depleted (and thus relatively stronger) in the WLV components, lies above the other two curves --- reflecting the effectively somewhat broader distribution of the dominant, depleted species.
The curve defined by the dominant, mildly depleted species falls in between the other two curves. 
Note that the strong lines of \ion{N}{1} and \ion{O}{1}, whose column densities were derived from weak semi-forbidden transitions, fall nicely along the predicted curve for dominant, mildly depleted species [near log($Nf\lambda$) $\sim$ 9.8--10.6 cm$^{-1}$], even though the curve is based on the component structure found for the significantly weaker \ion{Zn}{2} lines.
The three triangles nearest log($Nf\lambda$) $\sim$ 9 cm$^{-1}$ represent the \ion{S}{2} $\lambda$1250, $\lambda$1253, and $\lambda$1259 lines.
The strongest lines of \ion{Mg}{2} and \ion{C}{2} exhibit both damping wings and weak IV absorption, and so lie somewhat above the predicted curves based on \ion{Fe}{2} and \ion{Zn}{2}, respectively.

While differences in the curves defined by trace and dominant species have been noted before (e.g., Morton 1975; Spitzer \& Jenkins 1975), dominant species with significantly different levels of depletion have often been assumed to occupy the same curve.
Toward 23 Ori, that assumption could lead to errors of factors of 2--3 in inferred column density for some species, depending on the particular absorption lines available, even though the SLV and WLV depletions differ by only about 0.4 dex for most typically depleted elements.
Furthermore, the curves of growth for both trace neutral species and dominant, mildly depleted species exhibit ``kinks'', at  log($Nf\lambda$) $\sim$ 9.5 and $\sim$ 8.5 cm$^{-1}$, respectively --- i.e., neither curve can be characterized by a single effective $b$-value.
The potential presence of ``kinks'' in curves of growth complicates the use of such curves for determining ``mutually consistent'' $f$-values for lines of a particular species.

\subsection{Physical Conditions in SLV Gas}
\label{sec-condslv}

\subsubsection{SLV Temperature}
\label{sec-phst}

The strong absorption due to trace neutral species, the presence of moderate amounts of H$_2$, fairly large values for $N$(\ion{Na}{1})/$N$(\ion{Ca}{2}) $\sim$ 10-30, and indications that \ion{Ca}{2} is the dominant ionization state of Ca all suggest that the two strongest SLV components (17S and 18S, near +23 km s$^{-1}$) are relatively cold and dense.
The apparent equality in $b$ for \ion{K}{1} and \ion{Na}{1}, which differ by a factor 1.7 in atomic weight, also suggests that these components are due to relatively cold gas, with the absorption line widths dominated by turbulent broadening or bulk motions. 
The \ion{K}{1} $b$-values for the two strongest SLV components yield strict upper limits for $T$ of about 1870 K and 4200 K, respectively, though the temperatures are actually much smaller if turbulence dominates the broadening.
The slightly larger $b$(\ion{Ca}{2}) may indicate that \ion{Ca}{2} occupies a larger volume, characterized by larger temperature and/or turbulence (Welty et al. 1996).

For H$_2$ column densities greater than about 10$^{17}$ cm$^{-2}$, the relative populations of the $J$=0 and $J$=1 rotational levels of H$_2$ are thermalized via collisions with protons, and we can estimate the local kinetic temperature $T$ from the ratio $N$(1)/$N$(0):
\begin{displaymath}
\frac{N(1)}{N(0)} = \frac{g_1}{g_0}~{\rm exp}(-E_{01}/kT_{01}) = 9~{\rm exp}(-170~{\rm K}/T_{01}).
\end{displaymath}
The temperature inferred from $N$(1)/$N$(0) lies in the range 65-150 K, depending on the exact component structure assumed for the H$_2$ (cf. Frisch \& Jura 1980); we will adopt $T$ = 100 K for the two strongest SLV components.
The analogous ``temperature'' inferred from the $N$(4)/$N$(2) ratio is about 350 K --- similar to values found by Spitzer et al. (1974) for other lines of sight and consistent with the relative level populations predicted by Spitzer \& Zweibel (1974) for $n_{\rm H}$ = 10 cm$^{-3}$, $T$ = 80 K, and an ``average'' interstellar radiation field.

\subsubsection{SLV Thermal Pressure and Local Density}
\label{sec-phsp}

The relative populations of the fine structure levels in the ground electronic term of \ion{C}{1} can yield information concerning the product $n_{\rm H}T$ (i.e., the thermal pressure) in the SLV clouds (Jenkins \& Shaya 1979; Jenkins, Jura, \& Loewenstein 1983).
If we have independent constraints on the temperature, then we can determine the local hydrogen density.
Taking $N$(\ion{C}{1}) = 8.5$\pm$1.0 $\times$ 10$^{14}$  cm$^{-2}$, $N$(\ion{C}{1}*) = 9.4$\pm$1.8 $\times$ 10$^{13}$ cm$^{-2}$, and $N$(\ion{C}{1}**) = 1.4$\pm$0.2 $\times$ 10$^{13}$ cm$^{-2}$, we have $N$(\ion{C}{1}**)/$N$(\ion{C}{1})$_{\rm tot}$ = 0.015$\pm$0.003 and $N$(\ion{C}{1}*)/$N$(\ion{C}{1})$_{\rm tot}$ = 0.10$\pm$0.02. 
Those ratios imply log($n_{\rm H}T$) $\sim$ 3.1$\pm$0.1 cm$^{-3}$K for $T$ $\sim$ 100 K and the WJ1 radiation field --- at the low end of the range of values reported by Jenkins et al. (1983).
Inclusion of a ``turbulent'' pressure term, using the $v_t$ obtained from the \ion{K}{1} $b$-values for the two strongest SLV components, would increase the pressure by no more than a factor of about 2.
If $T$ $\sim$ 100 K, as derived from H$_2$, then $n_{\rm H}$ $\sim$ 10--15 cm$^{-3}$. 
The total thickness implied for the two main clouds is thus about 12--16 pc --- comparable to the apparent thickness of the 21 cm emission filaments, at SLV velocities near 23 Ori, which define the shell surrounding the Orion-Eridanus bubble in Brown et al. (1995; see discussion in Section~\ref{sec-region}).
A relatively long path length through the expanding shell might also explain why the $b$-values for SLV components 17S and 18S are significantly larger than the thermal widths at $T$ $\sim$ 100 K --- i.e., the apparent dominance of ``turbulent '' broadening could be due to slight differences in velocity through the shell.

\subsubsection{SLV Electron Density and Fractional Ionization}
\label{sec-phsne}

If we assume a ``typical'' interstellar radiation field (e.g., the WJ1 field) and $T$ $\sim$ 100 K, we can use the equation of ionization equilibrium, $\Gamma(X^0)~n(X^0)~=~\alpha(X^0,T)~n_e~n(X^+)$, to compute local electron densities $n_e$ for various pairs of neutral and singly ionized species (Table~\ref{tab:ne}).
We have assumed depletions of $-$0.5 dex for Na and K (slightly less severe than the cold cloud values in Table~\ref{tab:refab}) to estimate SLV column densities for \ion{Na}{2} and \ion{K}{2}. 
The $n_e$ estimated from Na and K are then consistent with those obtained from Mg and Si, whose ionization balance should be dominated by simple photoionization and radiative recombination in the SLV gas ($\S$~\ref{sec-ne}).
All other SLV column densities in Table~\ref{tab:ne} are measured values.
The photoionization rate $\Gamma$ depends on the ambient radiation field and the optical depth (e.g., $A_{\rm V}$) within the cloud, and can be parameterized as $\Gamma$ = $C$ exp($-\alpha~A_{\rm V}$ + $\beta~A^2_{\rm V}$), where $C$ is the unattenuated rate and $\alpha$ and $\beta$ (typically in the range 3 to 5) are fit coefficients derived for a chosen grain model (e.g., van Dishoeck 1988).
Toward 23 Ori, the small total $E(B-V)$ = 0.11 suggests that the attenuation will generally be small within the individual clouds along the line of sight --- probably not by more than 25$\pm$5\% (for different species) at the centers of the two strongest SLV clouds.
We have listed the unattenuated rates for the WJ1 and Draine (1978) radiation fields in Table~\ref{tab:ne}.
The radiative recombination coefficient $\alpha$ depends on the temperature, and can be approximated by $\alpha$ = $\alpha_o$ ($T/T_o$)$^{-\eta}$ for $T$ $\la$
10$^4$ K, where $\eta$ $\sim$ 0.6--0.7 for most species of interest (P\'{e}quignot \& Aldrovandi 1986).
While the true values for $n_e$ will depend on the actual temperature and radiation field, the values listed in Table~\ref{tab:ne} (for $T$ = 100 K and the WJ1 field) can be compared in a relative sense, due to the similarities in the response of different species to changes in those parameters.
[These $n_e$ are average values for the two strongest SLV components; recall that $n_e$(17S) $\sim$ 2.6 $n_e$(18S), as estimated from $N$(\ion{Ca}{1})/$N$(\ion{Ca}{2}).]
We find that the computed average values of $n_e$ range over a factor of about 25 --- from about 0.04--0.05 cm$^{-3}$ for S, Mn, and Fe to about 0.09--0.26 cm$^{-3}$ for C, Na, Mg, Si, and K to about 0.95 cm$^{-3}$ for Ca.
We will adopt for the present $n_e$ $\sim$ 0.15$\pm$0.05 cm$^{-3}$, from Na, Mg, Si, and K.
With the $n_{\rm H}$ derived above, the fractional ionization $n_e$/$n_{\rm H}$ $\sim$ 0.01 --- significantly larger than the value expected if most electrons come from photoionization of C (i.e., $\sim$ 2 $\times$ 10$^{-4}$) --- which suggests that hydrogen is partially ionized (at a level $\sim$ 1\%) in the SLV gas.
We will discuss both the large range in $n_e$ and the apparently high fractional ionization in the SLV gas further below ($\S$~\ref{sec-ne}).
As discussed in Section~\ref{sec-region}, it is unlikely that the ambient radiation field is weaker than the "typical" galactic field (in which case the true $n_e$ and $n_e$/$n_{\rm H}$ would be lower than the values computed here). 

We may use the $n_e$ inferred from photoionization equilibrium, together with the $T$ and $n_{\rm H}$ derived above, to predict the column densities of the excited fine-structure levels of \ion{C}{2} and \ion{Si}{2} in the predominantly neutral SLV gas.
The equation of excitation equilibrium, for element X = C or Si, is (Bahcall \& Wolf 1968; York \& Kinahan 1979; Fitzpatrick \& Spitzer 1997):
\begin{displaymath}
\frac{N(X~II^*)}{N(X~II)} = \frac{n_e \gamma_{12}(e) + n_{\rm H} \gamma_{12}(H)}{A_{21} + n_e \gamma_{21}(e) + n_{\rm H} \gamma_{21}(H)},
\end{displaymath}
Here the collisional excitation rate for electrons is
\begin{displaymath}
\gamma_{12}(e) = \frac{8.63 \times 10^{-6}}{g_1 T^{1/2}}~\Omega_{12}~e^{-E_{12}/kT}~cm^3 s^{-1}.
\end{displaymath}
For \ion{C}{2}, $g_1$ = 2, $E_{12}$ = 0.0079 eV, the collision strength $\Omega_{12}$ ranges from 1.80 at $T$ = 100 K to 2.26 at $T$ = 3000 K to 2.85 at $T$ = 8000 K (Hayes \& Nussbaumer 1984),
the collision rate for H atoms $\gamma_{12}$(H) ranges from about 5 $\times$ 10$^{-10}$ cm$^3$ s$^{-1}$ at $T$ = 100 K to about 1.2 $\times$ 10$^{-9}$ cm$^3$ s$^{-1}$ at $T$ = 3000 K (York \& Kinahan 1979), and the radiative decay rate $A_{21}$ = 2.29 $\times$ 10$^{-6}$ s$^{-1}$ (Mendoza 1983).
For \ion{Si}{2}, $g_1$ = 2, $E_{12}$ = 0.036 eV, $\Omega_{12}$
$\sim$ 5.66 (within $\pm$ 0.12 for $T$ between 100 K and 20,000 K; Keenan et al. 1985), 
$\gamma_{12}$(H) ranges from about 2 $\times$ 10$^{-11}$ cm$^3$ s$^{-1}$ at $T$ = 100 K to about 10$^{-9}$ cm$^3$ s$^{-1}$ at $T$ = 3000 K (York \& Kinahan 1979), and $A_{21}$ = 2.17 $\times$ 10$^{-4}$ s$^{-1}$ (Nussbaumer 1977).
For $n_e$ = 0.15 cm$^{-3}$, $n_{\rm H}$ = 12 cm$^{-3}$, and $T$ = 100 K, the excited state populations are dominated by electron collisional excitation and radiative decay.
Using the SLV column densities of \ion{C}{2} and \ion{Si}{2} listed in Table~\ref{tab:h1abund}, we would predict log[$N$(\ion{C}{2}*)] $\sim$ 15.3 cm$^{-2}$ and log[$N$(\ion{Si}{2}*)] $\sim$ 10.8 cm$^{-2}$.
The best fit to the observed \ion{C}{2}* $\lambda$1335 profile, using the component structure found for \ion{S}{2} and \ion{Zn}{2}, yields log[$N$(\ion{C}{2}*)] $\sim$ 14.2 cm$^{-2}$, though values up to about 15.0 cm$^{-2}$ may be possible if some residual background remains at SLV velocities.
The predicted SLV column density of \ion{C}{2}* is thus a factor of 2 to 13 larger than the observed column density.
The predicted SLV log[$N$(\ion{Si}{2}*)] is slightly greater than the upper limit (10.7 cm$^{-2}$) obtained from scaling the derived \ion{Si}{2} component structure to fit the weak $\lambda$1264 absorption.
If we reduce $n_e$ by a factor 10 (a possibility discussed in Section~\ref{sec-ne} below), the predicted SLV column densities become log[$N$(\ion{C}{2}*)] $\sim$ 14.6 cm$^{-2}$ and log[$N$(\ion{Si}{2}*)] $\sim$ 9.9 cm$^{-2}$, more consistent with the observed values.
Similarly, Spitzer \& Fitzpatrick (1993) found the $n_e$ inferred from \ion{C}{2}* fine structure excitation to be systematically lower than those determined from Na ionization equilibrium in several components toward HD 93521.
The \ion{C}{2}* observed at SLV velocities thus may arise in the predominantly cool, neutral (H at most $\sim$ 1\% ionized) SLV gas.

The physical conditions derived for the SLV gas (averaged over the four SLV components) are summarized in Table~\ref{tab:h1prop}.
Note that the analysis of the \ion{C}{1} fine structure levels (from which $n_{\rm H}$ was obtained) assumes that the relative column densities for all three \ion{C}{1} fine structure levels in components 16S--19S are the same as those found for \ion{K}{1} and that each of $T$ and $n_{\rm H}$ are equal in those clouds.
The fractional ionization would thus be different for components 17S and 18S by the factor 2.6 difference in $n_e$ ($\S$~\ref{sec-modcomp}).
To obtain equal $n_e$/$n_{\rm H}$, one could, in principle, adjust the relative column densities for \ion{C}{1}* and \ion{C}{1}**, so that instead $T$ and/or $n_{\rm H}$ would be different for those two components.
The required column density adjustments would yield velocity offsets for \ion{C}{1}* and \ion{C}{1}** that differ by about 0.4 km s$^{-1}$ from that of \ion{C}{1} for the $\lambda$1260 multiplet observed with ECH-A.

\subsection{Physical Conditions in WLV Gas}
\label{sec-condwlv}

The weaker absorption from most trace neutral species (\ion{C}{1}, \ion{Na}{1}, and perhaps \ion{Mg}{1} are the only such species detected), the smaller values for $N$(\ion{Na}{1})/$N$(\ion{Ca}{2}) $\sim$ 0.3--2.8, the generally somewhat milder depletions, and the larger $b$-values for the individual \ion{Ca}{2} components suggest that the WLV gas is warmer and/or of lower density than the SLV gas.
We assume that the WLV gas is predominantly neutral, as fits to the G160M profiles of the strong \ion{O}{1} $\lambda$1302 and \ion{N}{1} $\lambda$1200 lines seem to indicate that the WLV column densities of those two species are unlikely to be much smaller than the values listed in Table~\ref{tab:h1abund}, though uncertainties as to the exact velocity offsets make it difficult to determine strong constraints.
The ratio $N$(\ion{Mg}{1})/$N$(\ion{Na}{1}) $\sim$ 5 (larger by less than a factor 2 than for the SLV gas) suggests that $T$ $\la$ 5000 K, since \ion{Mg}{1} would be significantly enhanced at higher temperatures due to dielectronic recombination\footnotemark.
\footnotetext{The WLV $N$(\ion{Mg}{1}) is based on the possible weak absorption in the \ion{Mg}{1}  $\lambda$2026 profile in the same velocity range as the strongest WLV \ion{Na}{1} components, and is not likely to be significantly underestimated.}
In the analysis below, we will consider possible WLV temperatures of 100 K and 3000 K, representative of cold and warm gas, respectively.

Analysis of the WLV \ion{C}{1} absorption yields relative fine structure level populations $N$(\ion{C}{1}**)/$N$(\ion{C}{1})$_{\rm tot}$ = 0.16$\pm$0.04 and $N$(\ion{C}{1}*)/$N$(\ion{C}{1})$_{\rm tot}$ = 0.40$\pm$0.08. 
While the nominal ratios do not fall along the curves in Jenkins \& Shaya (1979) for uniform clouds --- which would suggest that the individual WLV components are characterized by different physical properties --- the region permitted by the uncertainties does substantially intersect the curves.
If the WLV gas is cold ($T$ $\sim$ 100 K), then log($n_{\rm H}T$) $\sim$ 4.3 cm$^{-3}$K and $n_{\rm H}$ $\sim$ 200 cm$^{-3}$.
For warmer gas ($T$ $\sim$ 3000 K), log($n_{\rm H}T$) $\sim$ 4.7--4.8 cm$^{-3}$K and $n_{\rm H}$ $\sim$ 15--20 cm$^{-3}$. 
Estimates of the average WLV $n_e$ from the ratios of $N$(\ion{C}{1})/$N$(\ion{C}{2}), $N$(\ion{Na}{1})/$N$(\ion{Na}{2}), and $N$(\ion{Mg}{1})/$N$(\ion{Mg}{2}) (assuming a depletion of $-$0.3 dex for Na) range from about 0.02 cm$^{-3}$ for $T$ $\sim$ 100 K to about 0.2 cm$^{-3}$ for $T$ $\sim$ 3000 K (Table~\ref{tab:ne}; bottom).
For $T$ = 100 K, we thus would have $n_e$/$n_{\rm H}$ $\sim$ 0.0001 --- i.e., lower than the minimum value expected from photoionization of C alone.
For $T$ = 3000 K, we would have $n_e$/$n_{\rm H}$ $\sim$ 0.01 --- comparable to the fractional ionization inferred for the SLV components.
Since the higher temperature also yields better agreement with the observed \ion{C}{2}* and \ion{Si}{2}* absorption (below), we will adopt $T$ $\sim$ 3000 K, but note that the gas could be somewhat cooler.
For $T$ $\sim$ 3000 K, the density of the WLV gas would thus be slightly greater than that of the colder, higher column density SLV gas, the thermal pressure would be significantly higher, and the total thickness (0.7--0.9 pc) would be much smaller.
If the temperatures and radiation fields are similar among the WLV components, the ratios $N$(\ion{Na}{1})/$N$(\ion{Zn}{2}) suggest that $n_e$ may be higher by about a factor 3 for components 11W and 12W than for the other WLV gas (Fig.~\ref{fig:relind}).

As for the SLV gas, we may compare the observed \ion{C}{2}* and \ion{Si}{2}* absorption at WLV velocities with predictions based on the WLV $n_e$ inferred for different $T$.
Fits to the ECH-A \ion{C}{2}* $\lambda$1335 profile, using the WLV and SLV component structure found for \ion{S}{2} and \ion{Zn}{2}, yield log[$N$(\ion{C}{2}*)] $\sim$ 14.3--14.6 cm$^{-2}$ for the WLV gas\footnotemark.
\footnotetext{The column densities of \ion{C}{2}* for several of the WLV components are rather uncertain, due to blending with the weaker member of the $\lambda$1335 doublet for the SLV components 17S and 18S, whose \ion{C}{2}* column densities may be significantly larger (as discussed above).}
Using the WLV $N$(\ion{C}{2}) in Table~\ref{tab:h1abund}, for $T$ = 3000 K, $n_e$ = 0.2 cm$^{-3}$, and $n_{\rm H}$ = 20 cm$^{-3}$, we predict log[$N$(\ion{C}{2}*)] $\sim$ 14.25 cm$^{-2}$, while for $T$ = 100 K, $n_e$ = 0.02 cm$^{-3}$, and $n_{\rm H}$ = 200 cm$^{-3}$ we predict log[$N$(\ion{C}{2}*)] $\sim$ 14.5 cm$^{-2}$.
For the same two combinations of $T$, $n_e$, and $n_{\rm H}$, we predict log[$N$(\ion{Si}{2}*)] $\sim$ 11.3 cm$^{-2}$ and 10.0 cm$^{-2}$, respectively, compared to the observed log[$N$(\ion{Si}{2}*)] $\sim$ 11.5 (most of the observed \ion{Si}{2}* absorption appears to be at WLV velocities).
If the gas is cold, the dominant processes are excitation by H atoms (because of the high $n_{\rm H}$) and radiative decay;
collisions with electrons dominate the excitation in warmer gas.
These comparisons indicate that the \ion{C}{2}* and \ion{Si}{2}* absorption at WLV velocities could arise in the predominantly neutral WLV gas, if that gas is warm ($T$ $\sim$ 500--5000 K).
While $N$(\ion{C}{2}*)/$N$(\ion{C}{2}) is of order 0.01 for most of the SLV and WLV components [assuming $N$(\ion{C}{2})/$N$(\ion{S}{2}) $\sim$ 10], the ratio appears to be significantly higher for the WLV components 10W, 11W, and (perhaps) 12W --- which may indicate a higher $n_e$ in those components, as suggested also by the $N$(\ion{Na}{1})/$N$(\ion{Zn}{2}) ratio (above).

The physical conditions derived for the WLV gas (averaged over the seven WLV components, for the two representative temperatures) are summarized in Table~\ref{tab:h1prop}.
Much of the characterization of the WLV components depends on the \ion{C}{1} fine-structure level populations, which were derived from fits to the apparent blueward wing in the G160M profiles of the $\lambda$1560 and $\lambda$1656 multiplets.
The WLV column densities are consistent with the limits obtained from the weaker $\lambda$1260 multiplet observed with ECH-A, but higher resolution data for one of the stronger multiplets are needed to confirm the relatively high \ion{C}{1} excited state populations in the WLV gas.
We note, however, that the components near 0 km s$^{-1}$ toward $\zeta$ Ori (where we have high S/N ECH-A spectra) also have relatively high excited state populations, and pressures significantly higher than those in the higher column density components near +25 km s$^{-1}$ along that line of sight (Welty et al. 1999c).
Higher resolution spectra of \ion{N}{1}, \ion{O}{1}, \ion{Na}{1}, and \ion{Mg}{1} $\lambda$2853 would also be useful for more precise characterization of the individual WLV components.

\section{Predominantly Ionized Gas}
\label{sec-ionized}

\subsection{Ionized Gas Component Structure}
\label{sec-compion}

In addition to the strong absorption at low velocities, the strongest lines in the UV spectra also show evidence for the presence of additional lower column density components not seen in the optical lines --- at low, intermediate, and high velocities (Figures~\ref{fig:uv} and~\ref{fig:highv}).
As discussed below, the gas at high and intermediate velocities is predominantly ionized; as noted above, there is some ionized gas at low velocities in addition to the dominant neutral gas.
The HV gas (``Orion's Cloak''), at $-$108 km s$^{-1}$ $\la$ $v$ $\la$ $-$83 km s$^{-1}$, is seen in the strong lines of \ion{Mg}{2}, \ion{Si}{2}, \ion{Si}{3}, \ion{Al}{2}, \ion{N}{2}, and \ion{C}{2}; weak absorption is also detected for \ion{C}{2}*, \ion{Al}{3}, and perhaps \ion{S}{2} (Figure~\ref{fig:highv}). 
Four components (1H--4H) are present in the echelle profiles of \ion{C}{2}, \ion{Si}{2}, and \ion{Mg}{2}.
The IV gas, at $-$43 km s$^{-1}$ $\la$ $v$ $\la$ $-$4 km s$^{-1}$ (components 5I--9I), is seen most prominently in the strong lines of \ion{Si}{3}, \ion{C}{2}, \ion{Mg}{2}, and (probably) \ion{N}{2}.
The component parameters listed in Table~\ref{tab:h2cmp} for the generally  weaker IV and HV components for several species were derived via a simultaneous fit to the echelle profiles of the stronger lines.
That same component structure was used to fit the weak HV and/or IV absorption from \ion{Al}{3}, \ion{S}{2}, and \ion{C}{2}*.
For the high and intermediate velocity gas observed with G160M, we adopted the component structure determined from the echelle profiles of \ion{C}{2}, \ion{Mg}{2}, and \ion{Si}{2}.
Table~\ref{tab:h2abund} gives the resulting total column densities for the ionized HV and IV gas. 

Unfortunately, however, we have no high resolution spectra to define the component structure for the ionized gas at low velocities (apart from the ECH-B spectra of \ion{Al}{3}, which shows weak components at $-$6 and +20 km s$^{-1}$).
This lack of component information particularly hinders our analysis of the strong \ion{N}{2} $\lambda$1083 and \ion{Si}{3} $\lambda$1206 lines.
For \ion{Si}{3}, we may obtain a lower limit to the IV + LV column density, log[$N$(\ion{Si}{3})] $>$ 13.4 cm$^{-2}$, by integrating the ``apparent'' optical depth over the line profile (Hobbs 1971; Savage \& Sembach 1991), but similar limits for \ion{N}{2} are uncertain, because the true background level in the {\it Copernicus} spectrum is poorly known.
We attempted to fit the \ion{Si}{3} and \ion{N}{2} absorption by using the HV and IV component structure found for \ion{C}{2} (which, like \ion{Si}{3} and \ion{N}{2}, seems to be the dominant ion in the ionized gas toward 23 Ori) together with either a ``minimal'' low velocity component structure based on the components found for \ion{Al}{3} and \ion{Si}{2}* or the WLV + SLV structure found for \ion{C}{2}.
The resulting HV column densities for \ion{Si}{3} and \ion{N}{2} are fairly well determined; the IV column densities are somewhat less so. 
For both species, $N$(WLV) appears to be greater than $N$(SLV), but the total $N$(WLV + SLV) is very uncertain, so we have listed only (fairly conservative) upper limits in Table~\ref{tab:h2abund}.
The relatively weak absorption at low velocities ($-$5 km s$^{-1}$ $\la$ $v$ $\la$ +6 km s$^{-1}$) seen in the G160M spectra of \ion{Si}{2}* and \ion{S}{3} (blended with \ion{C}{1} and \ion{Si}{2}) and of the higher ions \ion{C}{4}, \ion{Si}{4}, and \ion{N}{5} (Figure~\ref{fig:highv}) can be acceptably fitted in each case with a single broad component (Table~\ref{tab:high}) --- though differences in central velocity and the relatively large $b$-values suggest that additional substructure may be present.
The weak absorption from the higher ions was not detected in the {\it Copernicus} spectra reported by Cowie et al. (1979), due primarily to the lower S/Ns achieved for those earlier spectra.
In principle, \ion{P}{3} could be present in the wing of the strong \ion{C}{2} $\lambda$1334 line, but uncertainties as to the appropriate velocity and line width [and uncertainties in the SLV $N$(\ion{C}{2})] make it difficult to determine reliable and interesting limits.

\subsection{High-Velocity Gas}
\label{sec-hvgas}

\subsubsection{HV Gas Abundances and Depletions}
\label{sec-abundhv}

No absorption from \ion{O}{1} or \ion{N}{1} is seen at high velocities, so the HV gas is primarily ionized (Trapero et al. 1996).
The limit on $N$(\ion{H}{1}) in Table~\ref{tab:h2abund} assumes that \ion{O}{1} is dominant in any neutral gas [which appears to be valid whether or not the gas is in equilibrium (e.g., Schmutzler \& Tscharnuter 1993)] and that O is undepleted there.
We do not detect high-velocity absorption from the higher ions \ion{Si}{4} and \ion{C}{4}, suggesting that only a narrow range of ionization states is significantly populated for any element.
Even so, determination of depletions is complicated by our inability to observe
all the potentially significant ionization states of many of the elements.
In principle, we would like to determine $N$(\ion{H}{2}) by summing the contributions from the relevant ionization states of typically undepleted elements (e.g., C, N, and S), and then compare with the abundances of typically depleted elements (e.g., Si, Mg, Fe, and Al) to determine the level of depletion.
We do not have data for \ion{C}{3}, \ion{N}{3}, \ion{Fe}{3}, \ion{Mg}{3}, or \ion{Al}{4}, however.
Only for Si do we have essentially complete coverage --- i.e., detections of \ion{Si}{2} and \ion{Si}{3} and significant limits on \ion{Si}{1} and \ion{Si}{4}.
The column densities of \ion{C}{2}, \ion{N}{2}, and \ion{Si}{2} + \ion{Si}{3} all yield estimates for log[$N$(\ion{H}{2})] of about 17.7 cm$^{-2}$, assuming no depletion for C, N, and Si. 
While a similar estimate from $N$(\ion{S}{2}) + $N$(\ion{S}{3}) is consistent with that value, the uncertainties would permit a somewhat higher $N$(\ion{H}{2}), by perhaps a factor $\sim$ 1.5, or 0.2 dex (assuming S undepleted)\footnotemark.
\footnotetext{We have combined our G160M spectra of the \ion{S}{3} $\lambda$1190 line with similar data from the {\it HST} archive to obtain a smaller upper limit on $N$(\ion{S}{3}) in the HV gas than the one given by Trapero et al. (1996).}
Some \ion{C}{3} and \ion{N}{3} could thus be present, as found for the HV gas in several other lines of sight in Orion (Cowie et al. 1979; Welty et al. 1999c).
In the HV gas toward $\zeta$ Ori, for example, Welty et al. (1999c) find $N$(\ion{C}{3})/$N$(\ion{C}{2}) $\sim$ 2.4 and $N$(\ion{N}{3})/$N$(\ion{N}{2}) $\sim$ 1.3.
The ratios of $N$(\ion{Al}{3})/$N$(\ion{Al}{2}), $N$(\ion{Si}{3})/$N$(\ion{Si}{2}), and $N$(\ion{S}{3})/$N$(\ion{S}{2}) are all significantly smaller in the HV gas toward 23 Ori than those found toward $\zeta$ Ori, however, so that there is likely to be relatively less \ion{C}{3} and \ion{N}{3} toward 23 Ori.
If log[$N$(\ion{H}{2})] is greater than 17.7 cm$^{-2}$, then Si would be slightly depleted (but by less than 0.2 dex); likewise, C could not be depleted by more than 0.2 dex [and $N$(\ion{C}{3}) must be less than $N$(\ion{C}{2})].
The ratio [$N$(\ion{Al}{2}) + $N$(\ion{Al}{3})]/$N$(\ion{C}{2}), however, is less than half the solar value, so that Al appears to be slightly depleted even for log[$N$(\ion{H}{2})] $\sim$ 17.7 cm$^{-2}$. 
(The relatively low ratios of $N$(\ion{Si}{3})/$N$(\ion{Si}{2}) and $N$(\ion{Al}{3})/$N$(\ion{Al}{2}) suggest that there is unlikely to be an appreciable amount of \ion{Al}{4} present.)
The depletion of Fe cannot be reliably estimated, since comparisons with the HV gas toward $\zeta$ Ori suggest that a substantial fraction of the gas phase Fe may be in \ion{Fe}{3} (Welty et al. 1999c).
Slight differences in various ion ratios for the individual HV components may be due to differences in ionization for those individual components, as \ion{C}{2} appears to be the dominant ion for C, but \ion{Si}{2} and \ion{Mg}{2} are not the dominant ions of Si and Mg.

\subsubsection{HV Gas Physical Conditions}
\label{sec-condhv}

Cowie et al. (1979) ascribed the high velocity gas seen with {\it Copernicus} toward a number of Orion stars (including 23 Ori) to a radiative shock due to one or more recent supernovae in that region.
They estimated a radius of about 120 pc (i.e., outside the expanding neutral shell) and an age of about 3 $\times$ 10$^5$ yr for this apparently large scale feature, as well as a pre-shock ambient density of about 3 $\times$ 10$^{-3}$ cm$^{-3}$.
Higher resolution GHRS spectra covering additional species have enabled a more detailed understanding of the composition and properties of this high-velocity gas (Trapero et al. 1996; Welty et al. 1999c).
Toward 23 Ori, the newly acquired ECH-A spectra of \ion{C}{2}, \ion{C}{2}*, and \ion{Si}{2} have yielded estimates for the temperatures and electron densities in the {\it individual} HV components.

If the HV gas were in collisional ionization equilibrium, the observed ratios of various ions would be roughly consistent with an equilibrium temperature of about 25,000 K (using, for example, the calculations of Sutherland \& Dopita 1993).
The widths of the individual components seen in \ion{Mg}{2}, \ion{Si}{2}, and \ion{C}{2}, however, yield firm upper limits $T_{\rm max}$ $\la$ 14,000 K in all cases, and comparisons of $b$(\ion{C}{2}) vs. $b$(\ion{Mg}{2}) and $b$(\ion{Si}{2}) suggest that the temperatures of the individual components are in fact 6,000--12,000 K (Table~\ref{tab:h2prop}).
The gas is therefore not in equilibrium, and has cooled more rapidly than it could recombine, presumably after having been shocked.
We have found, however, that shock models are not entirely successful in reproducing the column densities, ion ratios, and line profiles observed for the HV gas in the Orion region (Welty et al. 1999c).
For example, models with $v_{\rm shock}$ $\sim$ 100 km s$^{-1}$ generally produce significantly more HV \ion{Si}{4} than the limits observed toward both 23 Ori and $\zeta$ Ori.
Furthermore, models of cooling gas (e.g., Schmutzler \& Tscharnuter 1993) appear to predict significantly lower ionization for $T$ $<$ 10$^4$ K than is observed for the HV gas in those two sightlines --- though including the effects of an external radiation field may yield better agreement.

In this ionized HV gas, the excited state \ion{C}{2}* is populated primarily via collisions with electrons, and the ratio $N$(\ion{C}{2}*)/$N$(\ion{C}{2}) $\sim$ 0.03 (for the three strongest components) can be used to estimate $n_{\rm e}$ (and thus $n_{\rm H}$) from the equation of excitation equilibrium, since the temperature is known ($T$ $\sim$ 8000$\pm$2000 K):
$n_e$ = 0.186$T^{0.5}$[$N$(\ion{C}{2}*)/$N$(\ion{C}{2})].
The $n_{\rm e}$ = $n_{\rm H}$ derived for the three main HV components lies in the range 0.4--0.5 cm$^{-3}$, about 150 times the pre-shock density estimated by Cowie et al. (1979).
The thermal pressure, as measured by log(2$n_{\rm e}T$), is thus 3.7--4.0 cm$^{-3}$K in each component.
Estimating $N$(\ion{H}{2}) from $N$(\ion{C}{2}), assuming no depletion of C and no appreciable \ion{C}{3} [neither would increase $N$(\ion{H}{2}) by more than a factor 1.5], the individual component thicknesses range from about 0.005 pc to about 0.12 pc.

\subsection{Intermediate-Velocity Gas}
\label{sec-ivgas}

\subsubsection{IV Gas Abundances and Depletions}
\label{sec-abundiv}
 
As for the HV gas, \ion{O}{1} and \ion{N}{1} are not detected at intermediate velocities (the G160M profiles of the strong $\lambda$1302 and $\lambda$1200 lines can be completely accounted for by the WLV and SLV components), and $N$(\ion{Si}{3}) $>$ $N$(\ion{Si}{2}), indicating that the IV gas also is predominantly ionized.
Some of the absorption from the higher ions \ion{S}{3}, \ion{Si}{4}, \ion{C}{4}, and \ion{N}{5} may also be associated with the IV gas, though lower ionization states are dominant (Table~\ref{tab:h2abund}).
From $N$(\ion{Si}{2}) + $N$(\ion{Si}{3}) and $N$(\ion{S}{2}), we estimate that log[$N$(\ion{H}{2})] $\ga$ 17.7 cm$^{-2}$ for the IV components; the less certain $N$(\ion{N}{2}) suggests that log[$N$(\ion{H}{2})] could be as high as about 18.0 cm$^{-2}$.
Since the $N$(\ion{H}{2}) estimated from \ion{C}{2} alone is smaller by 0.2--0.5 dex than the above estimates, there may be some \ion{C}{3} in the IV gas, as found toward several other Orion stars (Cowie et al. 1979).
Comparisons of various ion ratios among those Orion lines of sight suggests, however, that $N$(\ion{C}{3}) is probably less than $N$(\ion{C}{2}) in the IV gas toward 23 Ori --- so that $N$(\ion{N}{2}) may be overestimated in Table~\ref{tab:h2abund}.
The ratio [$N$(\ion{Al}{2})+$N$(\ion{Al}{3})]/[$N$(\ion{Si}{2})+$N$(\ion{Si}{3})] is similar to the value found for the HV gas, suggesting that Al is comparably depleted in the IV gas [unless there is an appreciable amount of Al in (unobservable) \ion{Al}{4}].
Higher resolution spectra of \ion{N}{2}, \ion{Si}{3}, and \ion{S}{3}, as well as new spectra of \ion{C}{3}, \ion{N}{3}, and \ion{Fe}{3}, would enable more reliable estimates for the abundances and depletions in the IV gas.

\subsubsection{IV Gas Physical Conditions}
\label{sec-condiv}

Ionized gas at intermediate velocities is seen toward a number of stars in the Orion region (Cowie et al. 1979), suggesting that the IV gas is a large-scale, regional feature --- not intimately associated with the individual stars. 
Cowie et al. found the IV gas toward several of the Orion stars to be somewhat more highly ionized than the HV gas and so concluded that the IV gas is photoionized by (but at some distance from) the Orion association stars. 
In the IV gas toward 23 Ori, both the slightly higher $N$(\ion{Al}{3})/$N$(\ion{Al}{2}) and $N$(\ion{Si}{3})/$N$(\ion{Si}{2}) ratios (relative to those in the HV clouds; Table~\ref{tab:h2abund}) and the (possible) associated absorption from \ion{C}{3}, \ion{S}{3}, \ion{Si}{4}, \ion{C}{4}, and \ion{N}{5} would be consistent with somewhat higher ionization.
The $b$-values adopted for most of the IV \ion{C}{2}, \ion{Mg}{2}, and \ion{Si}{2} components suggest, however, that the temperatures are comparable to those in the HV gas.
Analysis of the \ion{C}{2} fine structure excitation, assuming $T$ $\sim$ 8000 K, yields $n_e$ $\sim$ 1.5--5.0 cm$^{-3}$, and thus log(2$n_e$$T$) $\sim$ 4.4--4.9 cm$^{-3}$K, for several of the IV components --- somewhat higher densities and thermal pressures than the values found for the HV gas (Table~\ref{tab:h2prop}).
Several of the IV components have significantly lower $N$(\ion{H}{2}), however, and so are much thinner ($\la$ 0.001 pc) than the HV components --- suggesting that they may arise in thin filaments or sheets.

\subsection{Ionized Gas at Low Velocity}
\label{sec-abhigh}

The strong absorption seen for \ion{N}{2} and \ion{Si}{3} and the weak, broad absorption due to \ion{S}{3}, \ion{C}{4}, \ion{Si}{4}, and \ion{N}{5} indicate that there is some ionized gas at low velocities (Figure~\ref{fig:highv} and Tables~\ref{tab:h2abund} and~\ref{tab:high}).
The somewhat narrower \ion{Si}{2}* absorption (centered at about +12 km s$^{-1}$) and the low-velocity \ion{C}{2}* absorption may be largely due to neutral gas, however ($\S\S$~\ref{sec-phsne} and~\ref{sec-condwlv}).
Because some of the broad, unresolved absorption from the higher ions may be associated with the IV gas, we have listed the total high ion column densities in a separate column in Table~\ref{tab:h2abund}.
As discussed above, the low-velocity column densities for \ion{N}{2} and \ion{Si}{3} are rather uncertain, but the limit for \ion{N}{2} suggests that log[$N$(\ion{H}{2})] $\la$ 19.3 cm$^{-2}$.
The observed column density of \ion{S}{3} provides a lower limit of  log[$N$(\ion{H}{2})] $\ga$ 18.2 cm$^{-2}$ (though some of this could be at intermediate velocities).
The $b$-values found from single component fits to the high ion profiles --- 16--17 km s$^{-1}$ for \ion{S}{3} and \ion{Si}{4}, 23--24 km s$^{-1}$ for \ion{C}{4} and \ion{N}{5}) --- provide upper limits for the temperature of 4-5 $\times$ 10$^5$ K.
The $b$-value for the +20 km s$^{-1}$ component of \ion{Al}{3}, however, implies $T$ $\la$ 3 $\times$ 10$^4$ K, and it seems likely that the low-velocity ionized gas is comprised of several components.
There is no obvious correlation between central velocity and ionization potential $\chi_{\rm ion}$ for these high ions.
We will briefly examine several possible origins for the high ions in Section~\ref{sec-high}.

\section{Discussion}
\label{sec-disc}

\subsection{The Orion Region}
\label{sec-region}

General surveys of interstellar \ion{Na}{1} and \ion{Ca}{2} absorption have revealed complex multi-component absorption-line profiles toward a number of the brighter stars in the Orion-Eridanus region (Hobbs 1969, 1978b; Marschall \& Hobbs 1973; Beintema 1975; Welty et al. 1994, 1996).
Frisch et al. (1990) obtained \ion{Ca}{2} and/or \ion{Na}{1} spectra of a number of stars in that region in order to locate the various velocity components along the lines of sight.
We will provide a brief update of the latter work, making use of additional (as yet unpublished) optical spectra and the parallaxes from {\it Hipparcos}.
In the southern part of this region (within about 5$\arcdeg$ of $\lambda$ Eri), weak \ion{Ca}{2} absorption at +20 km s$^{-1}$ $\la$ $v$ $\la$ +30 km s$^{-1}$ is seen in the spectra of several stars at distances $d$ $\sim$ 170 pc, but is not detected (at the level of several m\AA) toward stars at $d$ $\la$ 100 pc.
Nearer to 23 Ori, several weak components are seen in both \ion{Na}{1} and \ion{Ca}{2} toward $\gamma$ Ori (2$\fdg$9 N of 23 Ori), at $d$ $\sim$ 75 pc, but no \ion{Na}{1} is detected toward HD 35134 (0$\fdg$8 S), at $d$ $\sim$ 130 pc.
Stronger absorption at those velocities, with significantly larger $N$(\ion{Na}{1})/$N$(\ion{Ca}{2}) ratios, is seen over most of the region toward stars beyond about 270 pc --- including HD 34959, 25 Ori, 30 Ori, and 33 Ori, all within 2$\arcdeg$ of 23 Ori.
Weaker absorption at $-$10 km s$^{-1}$ $\la$ $v$ $\la$ +20 km s$^{-1}$, with lower $N$(\ion{Na}{1})/$N$(\ion{Ca}{2}) ratios, is also seen toward many stars beyond that point.
While there are significant differences in the column densities and velocities of individual components seen toward the stars beyond 270 pc, the overall general similarities among the profiles suggest that the bulk of the observed gas is not closely associated with the individual stars, but is likely to be located in the foreground, at 130 pc $\la$ $d$ $\la$ 270 pc.
The observed differences are not unexpected, in light of the variations commonly seen on much smaller spatial scales (Watson \& Meyer 1996) and of the energetic activity in the region.

Comparisons among the 21 cm, H$\alpha$, soft X-ray, and 100 $\mu$m emission maps of the Orion-Eridanus region have suggested the presence of a large, expanding shell, predominantly neutral but with an inner ionized zone, encompassing a bubble of hot gas at log($T$) $\sim$ 6.2 K (Heiles 1976; Reynolds \& Ogden 1979; Burrows et al 1993; Brown et al. 1995; Snowden et al. 1995; see also Cowie et al. 1979).
If that picture is accurate, 23 Ori would lie within the bubble, near its boundary at low Galactic latitudes.
The SLV (and perhaps the WLV) components could arise in the expanding shell, based on the distance estimates given above and on the general agreement in velocity with the shell's 21 cm emission near the line of sight to 23 Ori (Brown et al. 1995).
The WLV gas may represent the more ``active'' part of the shell, in view of its higher temperature and thermal pressure, more negative velocities, and less severe depletions.
The total \ion{H}{1} column density, 5.5 $\times$ 10$^{20}$ cm$^{-2}$, is also very similar to that inferred for the shell (Snowden et al. 1995).
The ionized low-velocity gas could then represent the inner ionized region of the shell, estimated to contain about 10-15\% of the total hydrogen in the shell (Reynolds \& Ogden 1979).
If 23 Ori itself is immersed in the hot, low density gas contained within the shell, we would not expect a localized, associated circumstellar \ion{H}{2} region --- consistent with the lack of absorption due to ionized gas observed within 10--15 km s$^{-1}$ of the stellar radial velocity.
(In any case, a B1 V star would typically have only a very small associated \ion{H}{2} region.)

If the gas seen in absorption toward 23 Ori (and toward other Orion OB1 stars) is not, in general, closely associated with the individual background stars, then the radiation field incident on the clouds will not be dominated by the flux from those individual stars, but will be due to the ensemble of stars in the region.
Wall et al. (1996) have estimated the radiation field in the Orion region by assuming all the stars in the Orion OB1 and $\lambda$ Ori associations to be located at a distance of 450$\pm$10 pc, with some allowance for extinction effects.
In that model, the field along the line of sight to 23 Ori is roughly twice the ``average'' interstellar field of Mathis, Mezger, \& Panagia (1983) at d $\sim$ 450 pc --- and presumably less than that for smaller distances.
Another estimate for the radiation field may be obtained by considering the relationship between $N$(H$_2$) and $N$(\ion{H}{1}), as a function of temperature, density, and radiation field (Reach, Koo, \& Heiles 1994).
Using equation (3) of Reach et al., with the $N$(H$_2$) = 2 $\times$ 10$^{18}$ cm$^{-2}$ and $N$(\ion{H}{1}) = 5.1 $\times$ 10$^{20}$ cm$^{-2}$ observed for the SLV clouds and the $T$ = 100 K and $n_{\rm H}$ $\sim$ 10-15 cm$^{-3}$ derived above ($\S$~\ref{sec-condslv}), we would find the radiation field to be enhanced by a factor $\chi$ $\sim$ 2.5-3.8 over that average interstellar field.
That equation, however, is applicable only if the H$_2$ has reached its equilibrium abundance.
Consideration of the H$_2$ formation time (Reach et al. 1994), for the $T$ and $n_{\rm H}$ obtained for the SLV clouds, yields $t_{\rm form}$ $\sim$ 7 $\times$ 10$^8$ yr (i.e., much longer than the apparent age of the Orion-Eridanus bubble) --- suggesting that the H$_2$ has not reached equilibrium.
If the eventual equilibrium $N$(H$_2$) is higher than the currently observed value, then the true radiation field is likely to be somewhat weaker, since $N$(H$_2$) $\propto$ $\chi^{-2}$.
We have assumed in discussions above that the radiation field incident on the clouds toward 23 Ori can be approximated by the ``average'' WJ1 interstellar field defined by de Boer, Koppenaal, \& Pottasch (1973), which agrees with the Mathis et al. field to within 20\% below 1300 \AA.
In light of the above estimates, the radiation field could be somewhat stronger, but probably not by more than a factor 2. 
The field could also be somewhat ``harder'', in view of the number of bright O stars in the region and the shallow far-UV extinction observed toward some Orion stars (e.g., Fitzpatrick \& Massa 1990).
Toward Orion, the field is unlikely to be weaker than the WJ1 field (except in significantly shielded regions).

\subsection{Ionization and Electron Densities in the SLV Gas}
\label{sec-ne}

Because the SLV components exhibit detectable absorption from an unusually large number of trace neutral species (including \ion{Si}{1}, \ion{Ca}{1}, \ion{Fe}{1}, and perhaps \ion{Mn}{1}), we have many independent diagnostics for the electron density in the SLV gas.
As described above ($\S$~\ref{sec-condslv}), the estimates for $n_e$ resulting from the assumption of simple photoionization and radiative recombination are generally relatively large (implying $n_e$/$n_{\rm H}$ $\sim$ 0.01), but cover a wide range, from about 0.04 cm$^{-3}$ to about 0.95 cm$^{-3}$ (Table~\ref{tab:ne}).
Detailed studies of other lines of sight show analogous, seemingly systematic offsets among the $n_e$ inferred from different diagnostics --- but with subtle variations among the element-to-element offsets for the different lines of sight (Cardelli, Sembach, \& Savage 1995; Fitzpatrick 1996, 1998; Snow et al. 1996; Fitzpatrick \& Spitzer 1997;  Welty et al. 1999a, 1999b).
Several of those studies also find evidence for some ionization of H in predominantly warm, diffuse, neutral gas. 
The relatively high $n_e$/$n_{\rm H}$ in the SLV components toward 23 Ori, however, apparently occurs in colder, denser gas.
Accurate, consistent estimates for both $n_e$ and $n_e$/$n_{\rm H}$ are necessary to understand what processes are responsible for the ionization (simple photoionization, cosmic ray ionization, charge exchange processes, decaying neutrinos, ...) and recombination (radiative, dielectronic, charge exchange, ...) in these interstellar clouds.
In studies of diffuse clouds, the fractional ionization is often assumed to be $\sim$ 3 $\times$ 10$^{-4}$ (due to photoionization of C and other heavy elements) in order to estimate $n_{\rm H}$ from $n_e$ --- but that assumption may be invalid in many cases.
We will discuss these issues in more depth in a separate paper (Welty et al. 1999b), but briefly present some of the possibly relevant processes here as they apply to the SLV gas toward 23 Ori.

\begin{itemize}

\item{The neutral species \ion{Ca}{1} and \ion{Mg}{1} can be enhanced via dielectronic recombination, for temperatures $\ga$ 3000 K and $\ga$ 6000 K, respectively --- which would then yield apparent overestimates of $n_e$ assuming only radiative recombination.
As discussed above ($\S$~\ref{sec-condslv}), however, the temperatures in the SLV components are likely to be $\sim$ 100 K.
While the $N$(\ion{Ca}{1})/$N$(\ion{Ca}{2}) ratio often yields much higher estimates for $n_e$ than do other neutral-first ion ratios, the \ion{Ca}{1} seems to arise predominantly in cool gas in the Galactic ISM, so its apparent enhancement does not, in general, seem to be due to dielectronic recombination (Welty et al. 1999b).}

\item{The derived column densities could be in error, due either to errors in the $f$-values or to problems with the profile-fitting analysis.
In many cases, however, the column densities were derived from several lines of different strength --- including some very weak lines --- which yielded consistent results.
In most cases, the nominal uncertainties in the $f$-values are less than about 0.1 dex (Morton 1991), and a number of the $f$-values have more recent, more accurate determinations (see the Appendix).}

\item{The photoionization and/or recombination rates for some species could be in error, due to inaccurate cross-sections or to differences between the true and adopted radiation fields.
A recent estimate of the rates for the \ion{Fe}{1}--\ion{Fe}{2} equilibrium, for example, suggests that $\Gamma$/$\alpha$ is roughly a factor 2 higher than previously thought --- apparently due largely to a more accurate and consistent accounting for the strong resonances in the photoionization cross-section of \ion{Fe}{1} (Nahar, Bautista, \& Pradhan 1997; see also Bautista, Romano, \& Pradhan 1998).
A higher $\Gamma$/$\alpha$ would increase the (currently low) $n_e$ inferred from the $N$(\ion{Fe}{1})/$N$(\ion{Fe}{2}) ratio.
For most of the species re-considered by Bautista et al., the new $\Gamma$'s are larger than previously estimated, but new, consistent $\alpha$'s are needed in order to obtain revised $\Gamma$/$\alpha$ ratios.
New estimates for $\alpha$ have been determined for C and Si (Nahar \& Pradhan 1997), yielding $\Gamma$/$\alpha$ ratios slightly smaller than previous values, as both $\Gamma$ and $\alpha$ increased by similar amounts.
Self-consistent re-appraisals of the rates for other elements would be very useful --- though the net effect on the $\Gamma$/$\alpha$ ratios may generally be minor, if the issue is primarily one of accounting for the resonances in the photoionization cross-section. 

In \ion{H}{1} regions, the photoionization rates for species with relatively large ionization potentials (\ion{C}{1}, \ion{S}{1}, \ion{P}{1}, \ion{Cl}{1}, \ion{Ca}{2}) depend on the (poorly known) radiation field in the narrow interval between $\chi_{\rm ion}$ and the Lyman limit at 13.6 eV.
The rates for those species would thus be somewhat higher if the field in the Orion region is harder than the WJ1 field --- as illustrated in Table~\ref{tab:ne} for the Draine (1978) radiation field, which is stronger than the WJ1 field below about 1400\AA.
The rates for other neutrals with lower $\chi_{\rm ion}$ are dominated by the field at longer wavelengths, and thus depend much less on the choice of radiation field.
The neutral species \ion{C}{1} and \ion{S}{1} do appear to be relatively enhanced for lines of sight characterized by moderate reddening [E(B-V) $\ga$ 0.3] and steep far-UV extinction, where the far-UV field is preferentially suppressed (Welty 1989; Welty et al. 1991; see also Jenkins \& Shaya 1979).
The expected magnitude of this effect depends on the grain scattering properties, and will be smaller if the grains are very forward scattering in the far UV.
This effect should be insignificant for the clouds toward 23 Ori, however, since the color excess is small and the far-UV extinction does not appear to be steeper than ``average'' (Savage et al. 1985; Papaj, Wegner, \& Krelowski 1991).
Furthermore, differences in the radiation field in the far-UV would tend to affect C and S similarly (relative to other elements) --- which does not appear to be the case toward 23 Ori --- though other processes may affect both (see below).}

\item{In principle, a dependence of depletion on local density $n_{\rm H}$ could produce differences in inferred $n_e$ between mildly and severely depleted elements.
If the density of trace neutral species is proportional to $n_{\rm H}^2$ (assuming, for example, $n_e$ $\propto$ $n_{\rm H}$; though see Welty et al. 1999b), then the trace neutrals will be more concentrated in the denser regions of a cloud.
(Note that this implies that all the $n_e$ estimates from trace/dominant ratios underestimate the $n_e$ in the denser regions.)
Integrated over the whole cloud, the ratios $N$(\ion{X}{1})/$N$(\ion{X}{2}) for depleted elements will be dominated by the less dense regions, while the ratios for less depleted elements will be more heavily weighted by the denser regions --- which could lead to systematic differences in derived $n_e$, if $n_e$ depends (in some way) on $n_{\rm H}$.
In the SLV clouds, however, there does not appear to be any relationship between elemental depletion $\delta_X$ (as determined from the dominant species) and the $n_e$ inferred from $N$(\ion{X}{1})/$N$(\ion{X}{2}).
For example, the lowest $n_e$ are derived from S, Mn, and Fe --- which have significantly different depletions [column 5 ($n_e$) versus column 6 ([X/C]$_{\rm II}$ = D$_{\rm X}$ $-$ D$_{\rm C}$) in Table~\ref{tab:ne}].}

\item{Fitzpatrick \& Spitzer (1997) derived systematically higher $n_e$ from the $N$(\ion{C}{1})/$N$(\ion{C}{2}) ratio (versus several other ratios) in the halo clouds toward HD 215733, under the assumption of simple photoionization and radiative recombination\footnotemark.
They conjectured that dissociative recombination of CH$^+$ might have enhanced the \ion{C}{1}, though CH$^+$ has not been observed toward HD 215733.
At $T$ $\sim$ 100 K, the radiative recombination rate for \ion{C}{2} to \ion{C}{1} is $\alpha$ $\sim$ 9 $\times$ 10$^{-12}$~cm$^3$s$^{-1}$~(Nahar \& Pradhan 1997) and the dissociative recombination rate for CH$^+$ is $\beta$ $\sim$ 3.3 $\times$ 10$^{-7}$~cm$^3$s$^{-1}$~(Mitchell \& McGowan 1978).
For the SLV clouds toward 23 Ori, log[$N$(\ion{C}{2})] $\sim$ 16.95 cm$^{-2}$ and log[$N$(CH$^+$)] $\sim$ 13.06 cm$^{-2}$.
We would thus find that dissociative recombination should dominate radiative recombination by a factor of about 4.8 --- which would imply that the $n_e$ derived from C should be reduced to about 0.04 cm$^{-3}$, similar to the values derived from S, Mn, and Fe.
That comparison assumes that the CH$^+$ coexists in the cool SLV clouds with the various atomic species. 
The CH$^+$ molecular ion, however, is rapidly destroyed (by H$_2$ and by H atoms as well as by electrons) in cold, dense regions.
Furthermore, if we fit the CH$^+$ profile (either ours or the higher resolution spectrum obtained by Crane et al. 1995) with two components corresponding to the two strongest SLV components, we obtain $b$-values that are larger than those that would result from $T$ $\sim$ 100 K and the turbulence implied by the SLV \ion{K}{1} or \ion{Ca}{2} line widths.
It is thus likely that the CH$^+$ is not completely coextensive with the various atomic species in these relatively cold clouds, even though the component velocities are similar.
The effect of dissociative recombination on the carbon ionization balance is thus somewhat uncertain.

Given the current limit log[$N$(SH$^+$)] $<$ 12.7 cm$^{-2}$ toward 23 Ori (Magnani \& Salzer 1991), dissociative recombination could also enhance $N$(\ion{S}{1}), if the rate is of order 10$^{-7}$ cm$^3$ s$^{-1}$ (Fitzpatrick 1998). 
The \ion{S}{1} column density is low, relative to the other trace neutral species, however, which may imply that the SH$^+$ abundance is significantly smaller than the current upper limit.}

\footnotetext{We note, however, that use of the revised $f$-value for the \ion{Mg}{2} $\lambda$1239 doublet (Fitzpatrick 1997; see the Appendix) brings the $n_e$ determined from C and Mg toward HD 215733 essentially into agreement.
Furthermore, if Ca is more depleted than Fe in those clouds, as we find for the WLV clouds toward 23 Ori, then the $n_e$ determined from \ion{Ca}{2}/\ion{Fe}{2} may also be in agreement.}

\item{We have not included chlorine in Table~\ref{tab:ne}, because its ionization balance is dominated by a rapid reaction between \ion{Cl}{2} and H$_2$, with rate constant $k$ = 7 $\times$ 10$^{-10}$ cm$^3$ s$^{-1}$, which ultimately leads to enhanced \ion{Cl}{1} (Jura 1974; Fehsenfeld \& Ferguson 1974; Jura \& York 1978).
For the SLV clouds, if we take $n$(H$_2$) $\sim$ $N$(H$_2$) / 14 pc (likely a lower limit to the H$_2$ density), the \ion{Cl}{2} radiative recombination rate $\alpha$ = 8 $\times$ 10$^{-12}$ cm$^3$ s$^{-1}$ (P\'{e}quignot \& Aldrovandi 1986), and $n_e$ $\sim$ 0.15 cm$^{-3}$, then [$n$(H$_2$) $k$] / [$\alpha$ $n_e$] $\ga$ 30, so the reaction with H$_2$ is much faster than radiative recombination.
If we estimate the chlorine depletion as D$_{\rm Cl}$ $\sim$ -0.4 dex (i.e., slightly less severe than the cold cloud value, as found for other lightly depleted elements in the SLV clouds), then log[$N$(\ion{Cl}{2})] $\sim$ log[$N$(\ion{Cl}{1})] $\sim$ 13.3 cm$^{-2}$, consistent with the very uncertain value given by Jenkins et al. (1986).
The photoionization rate $\Gamma$ for \ion{Cl}{1} can then be estimated from the equation $\Gamma$ $N$(\ion{Cl}{1}) = $k$ $n$(H$_2$) $N$(\ion{Cl}{2}).
The resulting value is about twice the $\Gamma$ = 1.6 $\times$ 10$^{-11}$ s$^{-1}$ given by P\'{e}quignot \& Aldrovandi (1986) for the WJ1 field --- perhaps suggestive of a somewhat harder far-UV radiation field ($\chi_{\rm ion}$ = 12.97 eV for \ion{Cl}{1}).}

\item{Lepp et al. (1988) suggested that charge exchange between singly ionized species and either neutral or negatively charged large molecules (e.g., PAH's) might significantly increase the relative fractions of neutral species in interstellar clouds --- thus leading to spuriously high $n_e$ inferred from simple ionization balance calculations.
The computed neutral species enhancements, for models of the main clouds toward $\zeta$ Per and $\zeta$ Oph, range from factors of about 10 to 15 for Mg, Ca, and Fe, depending on the details of the cloud model (and on the rather uncertain properties of the PAH's), as the PAH abundance increases from 10$^{-11}$ to 10$^{-6}$ (relative to hydrogen).
The enhancements appear to be about 30\% smaller for C and S, presumably due to their larger ionization potentials; estimates were not given for other elements.
Thus, while charge exchange with large molecules might provide a partial explanation for the apparently large overall $n_e$ (and $n_e$/$n_{\rm H}$) in the SLV clouds, it would not appear to account for the large differences in $n_e$ inferred from different elements.}

\item{P\'{e}quignot \& Aldrovandi (1986) suggested that charge exchange between neutral species and protons or other positive ions ($X^0$ + $H^+$ $->$ $X^+$ + $H^0$ + $\Delta$E) could affect the abundances of the neutral species in cool interstellar clouds.
They identified which reactions would be significant and made rough guesses as to the rates, based on the energy ($\Delta$E) released by the different possible reactions.
Where the fractional ionization is low [e.g., in the main $\zeta$ Oph (``B'') clouds considered by P\'{e}quignot \& Aldrovandi], reactions with C$^+$ can dominate; where the fractional ionization is high (as appears to be the case in the SLV clouds toward 23 Ori), reactions with protons should dominate, essentially to reduce some of the neutral species column densities.
In the latter case, the species \ion{Al}{1}, \ion{S}{1}, \ion{Ti}{1}, \ion{Mn}{1}, \ion{Fe}{1}, and \ion{Ni}{1} should be most affected; for all these, P\'{e}quignot \& Aldrovandi estimated a rate constant $\beta$ $\sim$ 3 $\times$ 10$^{-9}$ cm$^3$ s$^{-1}$.
[We note, however, that the relevant charge exchange rates have not been determined for most of the species in question for temperatures as low as 100 K (Kingdon \& Ferland 1996; Stancil et al. 1998).]
Interestingly, \ion{S}{1}, \ion{Mn}{1}, and \ion{Fe}{1} are the three apparently underabundant neutral species in the SLV clouds toward 23 Ori (and \ion{Ni}{1} may also be underabundant).
If we add a charge exchange term to the ionization balance equation, we have  $X^0$$\Gamma$ + $\beta$$X^0$$n_p$ = $\alpha$$n_e$$X^+$ (with the proton density $n_p$ = $n_e$).
If we assume $\beta$ = 3 $\times$ 10$^{-9}$ cm$^3$s$^{-1}$ and $T$ = 100 K, then we infer $n_e$ = 0.06, 0.18, and 0.10 cm$^{-3}$ for S, Mn, and Fe, respectively --- closer to the estimates derived from most other elements than the 0.04--0.05 cm$^{-3}$ obtained with no charge exchange.
If we instead fix $n_e$ = 0.15 cm$^{-3}$ (see below), then we would need rate coefficients $\beta$ = 10.3, 2.8, and 3.5 $\times$ 10$^{-9}$ cm$^3$s$^{-1}$ for S, Mn, and Fe to obtain that $n_e$.
Charge exchange between protons and neutral species thus appears to provide a possible explanation for the low $n_e$ derived from S, Mn, and Fe in the SLV clouds toward 23 Ori, if in fact $n_e$/$n_{\rm H}$ is as high as 0.01 (for protons to dominate) and if the rates are of order a few times 10$^{-9}$ cm$^3$s$^{-1}$.
Calculations by Kimura et al. (1997) indicate, however, that the cross-section for charge transfer ionization of \ion{S}{1} is smaller than the corresponding cross-section for \ion{Si}{1} (Kimura et al. 1996) --- suggesting that the rate coefficient for \ion{S}{1} is of order 10$^{-11}$ cm$^3$ s$^{-1}$ at 100 K, much smaller than would be needed to significantly affect the sulfur ionization balance (above; P\`{e}quignot \& Aldrovandi 1986; Fitzpatrick \& Spitzer 1997).}

\end{itemize}

The large range in $n_e$ inferred from different trace/dominant ratios in Table~\ref{tab:ne} and the uncertainties associated with the various additional processes that might affect the ionization balance for different elements imply that any estimate of $n_e$ will be somewhat uncertain.
We have adopted $n_e$ $\sim$ 0.15$\pm$0.05 cm$^{-3}$ (from Na, Mg, Si, and K) as an average value for the SLV clouds toward 23 Ori [and recall that $n_e$(17S) $\sim$ 2.6 $n_e$(18S)].
Charge exchange between neutral species and protons may account for the lower $n_e$ inferred here from S, Mn, and Fe under the assumption of simple photoionization and radiative recombination, though more accurate values for the charge exchange rates at low temperatures are needed to confirm this.
A harder far-UV radiation field would lead to lower $n_e$ estimated from C, P, and S (assuming the WJ1 field); dissociative recombination of CH$^+$, however, would increase the estimate from C.
At the low temperatures found for the SLV clouds, dielectronic recombination should not affect the $n_e$ estimates from Mg and Ca.
Errors in the adopted photoionization and/or recombination rates (e.g., for Fe) may also contribute to the range in $n_e$ derived from different diagnostic ratios.
The true $n_e$ could be higher if the actual radiation field is higher than the adopted ``average'' Galactic field; the actual field is unlikely to be lower than average.
The true $n_e$ could be lower, by perhaps an order of magnitude, however, if charge exchange between singly ionized species and large molecules is important at the moderate densities found for the SLV clouds.
A smaller $n_e$ would be more consistent with estimates based on the $N$(\ion{C}{2}*)/$N$(\ion{C}{2}) ratio, as discussed above ($\S$~\ref{sec-phsne}; see also Fitzpatrick 1998).
Even if that is the case, the $n_e$/$n_{\rm H}$ ratio would still imply some partial ionization of hydrogen in the SLV clouds.

For other lines of sight, we would emphasize that inferred $n_e$ are likely to be very unreliable, apart from a detailed understanding of the physical conditions and processes affecting the ionization balance.
In many cases, for example, the $n_e$ obtained from $N$(\ion{Ca}{1})/$N$(\ion{Ca}{2}) are higher than the $n_e$ inferred from other such ratios --- by an order of magnitude or more.
Such a systematic effect could be quite significant, as the $N$(\ion{Ca}{1})/$N$(\ion{Ca}{2}) ratio has been extensively used both to calculate electron densities and to infer abundances for elements (e.g., Na, K, Li) whose dominant ionization states cannot be observed.
Estimates of relative depletions based on ratios of trace neutral species (e.g., Snow 1984; Snow et al. 1996) will also be affected by any additional processes selectively affecting the ionization balance.

There are several possible explanations for the partial ionization of the hydrogen in the predominantly neutral SLV gas.  
The enhanced soft X-ray flux observed from the Orion-Eridanus region (Burrows et al. 1993), which may be due to thermal emission from hot gas [log($T$) $\sim$ 6.2 K] produced by Orion OB1 association stellar winds and/or supernovae, could ionize some of the \ion{H}{1} in the surrounding neutral clouds.
Snowden et al. (1995) estimated that an intervening \ion{H}{1} column density comparable to that observed toward 23 Ori could account for the smaller ratio of $\onequarter$ keV to $\threequarters$ keV flux toward Orion, relative to the ratio seen in the part of the bubble farther from the Galactic plane in Eridanus.
We may use equation 5-32 in Spitzer (1978) to estimate the ionization rate required to produce the observed fractional ionization in the SLV gas:
\begin{displaymath}
n_p = \frac{1}{2}~n_i~ \left\{ \left( 1~+~\frac{4~\zeta_{\rm H}~n({\rm H~I})}{\alpha^{(2)}~n_i^2} \right)^{1/2} ~-~1 \right\}.
\end{displaymath}
Here $\alpha^{(2)}$, the hydrogen recombination rate (excluding recombinations to the lowest level),  is $\sim$ 7 $\times$ 10$^{-12}$ s$^{-1}$ for $T$ $\sim$ 100 K, and we set $n_i$, the density of positive ions heavier than He, to 2 $\times$ 10$^{-3}$ cm$^{-3}$, based on the observed gas-phase heavy element abundances (primarily C) and the derived $n$(\ion{H}{1}) $\sim$ 10 cm$^{-3}$.
To obtain the inferred fractional ionization $n_e$/$n_{\rm H}$ $\sim$ 0.01, we thus need an ionization rate $\zeta_{\rm H}$ $\sim$ 7 $\times$ 10$^{-15}$ s$^{-1}$.
While this rate is two orders of magnitude larger than the cosmic ray ionization rate usually assumed for the local ISM, it is only a factor of 2 larger than the rate inferred for some of the clouds observed toward HD 93521 and HD 149881 (Spitzer \& Fitzpatrick 1993, 1995), and some enhancement might be expected in the Orion region, in view of the X-ray observations noted above.
If charge exchange with large molecules has enhanced the neutral species by a factor of 10, however, the inferred ionization fraction would be 0.001, requiring an ionization rate lower by a factor of 100, since $\zeta_{\rm H}$ $\sim$ ($\alpha^{(2)}$ $n_p^2$) / $n_{\rm H}$.
Quantitative estimates of the X-ray flux --- and the predicted ionization of hydrogen --- would be valuable.

If the SLV components arise in the expanding shell around the Orion-Eridanus bubble, the large apparent fractional ionization might also be due to partial mixing of the dominant neutral gas with the adjacent (interior) layer of ionized gas, which Reynolds \& Ogden (1979) estimate comprises about 10\% of the material in the shell.
As noted above, some ionized gas is present at similar velocities, and such a mixture would have only minor effects on the abundances and depletions in the SLV gas.
This possibility is intriguing, in light of the relatively high abundances of CH and CH$^+$, as turbulent boundary layers have been proposed as one possible source of CH$^+$ (e.g., Duley et al. 1992, and references therein).
Duley et al. present diagnostic ratios (e.g., ratios of the densities in several H$_2$ rotational levels to that of CH$^+$) for steady-state models characterized by several different temperatures, densities, and radiation fields.
The models with $T$ = 2000 K, $n$ = 20 cm$^{-3}$, $n$(CH$^+$) $\sim$ 1.5 $\times$ 10$^{-4}$ cm$^{-3}$, and radiation field 1--3 times the average IS field seem to agree best with the ratios of CH and of the $J$=3--5 levels of H$_2$, with respect to CH$^+$, toward 23 Ori.
If those models are appropriate for the 23 Ori line of sight, the CH, CH$^+$, and the higher rotational levels of H$_2$ might thus arise in a thin ($\sim$ 0.02 pc), warm layer adjacent to the colder SLV gas.
Observations of OH absorption toward 23 Ori could provide a stringent test for the Duley et al. models.

Either of these possible explanations for increased $n_e$/$n_{\rm H}$ toward 23 Ori would also apply to the corresponding strongest neutral components toward other stars in the Orion region.
Preliminary analyses of the strongest components at low velocity toward $\zeta$ Ori seem to indicate that both $n_e$ and the fractional ionization are somewhat lower than in the SLV clouds toward 23 Ori, but that the pattern of neutral/first ion ratios for different elements (and thus the inferred relative $n_e$) appears to be very similar.

\subsection{Elemental Abundances and Depletions}
\label{sec-depl}

\subsubsection{Previous Studies}
\label{sec-deplprev}

Spectra obtained with {\it Copernicus} and {\it IUE} provided total line-of-sight abundances and depletions for a number of elements (e.g., Jenkins 1987; de Boer et al. 1987), but the relatively low resolution of those spectra has generally precluded studies of individual interstellar clouds.
The average line-of-sight depletions typically exhibit a fairly smooth variation with average line-of-sight hydrogen density $<n_{\rm H}>$ = $N$(H)/$d$ (though with some scatter), becoming increasingly severe for larger values of $<n_{\rm H}>$.
Similar trends have been noted versus the fraction of hydrogen in molecular form $f$(H$_2$), which may be a better indicator of local physical conditions than $<n_{\rm H}>$ (e.g., Cardelli 1994). 
Spitzer (1985) proposed a simple, idealized model, in which a given line of sight intercepts a mixture of  several distinct cloud types, each having a characteristic overall level and pattern of depletions.
In that model, warm, low density clouds have relatively mild depletions, while colder clouds (both ``standard'' and ``large'') are characterized by more severe depletions.
Jenkins et al. (1986) used that model to interpret a large body of {\it Copernicus} abundance data for Mg, P, Cl, Mn, Fe, Cu, and Ni [see Jenkins (1987) for additional elements], and derived representative depletion values for the cold and warm clouds.
We have preserved some of this framework in the ``cold cloud,'' ``warm cloud,'' and ``halo cloud'' depletions listed in Table~\ref{tab:refab}, but we view the values as representing a likely continuum of depletions encountered in individual clouds characterized by different temperatures, local densities, and histories.
Joseph (1988, 1993) took the ``best'' of the Jenkins et al. (1986) data, and found the depletions of Mg, P, Cl, Mn, and Fe to be linearly correlated --- contrary to expectations, based on ideas of grain processing in the ISM, that some elements should deplete more readily than others (e.g., Barlow 1978).
Joseph conjectured that the relative depletions were fixed at grain formation, and that subsequent processing in the ISM (accretion, sputtering, etc.) preserved those relative depletions.
Several attempts to detect preferential depletions, by comparing several lines of sight in a common region but with different overall levels of depletion (Snow \& Jenkins 1980) or by using trace neutral species to probe denser regions (Snow 1984; Snow, Joseph, \& Meyer 1986) likewise failed to reveal any strong, consistent signature of differential depletion.
In retrospect, several factors --- blending and possible saturation effects in the relatively low-resolution UV spectra, a somewhat limited range in overall depletions for the elements considered, and other processes affecting the neutral species --- may have conspired to hide element-to-element differences in depletion behavior.
We note that higher resolution optical spectra of \ion{Ti}{2} (e.g., Stokes 1978) had already suggested that the Ti depletion exhibits a significantly larger range (compared to less severely depleted elements) for lines of sight largely overlapping those used in the UV surveys.

The higher spectral resolution achievable with the {\it HST} GHRS has brought us closer to the goal of determining the abundances, depletions, and physical properties --- and their possible interrelationships --- for individual interstellar clouds.
Savage \& Sembach (1996) have summarized the results of a number of GHRS studies of interstellar abundances and depletions; Fitzpatrick \& Spitzer (1997) presented a detailed study of the individual components found toward the halo star HD 215733.
In one of the first individual-line-of-sight  abundance studies using spectra from GHRS, Savage et al. (1992) compared the abundances and depletions for two component groups (the aforementioned groups A and B) seen toward $\zeta$ Oph, and found that the differences in depletion between the two groups were larger for the more severely depleted elements Fe, Cr, Ni, and Ti than for the less depleted O, Mg, Cu, and Mn (see Figure~\ref{fig:depl}) --- i.e., differential depletion.
Sofia et al. (1994), incorporating data for several additional lines of sight, presented evidence for the preferential depletion of Mg and Fe relative to Si, developed a core-mantle model for the dust grains, and conjectured that the total (gas + dust) abundances in the ISM might be sub-solar.
Snow et al. (1996) inferred preferential depletion of some elements in the denser regions of the clouds toward HD 154368 from comparisons of trace neutral and dominant singly ionized species, but ionization balance uncertainties ($\S$~\ref{sec-ne}) may invalidate those conclusions.
Many of the other lines of sight observed with the GHRS (including a number toward stars in the Galactic halo) sample primarily warm, fairly diffuse gas, and have provided constraints on the minimum depletions in the diffuse ISM, due perhaps to a population of relatively hardy grain cores (Fitzpatrick 1996).

\subsubsection{23 Ori vs. $\zeta$ Oph --- Density--Dependent Depletions?}
\label{sec-depl23z}

Comparisons among the depletions obtained for the 23 Ori SLV components, those found for the $\zeta$ Oph B components, and the values adopted for ``typical'' cold, dense clouds suggest that a range of depletions may be present for relatively cold, neutral clouds and that the differences in depletion may be related to differences in local physical conditions.
Except for the very lightly depleted C, N, O, S, Sn, and Kr, the SLV depletions are 0.3--0.6 dex less severe than the adopted cold cloud values (relative to H; i.e., [X/H]), whereas the $\zeta$ Oph B depletions are comparable to or slightly more severe than the cold cloud values.
This is the case for lightly depleted elements (P, Zn), for moderately depleted elements (Mg, Si, Mn, Cu, Ge, Ga), and for more severely depleted elements (Al, Ca, Ti, Cr, Fe, and Ni).
We note, however, that since Zn also is less depleted than in the cold cloud pattern, the SLV depletions of other elements, relative to Zn (i.e., [X/Zn]), are more similar to the cold cloud pattern.
Since the SLV depletions of C, N, O, and Kr are consistent with those obtained for other lines of sight [essentially independent of $<n_{\rm H}>$ or $f$(H$_2$)], and since several independent determinations find the total log[$N$(\ion{H}{1})] less than or equal to the adopted 20.74 cm$^{-2}$, it is unlikely that we have significantly underestimated the amount of neutral H in the SLV components.
Furthermore, there is apparently only a small amount of ionized H at those velocities.
Finally, the individual component values of $N$(\ion{Fe}{2})/$N$(\ion{Zn}{2}), for the two strongest SLV components, differ by only about 0.2 dex.
The differences in depletion between the 23 Ori SLV components and the $\zeta$ Oph B components thus do not appear to be due to errors in abundances, mixtures of very disparate components, or contamination from ionized gas --- but instead appear to reflect a range of at least a factor 2 in the depletions among different cold clouds (cf. also the comparisons between the depletions toward $\xi$ Per and $\zeta$ Oph in Savage et al. 1992 and toward 1 Sco and $\zeta$ Oph in Welty et al. 1995).

The relationship between average line-of-sight depletions and $<n_{\rm H}>$ recognized in the early UV data was suggestive of a dependence of depletions on local hydrogen density.
Joseph et al. (1986) analyzed {\it IUE} spectra for several lines of sight showing strong CN absorption (presumably indicative of dense gas), and found some evidence for preferential depletion of Mn and perhaps Cr, but a number of the abundances are rather uncertain.
Cardelli, Federman, \& Smith (1991) compared \ion{Ca}{1} abundances for lines of sight with different $N$(CN)/$N$(CH) ratios, and concluded that the Ca depletion is proportional to $n_{\rm H}^{-3}$.
All those studies, however, were based on integrated line-of-sight abundances, and actual values of $n_{\rm H}$ were not known in most cases.
It is therefore intriguing that the densities derived for the SLV components toward 23 Ori ($n_{\rm H}$ $\sim$ 10--15 cm$^{-2}$) are at least an order of magnitude smaller than those found for the $\zeta$ Oph B components, where the depletions are more severe by a factor of 2--4. 
For the main component (group) B toward $\zeta$ Oph, Lambert et al. (1994) obtained $n_{\rm H}$ $\sim$ 100--200 cm$^{-2}$ from \ion{C}{1} fine-structure excitation, Federman et al. (1994) found $n_{\rm H}$ $\sim$ 400 cm$^{-2}$ from an analysis of the carbon chemistry, and Lepp et al. (1988) used central densities of 350 and 800 cm$^{-2}$ in their models.
On the other hand, if the SLV components are located in a shell of gas swept up by the expanding Orion-Eridanus bubble, we might expect the depletions to be somewhat less severe --- due either to partial erosion of the associated grains or to incorporation of lower density gas with milder depletions.
Detailed studies of other lines of sight are needed to determine whether there is any general relationship between the depletions and the local densities for individual cold clouds.

Snow (1984; see also Snow et al. 1986, 1996) investigated the possible dependence of depletion on density by comparing the relative depletions inferred from ratios of trace, neutral species (which may be concentrated in the denser regions) with those obtained from the presumably more broadly distributed dominant species.
Significant differences in the relative depletions inferred from the trace and dominant species were found only toward $\zeta$ Oph and HD 154368.
As noted above ($\S$~\ref{sec-ne}), however, such comparisons assume that photoionization and radiative recombination dominate the ionization equilibria for the various elements considered.
In the last two columns of Table~\ref{tab:ne}, we list relative depletions for a number of elements --- inferred both from the dominant first ions and from the trace neutral species.
In both cases, the depletions are given relative to that of carbon, whose depletion appears to be relatively constant over wide ranges in both $<n_{\rm H}>$ and $f$(H$_2$) (Sofia et al. 1997).
For all elements except Ca, the relative depletions inferred from the neutral species are more severe than those determined from the dominant species.  
The differences are less than a factor 2 for Na, Mg, and K, but may be as much as factors of 5--7 for S, Mn, and Fe --- corresponding to the differences in $n_e$ inferred from those elements.
On the other hand, if the depletions are instead computed relative to that of sulfur (which is typically undepleted), then in many cases the relative depletions inferred from the neutral species are {\it less} severe than those determined from the dominant species.
Firm conclusions regarding the density dependence of depletions cannot be drawn from the neutral species ratios until the impact of the various other processes on the ionization balance ($\S$~\ref{sec-ne}) has been evaluated.

\subsubsection{SLV vs. WLV Depletions}
\label{sec-deplsw}

We have independent estimates for the WLV and SLV depletions only for Si, Fe, Al, and Cr, for which the differences in depletion (WLV minus SLV) are 0.37, 0.43, 0.35, and 0.46 dex, respectively.
While these differences are all comparable --- and thus reminiscent of the correlated depletions found by Joseph (1988) --- the small baseline in depletion precludes a definitive confirmation of Joseph's result.
Conversely, the Si and Fe depletions are also quite consistent with those plotted by Sofia et al. (1994) and by Fitzpatrick (1996) for a larger range in overall depletion --- where the difference in depletion between the two elements increases with increasing overall depletion.
[As already noted, Savage et al. (1992) found larger differences in depletion toward $\zeta$ Oph (component group A minus component group B) for the more severely depleted elements.]
Toward 23 Ori, the even more severely depleted Ca appears to exhibit a somewhat larger difference in depletion between the WLV and SLV components (0.58 dex), but the WLV Ca depletion is based on considerations of the Ca ionization balance, and is therefore somewhat uncertain.
Higher resolution observations of \ion{Ti}{2} absorption toward 23 Ori would provide a more reliable estimate of the difference in depletion (WLV minus SLV) for the most severely depleted elements. 

\subsubsection{Dust-Phase Abundances}
\label{sec-depldp}

A number of papers have examined the ``dust-phase abundances'' of the more cosmically abundant elements in attempts to identify possible mineral constituents of the dust at different levels of depletion or in different environments (e.g., Sofia et al. 1994; Fitzpatrick \& Spitzer 1997; Fitzpatrick 1997).
The dust-phase abundances, defined as $A_{\rm dust}$ = $A_{\rm cosmic}$ $-$ $A_{\rm gas}$, depend (obviously) on the adopted cosmic reference abundances and on the atomic data used to determine the gas-phase abundances.
The atomic data now seem to be fairly well known for most of the important lines of the most abundant elements (Savage \& Sembach 1996; Fitzpatrick 1997; see the Appendix). 
There are strong suggestions, however, that the total (gas + dust) abundances of at least some elements in the ISM may be only about 2/3 of the solar values (Sofia et al. 1992; Snow \& Witt 1996; Meyer et al. 1997); S and (perhaps) Sn may be notable exceptions.
If we use the solar reference abundances listed in Table~\ref{tab:refab}, then the dust-phase abundances of Mg, Si, and Fe are all about 30--32 per 10$^6$ H atoms in the SLV components, since their total abundances are very similar and since all three are largely depleted\footnotemark.
\footnotetext{We note that the relative gas-phase abundances of Mg in the WLV and SLV components were assumed equal to those for Si, in view of the similar average depletions of the two elements for the total line of sight and of the similar relative SLV--WLV abundances found for Al, Cr, and Fe.}
In the WLV components, $A_{\rm dust}$ $\sim$ 19, 25, and 32 per 10$^6$ H atoms for Mg, Si, and Fe, respectively.
If we assume that the total ISM abundances are 2/3 times the solar values, then $A_{\rm dust}$ $\sim$ 17--21 per 10$^6$ H atoms for the three elements in the SLV components, and $A_{\rm dust}$ $\sim$ 6, 13, and 20 per 10$^6$ H atoms, respectively, in the WLV components.
In either case, $A_{\rm dust}$(Fe+Mg) $\sim$ 2 $A_{\rm dust}$(Si) for both WLV and SLV components, consistent with most of the three elements being present in the dust as olivine [(Mg,Fe)$_2$SiO$_4$].
In contrast, Fitzpatrick (1997) found $A_{\rm dust}$(Fe+Mg) significantly greater than 2 $A_{\rm dust}$(Si) for clouds exhibiting the mildest depletions, and suggested that some of the Mg and Fe might be in non-silicate grains.
In the SLV components, $A_{\rm dust}$(Fe) may be slightly greater than $A_{\rm dust}$(Mg), but $A_{\rm dust}$(Fe) is significantly larger in the WLV components.

\subsubsection{Depletions in High-Velocity Gas}
\label{sec-deplhv}

Determination of column densities for the significant ionization states of various elements in the HV components affords an opportunity to gauge how much dust has survived in that ionized, presumably shocked, high-velocity gas.
As discussed above, the HV column densities of \ion{C}{2}, \ion{N}{2}, and \ion{Si}{2} + \ion{Si}{3} are consistent with solar relative abundances of C, N, and Si and an \ion{H}{2} column density of about 5 $\times$ 10$^{17}$ cm$^{-2}$.
The limit on $N$(\ion{S}{2}) + $N$(\ion{S}{3}) suggests that the total $N$(\ion{H}{2}) is not more than a factor 1.5 higher than that value, which would imply that $N$(\ion{C}{3}) $<$ $N$(\ion{C}{2}) and $N$(\ion{N}{3}) $<$ $N$(\ion{N}{2}) [assuming negligible $N$(\ion{S}{4}) and no depletion of S].
Furthermore, C thus cannot be depleted by more than 0.2 dex [compared to the 0.4 dex found for the predominantly neutral gas in sightlines characterized by a wide range in $f$(H$_2$) (Sofia et al. 1997)].
Likewise, Si cannot be depleted by more than 0.2 dex.
On the other hand, Al does appear to be depleted, by 0.4 to 0.6 dex --- so that some dust has apparently survived.
If Si is not depleted, then the depleted Al might be in the form of oxides (i.e., instead of in silicates) in the remaining dust.
The uncertainties in the Si abundance would allow the dust-phase Al to be incorporated in silicate grains, however.
More extensive data for $\zeta$ Ori (including column densities for \ion{C}{3}, \ion{N}{3}, \ion{S}{3}, and \ion{Fe}{3}) suggest that Si may be depleted by about 0.3 dex, while Al and Fe may be depleted by about 0.8 dex, in the high-velocity gas along that line of sight (Welty et al. 1999c).

\subsection{High Ions}
\label{sec-high}

The relatively low resolution obtained for spectra of the high ions \ion{C}{4}, \ion{N}{5}, \ion{Si}{3}, \ion{Si}{4}, and \ion{S}{3} makes the interpretation of the low-velocity absorption from those species rather uncertain ($\S\S$~\ref{sec-compion} and~\ref{sec-abhigh}). 
The observed $N$(\ion{C}{4})/$N$(\ion{N}{5}) = 2.6$\pm$0.2 is slightly lower than the median value 4.0$\pm$2.4 (rms deviation) found by Sembach, Savage, \& Tripp (1997) for 23 lines of sight with data from {\it IUE} and/or GHRS.
The observed $N$(\ion{C}{4})/$N$(\ion{Si}{4}) = 8.3$\pm$0.7, however, is significantly higher than the corresponding median value 3.8$\pm$1.9 for 31 lines of sight.
Three lines of sight in the Sembach et al. compilation have comparably high $N$(\ion{C}{4})/$N$(\ion{Si}{4}) ratios, but only one ($\alpha$ Vir) also has a low $N$(\ion{C}{4})/$N$(\ion{N}{5}) ratio.
Higher resolution spectra obtained with the GHRS echelle have revealed complex structure in the profiles of these high ions in several lines of sight --- for example, relatively narrow components (at similar velocities) in \ion{Si}{4} and \ion{C}{4}, intermingled with broader components seen in \ion{N}{5} and \ion{C}{4} (e.g., Savage, Sembach, \& Cardelli 1994; Fitzpatrick \& Spitzer 1997).
Some of the narrower high ion components may be associated with components seen at similar velocities in lower ionization stages.
While the absorption from \ion{C}{4}, \ion{N}{5}, \ion{Si}{4}, and \ion{S}{3} toward 23 Ori appears to be symmetric and featureless (in the G160M spectra), slight differences in central velocity among the four species (Table~\ref{tab:high}) suggest that multiple components, arising from different locations and due to different physical processes, may be present.
We briefly consider several possible origins for these high ions.

\begin{itemize}
\item{Hot gas in collisional ionization equilibrium (CIE) at log($T$) $\sim$ 6.2 K (as is thought to fill the Orion-Eridanus bubble) would include a small percentage of \ion{C}{4} and \ion{N}{5} (e.g., Sutherland \& Dopita 1993).
The observed $b$-values (23--24 km s$^{-1}$), however, indicate that those ions arise in gas with log($T$) $\la$ 5.7 K.
In CIE, the peak abundances of \ion{C}{4} and \ion{N}{5} occur at log($T$) $\sim$ 5.0 K and 5.25 K, respectively (Sutherland \& Dopita 1993).
The peak abundances in non-equilibrium situations, however, can occur at either higher $T$ (ionization lagging behind heating) or lower $T$ (recombination slower than cooling).
The calculations of Schmutzler \& Tscharnuter (1993), for example, indicate maximum abundances of \ion{C}{4} and \ion{N}{5} in isochorically (i.e., at constant density) cooling gas at log($T$) $\sim$ 4.25 K.
If the \ion{Al}{3} component at $-$6 km s$^{-1}$ is associated with the higher ions, its width implies that log($T$) $\la$ 4.8 K.}

\item{In principle, some of the observed \ion{Al}{3}, \ion{S}{3}, \ion{Si}{3}, and \ion{Si}{4} could be produced by photoionization.
The observed $N$(\ion{C}{4})/$N$(\ion{Si}{4}) ratio is larger than any of the values predicted by Gruenwald \& Viegas (1992) for spherical \ion{H}{2} regions, but could be consistent with the O4 star models if Si were slightly depleted.
It is unlikely, however, that any \ion{N}{5} could be due to photoionization, due to the large (77.5 eV) ionization potential of \ion{N}{4}.}

\item{Borkowski, Balbus, \& Fristrom (1990) described calculations of the high ion populations produced in conduction/evaporation fronts at the interface between hot (10$^6$ K) and cool ($\la$ 10$^4$ K) gas --- as might be expected, for example, between the hot, diffuse gas within the Orion-Eridanus bubble and either the walls of the bubble or any cool, neutral clouds remaining within the bubble.
The observed $N$(\ion{C}{4})/$N$(\ion{N}{5}) ratio is consistent with the values predicted for times 10$^5$ yr $\la$ $t$ $\la$ 10$^6$ yr, if the magnetic field is essentially perpendicular to the interface, or for brief periods slightly before 10$^5$ yr for fields at smaller angles relative to the interface.
The predicted ratios are too large for earlier times (for any field orientation) and too small for later times.
The observed $N$(\ion{C}{4})/$N$(\ion{Si}{4}) ratio is generally lower than the predicted values, but adding photoionization of \ion{Si}{3} to the models may remove that discrepancy.
For $t$ $\sim$ 10$^5$--10$^6$ yr, the models predict somewhat lower overall column densities than those observed, as well as somewhat narrower absorption line profiles [dominated by thermal broadening at log($T$) $\sim$ 5.0--5.5 K].
Consistent column densities and line widths might be achieved if several such interface regions, at somewhat different velocities, were present along the line of sight.}

\item{Slavin \& Cox (1992) calculated the long-term evolution of the structure and high ion content of spherical supernova remnant (SNR) bubbles expanding into a uniform, low density medium, with and without the effects of magnetic pressure.
The inclusion of a 5 $\mu$G magnetic field significantly affects the structure of the bubble --- producing a thicker, less compressed shell and a smaller interior volume of hot gas.
The predicted average column densities for \ion{C}{4}, \ion{N}{5}, and \ion{Si}{4} exhibit roughly constant ratios for times $t$ $\sim$ 0.4--5.0 $\times$ 10$^6$ yr. 
Within that time range, the predicted $N$(\ion{C}{4})/$N$(\ion{N}{5}) ratio is typically $\sim$ 2 (slightly smaller than the observed ratio), and $N$(\ion{C}{4})/$N$(\ion{Si}{4}) is $\sim$ 15 (a factor $\sim$ 2 larger than observed).
Photoionization of \ion{Si}{3} by radiation from within the remnant would increase the predicted $N$(\ion{Si}{4}), however.
The models predict comparable widths for the \ion{C}{4} and \ion{Si}{4} lines, but the \ion{Si}{4} lines toward 23 Ori appear to be somewhat narrower.
The predicted \ion{C}{4} line widths are also smaller than those observed.}

\item{Slavin, Shull, \& Begelman (1993) modelled the steady-state behavior of ``intermediate temperature'' gas [log($T$) $\sim$ 5.0--5.5 K] produced in turbulent mixing layers formed at the boundaries between hot and cool gas.
The predicted $N$(\ion{C}{4})/$N$(\ion{N}{5}) ratios are typically greater than 10 for any transverse velocity 25 km s$^{-1}$ $\la$ $v_t$ $\la$ 100 km s$^{-1}$ between the hot and cool gas --- much higher than the value observed toward 23 Ori.}

\end{itemize}

While none of the above models can completely reproduce the high-ion column density ratios and line widths observed toward 23 Ori, the observed high ions may plausibly be attributed to a combination of a SNR bubble (Orion-Eridanus) and one or more conductive interfaces (between the hot bubble gas and cooler gas either within the bubble and/or in the shell --- perhaps the WLV and/or the IV gas, which are at similar velocities to the high ions).
Some additional photoionization of \ion{Si}{3}, reasonable in either case, would be needed to produce the observed \ion{Si}{4}.
Higher resolution spectra of the high ions would allow more definitive tests. 
It would also be interesting to explore models in which multiple supernovae re-heat and further expand existing cavities (as is likely to have happened in the Orion region), and to predict the abundances of additional observed species such as \ion{Al}{3}, \ion{Si}{3}, and \ion{S}{3} (as well as others, such as \ion{C}{3}, \ion{N}{3}, and \ion{S}{4}, which will be observable with {\it FUSE}).

\section{Conclusions / Summary}
\label{sec-sum}

We have used spectra obtained with the {\it Hubble Space Telescope} GHRS together with high-resolution optical spectra and UV spectra from {\it Copernicus} to study the abundances and physical conditions in the diffuse interstellar clouds seen along the line of sight to the star 23 Ori.
High-resolution (0.3--1.5 km s$^{-1}$) optical spectra of \ion{Ca}{1}, \ion{Ca}{2}, \ion{Na}{1}, and \ion{K}{1} were used to define the underlying component structure for the neutral gas that is unresolved in the UV spectra, enabling more accurate abundances and physical conditions to be derived for individual clouds from the numerous transitions found in the UV.

Multiple absorption components are present for each of several distinct types of gas, which are
characterized by different relative abundance and depletion patterns and physical conditions.

\begin{itemize}

\item{Strong low-velocity (SLV) absorption, due to cool, moderately dense neutral gas and representing about 92\% of the total $N$(\ion{H}{1}), is seen for various neutral and singly ionized species at +20 km s$^{-1}$ $\la$ $v_{\odot}$ $\la$ +26 km s$^{-1}$.
For these neutral SLV components, most typically severely depleted species are less depleted by factors of 2--4, compared to the ``cold, dense cloud'' pattern found, for example, in the main components toward $\zeta$ Oph.
For the blend of the two strongest SLV components, the relative populations of the $J$=0 and $J$=1 rotational levels of H$_2$ imply $T$ $\sim$ 100 K; the \ion{C}{1} fine structure excitation then implies a thermal pressure log($n_{\rm H} T$) $\sim$ 3.1 cm$^{-3}$K, and thus $n_{\rm H}$ $\sim$ 10---15 cm$^{-3}$ and a total thickness of 12--16 pc.
The average SLV electron density, derived independently from nine pairs of neutral and singly ionized species, ranges from about 0.04 cm$^{-3}$ to about 0.95 cm$^{-3}$; we adopt $n_e$ = 0.15$\pm$0.05 cm$^{-3}$.
The relatively large $n_e$/$n_{\rm H}$ $\sim$ 0.01 implies some ionization of hydrogen in these predominantly neutral components.} 

\item{Weaker low-velocity (WLV) absorption, probably due to warmer neutral gas, is seen primarily for various singly ionized species at 0 km s$^{-1}$ $\la$ $v_{\odot}$ $\la$ +30 km s$^{-1}$.
The depletions in the neutral WLV gas are typically less severe by another factor of 2--3 than in the SLV gas, and are somewhat similar to the ``warm cloud'' pattern seen in lines of sight with low reddening, low mean density, and/or low molecular fraction.
If $T$ $\sim$ 3000 K for the WLV components, then we have log($n_{\rm H} T$) $\sim$ 4.7--4.8 cm$^{-3}$K, $n_{\rm H}$ $\sim$ 15--20 cm$^{-3}$,
$n_e$ $\sim$ 0.2 cm$^{-3}$, $n_e$/$n_{\rm H}$ $\sim$ 0.01, and a total thickness of 0.7--0.9 pc.}

\item{Weak intermediate-velocity (IV) absorption from ionized gas is seen in \ion{C}{2}, \ion{Mg}{2}, and \ion{Si}{3} at $-$43 km s$^{-1}$ $\la$ $v_{\odot}$ $\la$ $-$4 km s$^{-1}$.
This gas is denser and slightly more ionized than the high velocity gas, though the depletions may be comparable, and appears to arise in thin filaments or sheets (thicknesses $\la$ 0.001 pc to 0.04 pc).} 

\item{Absorption from a number of singly and doubly ionized species, perhaps due to a radiative shock (``Orion's Cloak'', Cowie et al. 1979), is seen at $-$108 km s$^{-1}$ $\la$ $v_{\odot}$ $\la$ $-$83 km s$^{-1}$.
While the depletions in the ionized high-velocity (HV) components are uncertain, due to unobserved ionization stages, Al is probably depleted by a factor $\sim$ 3, even at cloud velocities in excess of 100 km s$^{-1}$; C and Si are depleted by no more than 0.2 dex.
The individual ionized high-velocity components typically have $T$ $\sim$ 8000$\pm$2000 K, $n_e$ = $n_{\rm H}$ $\sim$ 0.4--0.5 cm$^{-3}$, thermal pressure log(2$n_e T$) $\sim$ 3.7--4.0 cm$^{-3}$K, and thicknesses of order 0.1 pc.}

\item{No obvious absorption is discerned from a circumstellar \ion{H}{2} region around this B1 V star, though none would be expected if 23 Ori is located within the hot, ionized Orion-Eridanus bubble.}

\item{Weak absorption from the more highly ionized species \ion{S}{3}, \ion{C}{4}, \ion{Si}{4}, and \ion{N}{5} is seen in G160M spectra, centered at $-$5 km s$^{-1}$ $\la$ $v_{\odot}$ $\la$ +6 km s$^{-1}$.
The relative abundances of the high ions are consistent with an origin in a supernova remnant bubble (Orion-Eridanus), with additional contributions from interface regions between hot and cool gas within the bubble.}

\end{itemize}

The large range in $n_e$ derived from different pairs of neutral and singly ionized species in the SLV gas suggests that additional processes besides simple photoionization and radiative recombination affect the ionization balance.
Charge exchange with protons may reduce the abundances of \ion{S}{1}, \ion{Mn}{1}, and \ion{Fe}{1} (though the relevant rates generally are not well known at $T$ $\sim$ 100 K). 
Dissociative recombination of CH$^+$ may help to enhance \ion{C}{1}.
A slightly harder far-UV field in the Orion region could selectively reduce (somewhat) the abundances of neutral species with ionization potentials near that of hydrogen.
There is no apparent relationship between $n_e$ and the elemental depletion determined from the dominant singly ionized species.
The apparent systematic enhancement of \ion{Ca}{1} found for many lines of sight remains unexplained.
Previous determinations of $n_e$ for other lines of sight, typically based on limited data and limited knowledge of the potentially relevant physical processes, must be viewed as very uncertain.

We have briefly considered several possible explanations for the large apparent fractional ionization ($n_e$/$n_{\rm H}$ $\sim$ 0.01) in the SLV gas.
Such a fractional ionization would seem to require an ionization rate $\zeta_{\rm H}$ two orders of magnitude higher than the value typically adopted for the local ISM.
An enhanced flux of X-rays, which appears to be present in the Orion region, could in principle ionize some of the hydrogen in the SLV clouds, but more quantitative estimates of the potential ionization rate are needed.
Alternatively, if the SLV components are located at the boundary of the Orion-Eridanus bubble, mixing with adjacent ionized gas (estimated to have of order one-tenth the column density) might yield the apparent high $n_e$. 
A third possibility is that charge exchange between singly ionized atomic species and large molecules (if present at a fractional abundance of order 10$^{-7}$--10$^{-6}$) could enhance the abundances of the neutral species --- lowering $n_e$ and $\zeta_{\rm H}$ by factors of up to 10 and 100, respectively.
A smaller $n_e$ would be more consistent with the values derived from the \ion{C}{2} fine-structure excitation equilibrium.
In any case, $n_e$/$n_{\rm H}$ is higher than the value 2 $\times$ 10$^{-4}$ that would obtain if the electrons were due only to photoionization of carbon.

Comparisons of the SLV depletions and $n_{\rm H}$ with those found for the strong ``component B'' blend toward $\zeta$ Oph (where the densities are higher by at least a factor 10 and the depletions of many elements are more severe by factors of 2--4) hint at a possible relationship between depletion and {\it local} density for relatively cold clouds.
Similar data for individual clouds along other lines of sight are needed to confirm any such relationship, however, as the SLV depletions may instead reflect the history and location of the SLV components in a swept-up shell.
While ratios of some trace neutral species might seem to suggest more severe depletions for some elements in the denser parts of the SLV clouds, uncertainties as to the importance of various processes which may affect the ionization balance of individual elements preclude firm conclusions.
Considerations of the ionization balance in the WLV and $\zeta$ Oph A components suggests that Ca may generally be significantly more depleted than Fe in warm, low density gas.
In both the SLV and WLV clouds, the dust-phase abundance of Mg+Fe is roughly twice that of Si, whether solar or B-star abundances are adopted as the reference.
The dust-phase abundance of Fe appears to be comparable to that of Mg in the SLV components, but is significantly larger in the WLV components.
The apparent residual depletion of Al in the high-velocity gas (and similar data for the HV gas in several other lines of sight) provides constraints on the efficiency of dust destruction in shocks.

Higher spectral resolution data for some of the species observed with G160M and/or {\it Copernicus} would enable some of the remaining uncertainties regarding individual cloud properties to be addressed:
\begin{itemize}
\item{New spectra of H$_2$ would yield more accurate column densities and temperatures for the individual SLV components.}
\item{Spectra of \ion{Mg}{1} $\lambda$2853, \ion{C}{1} $\lambda$1656, \ion{N}{1}, \ion{O}{1}, \ion{Si}{2}* $\lambda$1265, and \ion{Ti}{2} would yield a better characterization of the individual WLV components (ionization state, temperature, density, electron fraction, and depletion); the \ion{C}{1} data might also provide constraints on the relative $n_{\rm H}$ in the two strongest SLV components.}
\item{Spectra of \ion{C}{3}, \ion{N}{2}, \ion{N}{3}, \ion{Si}{3}, and \ion{Fe}{3} would yield more accurate abundances and depletions for individual ionized components at low, intermediate, and high velocities.}
\item{Higher resolution spectra of the higher ions \ion{C}{4}, \ion{N}{5}, \ion{Si}{3}, \ion{Si}{4}, and \ion{S}{3} would yield a clearer picture of the relationships among those ions and of the processes responsible for their creation.}
\end{itemize}
Observations of H$\alpha$ emission (e.g., with WHAM) should provide useful constraints on the ionized gas at low velocities.
Finally, determinations of charge exchange rates at low temperatures are needed to gauge the possible effects of charge exchange processes on the ionization balance --- to better constrain the ionization rate required in the SLV clouds and to evaluate the possibility of enhanced depletions of some elements at higher densities.

\acknowledgments

Most of the observations reported here were obtained with the NASA/ESA {\it Hubble Space Telescope}
at the Space Telescope Science Institute, which is operated by the
Association of Universities for Research in Astronomy, Inc., under NASA
contract NAS5-26555.
We thank D. Doss (McDonald), J. Spyromilio (AAO), and D. Willmarth (KPNO) for assistance in setting up the various ground-based spectrographs, J. Salzer for providing \ion{Ti}{2} data, and E. Fitzpatrick for obtaining the KPNO \ion{Na}{1} spectrum.
Support for this work was provided by NASA through grants
GO-2251.01-87A, GO-5882.01-94A, and GO-05896.01-94A (administered by STScI) and through LTSA grant NAGW-4445.

\appendix

\section{Atomic Data}

In general, wavelengths and oscillator strengths were taken from Morton (1991).
In a number of cases, however, different values were adopted --- in light of more recent work or in order to achieve a consistent set of component parameters for all lines of a given species.

\ion{C}{1}:  For all transitions of C, N, and O (except the intersystem transitions of \ion{C}{2}, \ion{N}{1}, and \ion{O}{1} described below), we have adopted $f$-values from 
the recent compilation by Wiese, Fuhr and Deters (1996).
In general, the new values differ little from those in Morton (1991), but they represent a new standard.
For \ion{C}{1}, there are some significant 
differences shortward of 1200 \AA, where Morton depended on observed
interstellar lines or calculations by Kurucz \& Peytremann (1975).
We have had to retain those older calculations for the lines at 1138.6, 1140.4, and 1158.7 \AA.
Our data for 23 Ori suggest that the $f$-values for \ion{C}{1} $\lambda$1276.5, \ion{C}{1}* $\lambda$1157.8, $\lambda$1189.2, and $\lambda$1193.7 may be too high, by about 0.4, 0.5, 0.4, and 0.2 dex, respectively. 
The $f$-values for the $\lambda$1276.5, $\lambda$1193.7, and $\lambda$1157.8 lines therefore appear to be more consistent with the values derived by Zsarg\'{o}, Federman, \& Cardelli (1997) from interstellar curves of growth.

\ion{C}{2}:  Fang et al. (1993) measured the lifetime of the upper level of the
important intersystem transition at 2325.405 \AA~and used a branching fraction from theory to find $f$ = 5.78 $\times$ 10$^{-8}$.

\ion{N}{1}:  The $f$-values for the intersystem lines at 1159.817 and 1160.937 \AA~are from the calculations of Hibbert et al. (1985).

\ion{O}{1}:  The mean of the lifetimes by Johnson (1972), Wells \& Zipf (1974), Nowak, Borst, \& Fricke (1978), and Mason (1990) for the intersystem transition at 1355.598 \AA~is 180$\pm$7 $\mu$sec.  
Bi\'{e}mont \& Zeippen (1992) calculated 200 $\mu$sec, and a branching fraction of 0.755 for the $\lambda$1356 transition.
We prefer to use this branching fraction with the experimental mean to derive $f$ =
1.16 $\times$ 10$^{-6}$, compared with 1.25 $\times$ 10$^{-6}$ in Morton (1991).

\ion{Mg}{2}:  Fleming et al. (1998) have published a careful calculation for the weak doublet at 1240 \AA~which addresses the severe cancellation of terms that had troubled previous attempts.
For the multiplet, they find $f$ = 8.3 $\times$ 10$^{-4}$, which in LS coupling gives $f$(1239.9) = 5.54 $\times$ 10$^{-4}$ and $f$(1240.4) = 2.77 $\times$ 10$^{-4}$.
We have adopted these values, pending experimental confirmation.
Fitzpatrick (1997) compared these lines with the damped profiles at 2796 and 2802 \AA~
toward $\zeta$ Oph, $\xi$ Per, and HD 93521 by using a detailed model for the complex component structure for each line of sight, and concluded that
the values for the weaker doublet should be $f$(1239) = 6.4 $\times$ 10$^{-4}$ and $f$(1240) = 3.2 $\times$ 10$^{-4}$, close to the adopted values.
Our similar comparisons for 23 Ori and 1 Sco are consistent with Fitzpatrick's results.
Fitzpatrick (1997) has discussed the implications of the revised \ion{Mg}{2} $f$-values for studies of interstellar electron densities and dust composition.

\ion{Si}{1}: O'Brian \& Lawler (1991) combined their accurate lifetimes with branching
fractions from Smith et al. (1987) to obtain $f$ = 0.210, with a 1-$\sigma$ error
of 0.022 dex, for the transition at 2515.0725 \AA.  
Another of their lifetimes implies that $f$ (1845) = 0.207, but it suffers from considerable uncertainty due to unknown infrared decays.
The $f$-value for the $\lambda$1562 line appears to be too high by at least 0.4 dex.

\ion{Si}{2}:  Dufton et al. (1992) performed a careful calculation which resulted in
$f$(1808) = 0.0020, with an estimated uncertainty of 25\%.  
Then Bergeson \& Lawler (1993) measured a lifetime for the upper level which gives $f$ =
0.00208, with a 1-$\sigma$ error of 10\%.  
We have adopted the latter value.
Spitzer \& Fitzpatrick (1993) revised the $f$-values for the lines at 1304 and 1526 \AA~
by comparing their profiles towards HD 93521 with that of $\lambda$1808 and their
adopted $f$(1808) = 0.0020.  
The increase to 0.00208 results in $f$(1304) = 0.0894 and $f$(1526) = 0.114.
For the latter transition, however, we prefer $f$(1526) = 0.127 from the measurements of Schectman, Povolny, \& Curtis (1998).

\ion{S}{1}:  Bi\'{e}mont et al. (1998) and Beideck et al. (1994) have listed new $f$-values for
a number of lines based on measured lifetimes and either experimental or
theoretical branching fractions.
The 1296.2 line appears to be stronger than expected, and may be contaminated by a residual detector feature.

\ion{Cl}{1}:  Theoretical calculations by Bi\'{e}mont, Gebarowski \& Zeippen (1994) support
the measurements of Schectman et al. (1993), which we have adopted, so that
$f$(1347) = 0.153.
The $f$-value for the 1167.1 line appears to be too high by at least 0.3 dex. 

\ion{Cr}{2}:  Bergeson \& Lawler (1993) have derived $f$(2056) = 0.106 and $f$(2062) = 0.0778 from measured lifetimes and branching fractions.

\ion{Mn}{2}:  We derived new values $f$(1162) = 0.024, $f$(1163) = 0.018, and $f$(1164) = 0.011 by comparison with the multiplet near 1197 \AA~towards 23 Ori and the assumption of LS coupling ratios for $\lambda$1164, which is blended with \ion{Ge}{2}.
The $f$-values for the latter multiplet depend on a curve-of-growth fitting of these interstellar lines with the stronger multiplet near 2600 \AA~by Lugger et al. (1982).  
Our values are close to those
which Allen, Snow, \& Jenkins (1990) derived in a similar way for $\sigma$ Sco, when allowance is made for differences in the $f$-values adopted for the $\lambda$1197 multiplet.

\ion{Fe}{2}:  Improved $f$-values based on measured lifetimes and branching fractions are
now available for many transitions, including $f$(2250) = 0.00182 from
Bergeson, Mullman, \& Lawler (1994) and $f$(2344) = 0.114, $f$(2374) =
0.0313 from Bergeson et al. (1996).  
Mullman, Sakai, \& Lawler (1997) have extended these results to $f$(1608) = 0.058 by absorption ratios.  
The values $f$(1142) = 0.00247 and $f$(2368) = 6.28 $\times$ 10$^{-5}$ are from a comparison of interstellar lines by Cardelli \& Savage (1995), who tied their scale to
the laboratory measurements for $\lambda$2250 and $\lambda$2260.  
The factor 2 decrease in $f$(1142) leaves uncertain the quoted $f$-values for the lines at 1143.2 and 1144.9 \AA~in the same multiplet.
Cardelli \& Savage also derived $f$(1611) = 0.00102, but our comparison with other lines toward 23 Ori and 1 Sco implies $f$(1611) $\sim$ 0.0013, roughly 30\% larger.
This difference for $f$(1611) suggests that $f$(2368) should be higher as well.

\ion{Co}{2}:  Mullman, Cooper, \& Lawler (1998) have combined experimental lifetimes and
branching fractions to obtain $f$(2059) = 0.00751 and $f$(2026) = 0.0125.
Mullman et al. (1998) have extended these data with absorption measurements to obtain $f$(1553) = 0.0116.
We still must depend on the calculations of Kurucz for the lines at 1187.4 and 1547.9 \AA.

\ion{Ni}{2}:  Zsarg\'{o} \& Federman (1998) adopted $f$(1412) = 0.00665 from the calculations of Kurucz and used observations of interstellar absorption lines to estimate $f$(1370) = 0.144, $f$(1393) = 0.0189, $f$(1710) = 0.0666, and $f$(1742) = 0.0776 (and others that we have not observed toward 23 Ori).
Recent accurate measurements of lifetimes and branching fractions by Fedchak \& Lawler (1999) give $f$(1710) = 0.0348 and $f$(1742) = 0.0419, indicating that the values of Kurucz and of Zsarg\'{o} \& Federman may be too large by 0.275 dex.
Pending measurements of the shorter wavelength lines by Fedchak \& Lawler, we have decreased all the values from Kurucz or Zsarg\'{o} \& Federman in Table~\ref{tab:ew} by 0.275 dex.
The new $f$-values imply that the depletion of Ni is generally very similar to that of Fe (Table~\ref{tab:refab}) --- so that the [Ni/Fe] ratio is not useful for determining depletions in QSO absorption-line systems (confirming the conjecture of Lu et al. 1996; see also Welty et al. 1997).

\ion{Zn}{2}:  The rest wavelengths of the $\lambda$2026 and $\lambda$2062 lines seem to be inconsistent by $\sim$ 1 km s$^{-1}$ for a number of lines of sight; both lines were observed together with lines of other species (\ion{Mg}{1} and \ion{Cr}{2}), so the inconsistency cannot be attributed solely to uncertainties in the GHRS velocity calibration.

\ion{Ge}{2}:  These $f$-values are based on new calculations by Bi\'{e}mont, Morton, \& Quinet (1998),
which supercede the values listed by Morton (1991).  
In the 1991 paper $f$(1237) appeared inadvertently too small by a factor 10.

\ion{Kr}{1}:  These $f$-values are from recent accurate measurements by Lang et al. (1998).

\ion{Sn}{2}:  Migdalek (1976) derived $f$(1400) = 0.798 from a relativistic calculation including core polarization. 
If his relative $f$-values are used to scale the beam-foil lifetime of Andersen \& Lindgard (1977), then $f$(1400) = 1.03, as listed in Table 3. 
A new experiment is needed before a definitive $f$-value can be assigned.

Further data on wavelengths and oscillator strengths for the resonance lines of elements Ge and heavier are available in a new compilation by Morton (1999).

\newpage

\begin{figure}
\epsscale{0.4}
\plotone{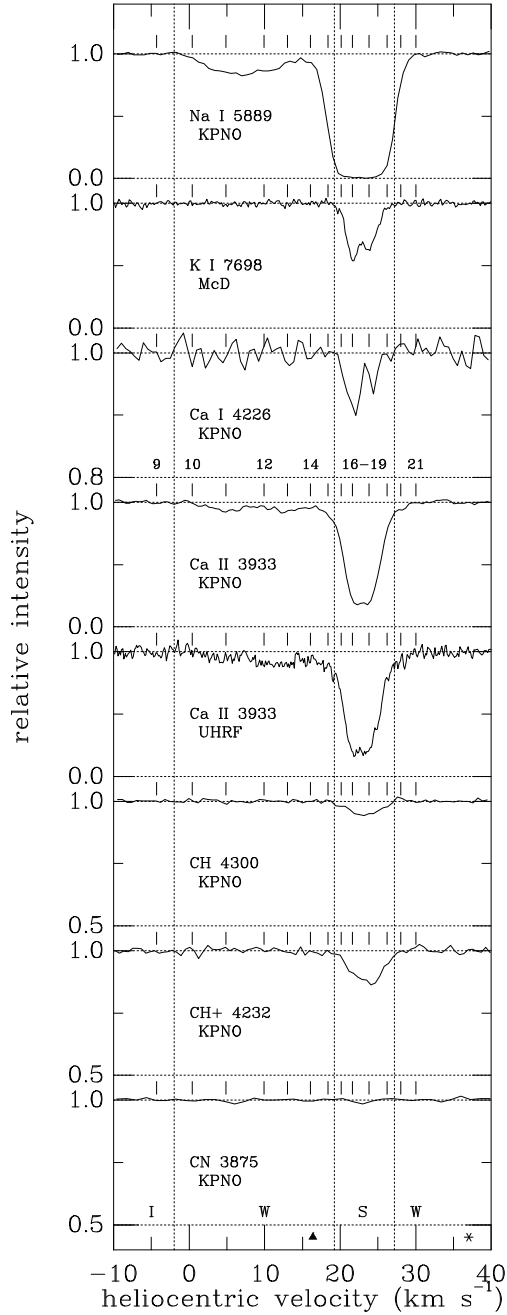}
\caption{High-resolution optical spectra of 23 Ori, showing the strong and weak low velocity gas (0 km s$^{-1}$ $\la$ $v$ $\la$ 30 km s$^{-1}$).
The tick marks denote the components derived from fits to the various optical and UV spectra; some components are numbered above the KPNO Ca II spectrum.
Note the expanded vertical scale for Ca I and the molecular species.
The vertical dotted lines separate the IV, WLV, and SLV components.
The triangle at +16.4 km s$^{-1}$ denotes the zero point for LSR velocities; the * at +37 km s$^{-1}$ denotes the stellar radial velocity.}
\label{fig:lowv}
\end{figure}

\begin{figure}
\epsscale{0.8}
\plotone{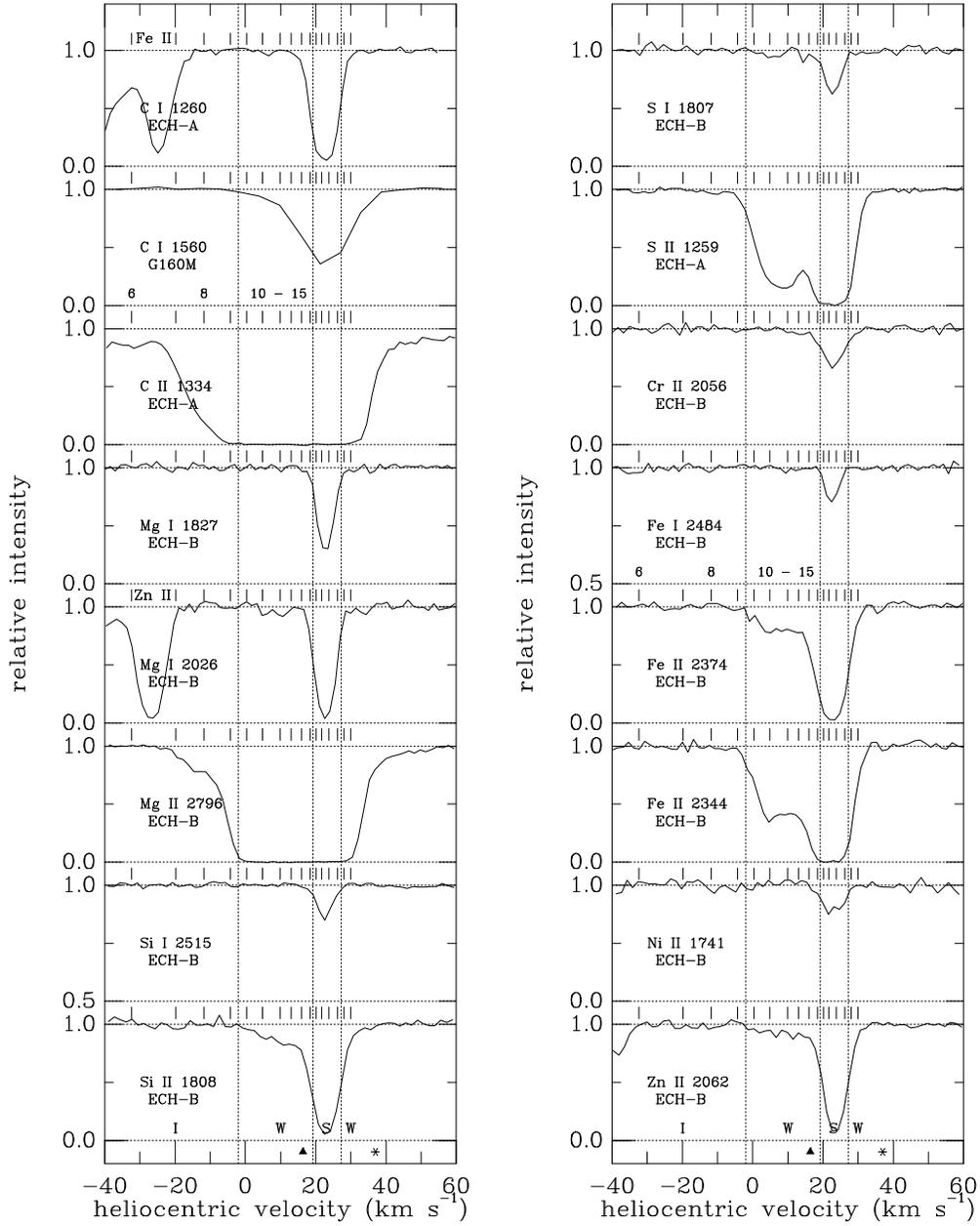}
\caption{Selected UV spectra of 23 Ori (most obtained with the GHRS ECH-A or ECH-B at resolutions of about 3.5 km s$^{-1}$), showing primarily the strong and weak low-velocity gas.
The tick marks denote the components derived from the profile fits; some components are numbered above the C II and Fe II $\lambda$2374 spectra.
Note the expanded vertical scale for Si I and Fe I.
Absorption due to other species is noted above several spectra.
The vertical dotted lines separate the IV, WLV, and SLV components.}
\label{fig:uv}
\end{figure}

\begin{figure}
\epsscale{0.8}
\plotone{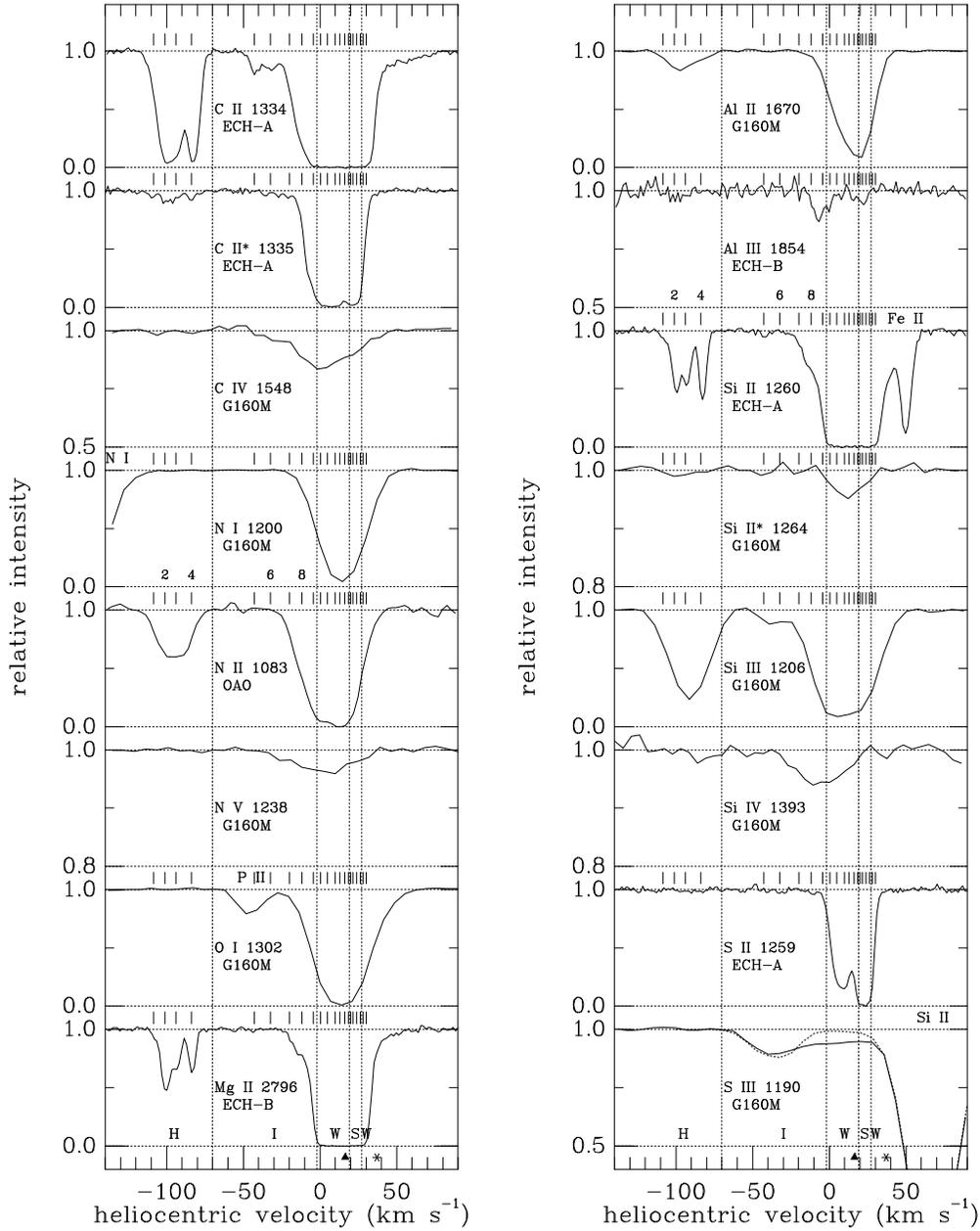}
\caption{Selected UV spectra of 23 Ori (most obtained with the GHRS echelle at resolutions of about 3.5 km s$^{-1}$ or with the G160M grating at resolutions of 13-20 km s$^{-1}$), covering a wider velocity range --- to show the high- and intermediate-velocity components and the absorption from more highly ionized species.
The tick marks denote the components derived from the profile fits; some components are numbered above the N II and Si II spectra.
Note the expanded vertical scale for some weak lines.
Absorption due to other species is noted above several spectra.
The vertical dotted lines separate the HV, IV, WLV, and SLV components.
The dotted line superposed on the S III spectrum (lower right) represents the absorption due to C I (SLV) and Si II (HV, WLV, SLV).}
\label{fig:highv}
\end{figure}

\begin{figure}
\epsscale{0.5}
\plotone{23orif4.eps}
\caption{Schematic diagram of the various components seen toward 23 Ori.
The asterisks denote discernible components (with corresponding component numbers); the solid lines show the range of absorption for most lines; the dotted lines show the range of absorption due to the wings of the strongest lines.
The weak (WLV) and strong (SLV) low velocity gas is primarily neutral, while the gas at $v$ $\la$ 0 km s$^{-1}$ (IV, HV) is primarily ionized.}
\label{fig:schem}
\end{figure}

\begin{figure}
\epsscale{0.9}
\plotone{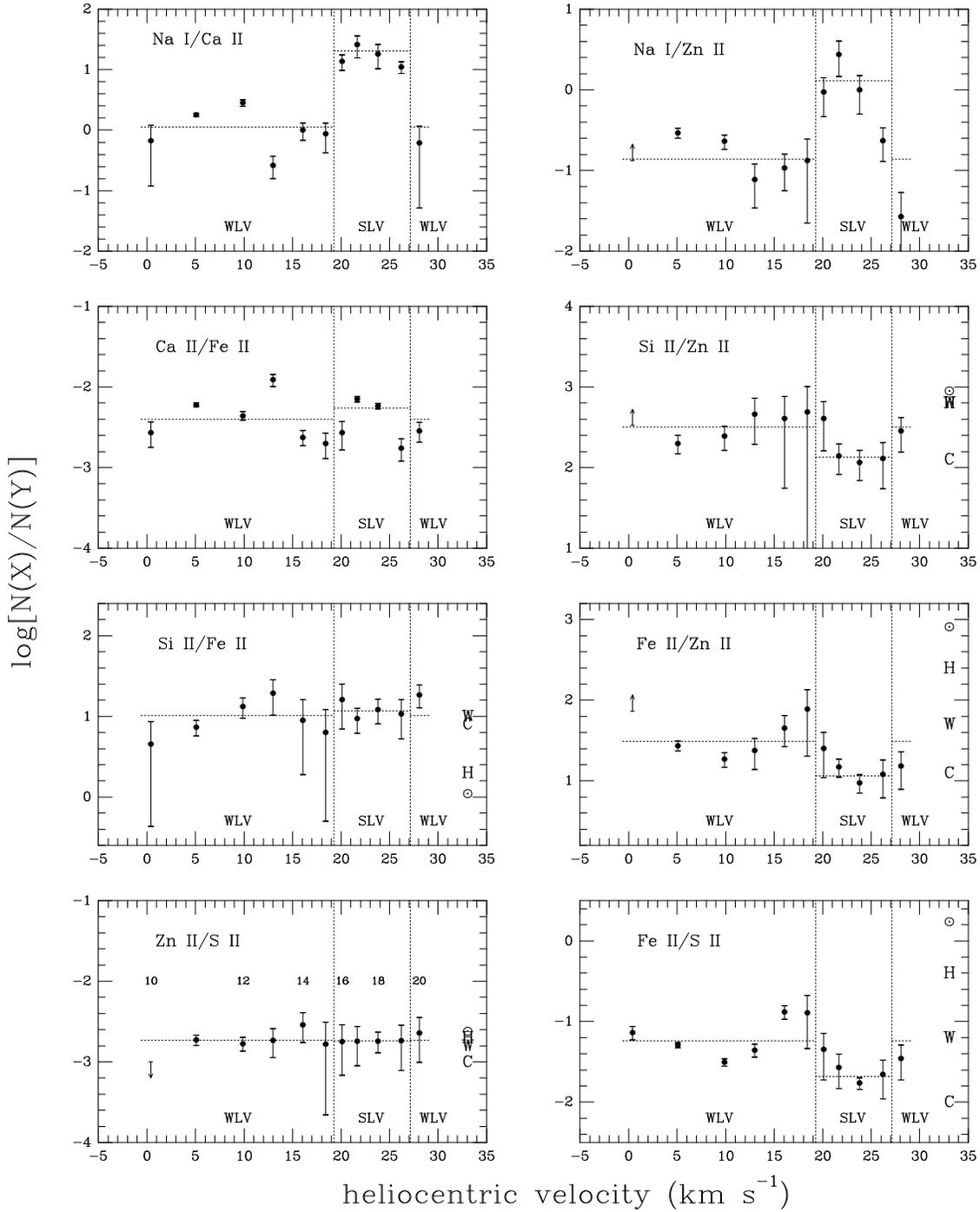}
\caption{Relative abundances for individual WLV and SLV components, as derived from the profile fits.
Uncertainties are 1$\sigma$; limits are 2$\sigma$; some components are numbered in the Zn II/S II panel at lower left.
The horizontal dotted lines give the average values for the WLV and SLV components.
At the right, the solar system (meteoritic) ratio ($\odot$) and the values found for Galactic cold (C), warm (W), and halo (H) clouds are noted.}
\label{fig:relind}
\end{figure}

\begin{figure}
\epsscale{0.8}
\plotone{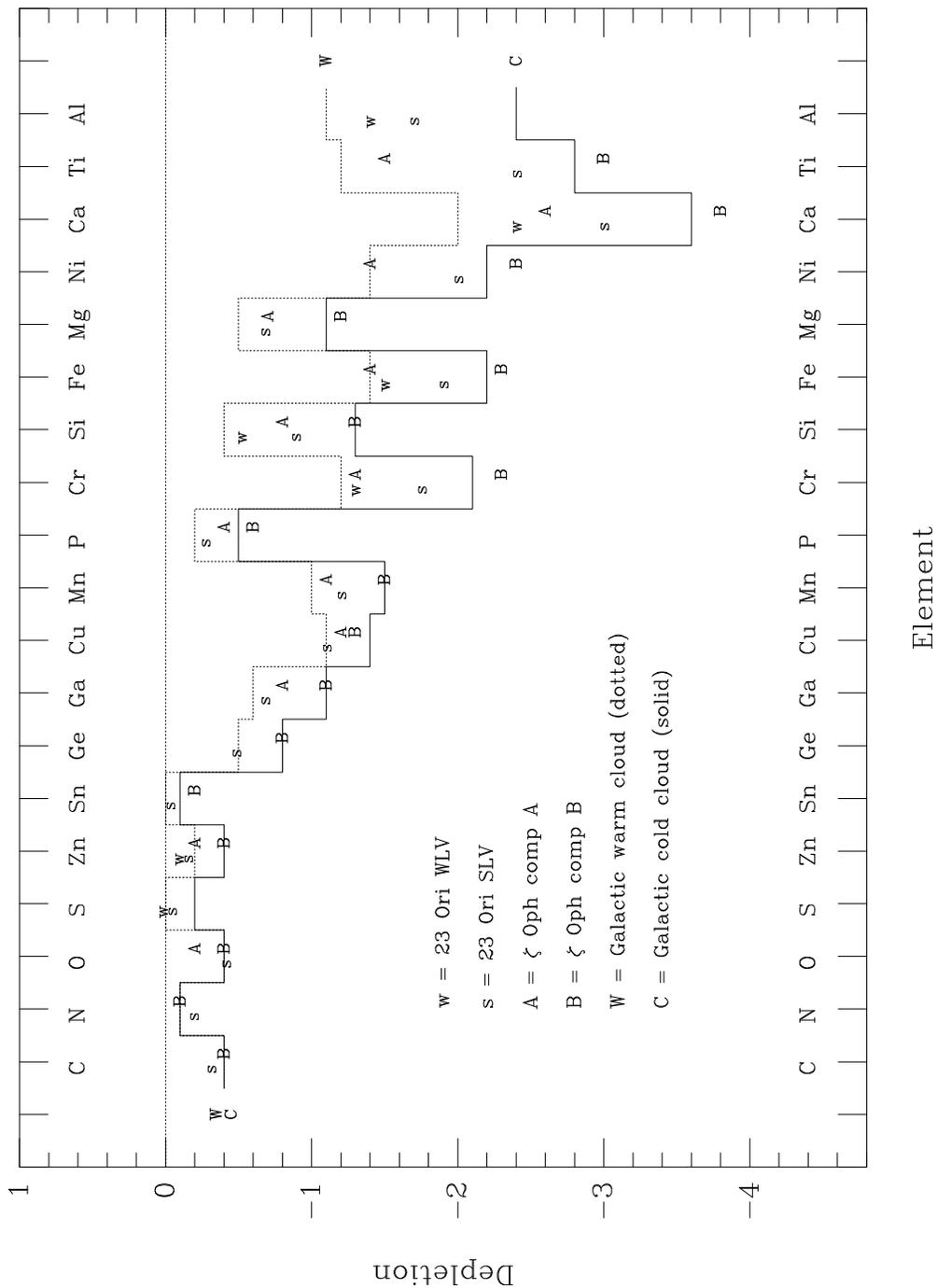}
\caption{Depletions for the weak (WLV) and strong (SLV) low velocity gas (left, 
for each element), ordered by elemental condensation temperature (Wasson 1985).
The Galactic warm and cold cloud depletions (dotted and solid lines) have been adapted and updated from Jenkins (1987) (see Table 5); the cold cloud depletion of sulfur is assumed here to be $-$0.2 dex.
The $\zeta$ Oph component (blend) A and B depletions (right; see $\S$ 3.2 for definitions) are from references listed in the text, adjusted for our adopted $f$-values and solar abundances.
The SLV gas toward 23 Ori seems less depleted (for most moderately to heavily depleted elements), by a factor 2--4, compared to the typical Galactic cold cloud or $\zeta$ Oph comp B pattern.
The WLV depletions are broadly consistent with the Galactic warm cloud and $\zeta$ Oph comp A pattern.}
\label{fig:depl}
\end{figure}

\begin{figure}
\epsscale{0.8}
\plotone{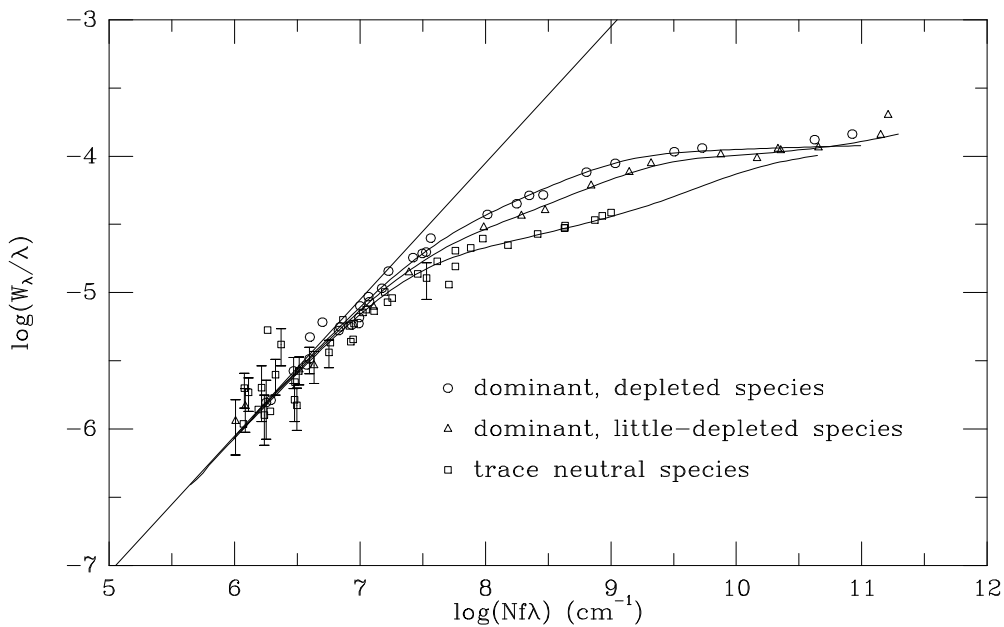}
\caption{Empirical curves of growth for trace neutral species (squares), dominant, little-depleted species (triangles), and dominant, depleted species (circles) in the H I gas toward 23 Ori.
The equivalent widths and oscillator strengths are given in Table 3; the column densities are given in Table 7; only detected, unblended lines are used.
Solid lines are predicted curves based on adopted component structures for Na I, Zn II, and Fe II.
The strongest lines of Mg II and C II exhibit both damping wings and intermediate-velocity absorption, and thus lie slightly above the predicted curves.
Uncertainties (1-$\sigma$) are not plotted where they are less than $\pm$0.1 dex.
Several lines have apparently discrepant $f$-values, as discussed in the Appendix.}
\label{fig:cog}
\end{figure}

\clearpage

\begin{deluxetable}{llcclc}
\scriptsize
\tablecolumns{6}
\tablenum{1}
\tablecaption{23 Ori --- Optical Spectra \label{tab:opt}}
\tablewidth{0pt}

\tablehead{
\multicolumn{1}{c}{Date}&
\multicolumn{1}{c}{Facility}&
\multicolumn{1}{c}{FWHM}&
\multicolumn{1}{c}{S/N}&
\multicolumn{1}{c}{Species Observed}&
\multicolumn{1}{c}{Ref}\\
\multicolumn{2}{c}{ }&
\multicolumn{1}{c}{(km s$^{-1}$)}}
\startdata
1978 Jan&McD 2.7m $+$ PEPSIOS&2.0&45&Na I&1\nl
1992 Sept&KPNO coud\'{e} feed + cam 5&2.5--3.0&90--180&K I, Ca II, Ca I, CN, CH, CH$^+$\nl
1993 Oct&KPNO coud\'{e} feed + cam 6&1.2--1.4&65--110&K I, Ca II, Ca I, CN, CH$^+$&2\nl
1994 Feb&AAT $+$ UHRF&0.32&30&Ca II&2\nl
1995 Oct&KPNO coud\'{e} feed + cam 6&1.3--1.4&60--125&Ca II, Ca I, CN, CH, CH$^+$&2\nl
1995 Oct&McD 2.7m coud\'{e}&0.56&60&K I&3\nl
1996 Apr&KPNO coud\'{e} feed + cam 6&1.5&95&Na I&4\nl
\enddata
\tablerefs{1 = Hobbs (1978b); 2 = Welty et al. (1996); 3 = Welty et al. (1999d); 4 = Fitzpatrick \& Welty (1999)}
\end{deluxetable}

\clearpage

\begin{deluxetable}{lcccccc}
\scriptsize
\tablecolumns{7}
\tablenum{8}
\tablecaption{23 Ori --- Relative Electron Densities and Abundances (H I Regions)\tablenotemark{a} \label{tab:ne}}
\tablewidth{0pt}
\tablehead{
\multicolumn{1}{c}{Element}&
\multicolumn{1}{c}{log[N(X I)]}&
\multicolumn{1}{c}{log[N(X II)]}&
\multicolumn{1}{c}{$\Gamma$/$\alpha$\tablenotemark{b}}&
\multicolumn{1}{c}{n$_{\rm e}$\tablenotemark{c}}&
\multicolumn{1}{c}{[X/C]$_{\rm II}$\tablenotemark{d}}&
\multicolumn{1}{c}{[X/C]$_{\rm I}$\tablenotemark{e}}\\
\multicolumn{1}{c}{ }&
\multicolumn{1}{c}{(cm$^{-2}$)}&
\multicolumn{1}{c}{(cm$^{-2}$)}&
\multicolumn{1}{c}{(cm$^{-3}$)}&
\multicolumn{1}{c}{(cm$^{-3}$)}&
\multicolumn{1}{c}{ }&
\multicolumn{1}{c}{ }}
\startdata
\multicolumn{7}{c}{SLV Components}\\
\tableline
\multicolumn{3}{c}{ }&
\multicolumn{1}{c}{(WJ1/D)}&
\multicolumn{1}{c}{(WJ1/D)}&
\multicolumn{1}{c}{ }&
\multicolumn{1}{c}{(WJ1)}\\
\enddata
\tablenotetext{a}{Values in parentheses are estimates based on assumed depletions (Na II and K II) or component structure (WLV C II and Mg II).  For WLV Ca$^+$, entries are for Ca II and Ca III, and D is the logarithm of the Ca depletion ($\delta$).}
\tablenotetext{b}{$\Gamma$/$\alpha$ from Aldrovandi \& P\'{e}quignot (1973; 1974), P\'{e}quignot \& Aldrovandi (1986), and Shull \& van Steenberg (1982).  For SLV, entries are for WJ1 and Draine radiation fields, both with $\alpha$ at T=100 K; for WLV, entries are for WJ1 field, for $\alpha$ at T=100 K and 3000 K.}
\tablenotetext{c}{SLV electron densities are average values over 4 SLV components [dominated by two strongest components (17 and 18), for which $n_e$(17) $\sim$ 2.6 $n_e$(18) ($\S$ 3.1.2)].
WLV electron densities are average values over 7 WLV components.}
\tablenotetext{d}{log(X II/C II) $-$ (A$_{\rm X}$ $-$ A$_{\rm C}$), where A$_{\rm X}$ is log(meteoritic abundance).}
\tablenotetext{e}{Relative abundance calculated from neutral species ratio, adjusted for solar abundances and ($\Gamma$/$\alpha$) ratios (for WJ1 field).}
\end{deluxetable}

\clearpage
 
\begin{deluxetable}{lcccc}
\scriptsize
\tablecolumns{5}
\tablenum{9}
\tablecaption{23 Ori --- Physical Conditions in SLV and WLV Gas \label{tab:h1prop}}
\tablewidth{0pt}
 
\tablehead{
\multicolumn{1}{c}{Property}&
\multicolumn{2}{c}{-~-~-~SLV~-~-~-}&
\multicolumn{2}{c}{-~-~-~WLV~-~-~-}\\
}
\startdata
$N$(H I) (cm$^{-2}$) & \multicolumn{2}{c}{20.71} & \multicolumn{2}{c}{19.61} \nl
$N$(C I*)/$N$(C I)$_{\rm tot}$ & \multicolumn{2}{c}{0.10$\pm$0.02} & \multicolumn{2}{c}{0.40$\pm$0.08} \nl
$N$(C I**)/$N$(C I)$_{\rm tot}$ & \multicolumn{2}{c}{0.015$\pm$0.003} & \multicolumn{2}{c}{0.16$\pm$0.04} \nl
$T$ (K) & \multicolumn{2}{c}{100$^{+50}_{-35}$} & (100)\tablenotemark{a} & (3000)\tablenotemark{a} \nl
log($n_{\rm H}T$) (cm$^{-3}$K) & \multicolumn{2}{c}{3.1$\pm$0.1} & 4.3$\pm$0.2 & 4.7--4.8 \nl
$n_{\rm H}$ (cm$^{-3}$) & \multicolumn{2}{c}{10--15} & 200 & 15--20 \nl
$n_e$ (cm$^{-3}$) & 0.15$\pm$0.05 & (0.015)\tablenotemark{b} & 0.02 & 0.20 \nl
$n_e$/$n_{\rm H}$ & 0.01 & 0.001 & 0.0001 & 0.01 \nl
$N_{\rm H}$/$n_{\rm H}$ (pc) & \multicolumn{2}{c}{12--16} & 0.1 & 0.7--0.9 \nl
 & \nl
log[$N$(C II*)]$_{\rm obs}$ (cm$^{-2}$) & \multicolumn{2}{c}{14.2--15.0} & \multicolumn{2}{c}{14.3--14.6} \nl
log[$N$(C II*)]$_{\rm pred}$ (cm$^{-2}$) & 15.3 & 14.6 & 14.5 & 14.25 \nl
log[$N$(Si II*)]$_{\rm obs}$ (cm$^{-2}$) & \multicolumn{2}{c}{$<$10.7} & \multicolumn{2}{c}{$\sim$11.5} \nl
log[$N$(Si II*)]$_{\rm pred}$ (cm$^{-2}$) & 10.8 & 9.9 & 10.0 & 11.3 \nl
\enddata
\tablenotetext{a}{Assumed representative temperatures for cold or warm gas.}
\tablenotetext{b}{Assumed lower $n_e$ --- if, for example, charge exchange with large molecules is significant.}
\end{deluxetable}
 
\clearpage

\begin{deluxetable}{cccccccc}
\scriptsize
\tablecolumns{7}
\tablenum{10}
\tablecaption{23 Ori --- H II Region (HV, IV) Component Structure\tablenotemark{a} \label{tab:h2cmp}}
\tablewidth{0pt}
\tablehead{
\multicolumn{1}{c}{Comp}&
\multicolumn{1}{c}{v$_\odot$}&
\multicolumn{1}{c}{C II}&
\multicolumn{1}{c}{C II*}&
\multicolumn{1}{c}{Mg II}&
\multicolumn{1}{c}{Al III}&
\multicolumn{1}{c}{Si II}\\
\multicolumn{1}{c}{ }&
\multicolumn{1}{c}{(km s$^{-1}$)}&
\multicolumn{1}{c}{1334}&
\multicolumn{1}{c}{1335}&
\multicolumn{1}{c}{2796}&
\multicolumn{1}{c}{1854}&
\multicolumn{1}{c}{1260}\\
\multicolumn{4}{c}{ }&
\multicolumn{1}{c}{2803}&
\multicolumn{1}{c}{1862}&\\
}
\startdata
 1H & -108.0$\pm$0.4 & 0.7$\pm$0.1e13 & 0.6$\pm$0.2e12 & 0.7$\pm$0.1e11 & 
     $<$2.0e10 & 1.1$\pm$0.2e11 \\
& & (3.5) & (3.5) & (2.5) & (2.4) & (2.35) \nl
 & \nl
 2H & -100.5$\pm$0.1 & 6.4$\pm$0.4e13 & 1.8$\pm$0.2e12 & 11.5$\pm$0.3e11 & 
     9.6$\pm$2.3e10 & 14.6$\pm$0.5e11 \\
& & 4.2$\pm$0.2 & (4.2) & 2.9$\pm$0.1 & (2.9) & 2.9$\pm$0.2 \nl
 & \nl
 3H &  -93.3$\pm$0.1 & 3.2$\pm$0.1e13 & 1.0$\pm$0.2e12 & 5.9$\pm$0.3e11 & 
     4.9$\pm$2.3e10 & 11.6$\pm$0.5e11 \\
& & 3.4$\pm$0.3 & (3.4) & 2.8$\pm$0.2 & (2.8) & 2.75$\pm$0.15 \nl
 & \nl
 4H &  -83.3$\pm$0.1 & 5.6$\pm$0.3e13 & 1.4$\pm$0.2e12 & 6.0$\pm$0.1e11 & 
     2.2$\pm$2.0e10 & 16.4$\pm$0.3e11 \\
& & 3.8$\pm$0.1 & (3.8) & 2.2$\pm$0.1 & (2.2) & 2.25$\pm$0.07 \nl
 & \nl
 5I &  -42.6$\pm$0.3 & 3.2$\pm$0.2e12 & $<$0.4e12 & $<$1.3e10 & 
     \nodata & 9.2$\pm$1.6e10 \\
& & (3.6) & \nodata & (2.35) & \nodata & (2.35) \nl
 & \nl
 6I &  -32.4$\pm$0.4 & 3.1$\pm$0.2e12 & 0.7$\pm$0.2e12 & $<$1.4e10 & 
     \nodata & 2.8$\pm$1.7e10 \\
& & (4.5) & (4.5) & (3.2) & \nodata & (3.2) \nl
 & \nl
 7I &  -19.8$\pm$0.8 & 3.3$\pm$0.9e12 & 1.0$\pm$0.2e12 & 3.5$\pm$0.2e10 & 
     \nodata & 1.0$\pm$0.4e11 \\
& & (4.5) & (4.5) & (3.2) & \nodata & (3.2) \nl
 & \nl
 8I &  -11.8$\pm$0.4 & 3.4$\pm$0.2e13 & (0.6e12) & 5.4$\pm$0.3e11 & 
     \nodata & 12.1$\pm$0.6e11 \\
\enddata
\tablenotetext{a}{Entries are N (cm$^{-2}$) and b (km s$^{-1}$).   HV, IV values in parentheses were fixed in the fitting.  Uncertainties are 1-$\sigma$; limits are 2-$\sigma$.  Components (1--4) are HV; components (5--9) are IV.  WLV and SLV values in parentheses refer to predominantly neutral gas (Table 7).}
\end{deluxetable}

\clearpage

\begin{deluxetable}{llrrr}
\scriptsize
\tablecolumns{5}
\tablenum{12}
\tablecaption{23 Ori --- Weak H II and High Ion Component Structure \label{tab:high}}
\tablewidth{0pt}
\tablehead{
\multicolumn{1}{c}{Ion}&
\multicolumn{1}{c}{$\lambda$}&
\multicolumn{1}{c}{$N$\tablenotemark{a}}&
\multicolumn{1}{c}{$b$\tablenotemark{a}}&
\multicolumn{1}{c}{$v$\tablenotemark{a}}\\
\multicolumn{1}{c}{ }&
\multicolumn{1}{c}{(\AA)}&
\multicolumn{1}{c}{(cm$^{-2}$)}&
\multicolumn{1}{c}{(km s$^{-1}$)}&
\multicolumn{1}{c}{(km s$^{-1}$)}}
\startdata
Si II*\tablenotemark{b} & 1264.7377  & 3.6$\pm$0.4e11 & 6$\pm$3 & 12$\pm$1 \\
 & \nl
Al III & 1854.7164 & 5.3$\pm$0.5e11 & 6.3$\pm$0.8 & $-$5.7$\pm$0.5 \\
         &                   & 1.7$\pm$0.4e11 & 4.7$\pm$1.7 &    20.5$\pm$1.1 \\
         & 1862.7895 & 4.3$\pm$0.6e11 & 5.6$\pm$1.1 & $-$4.2$\pm$0.7 \\
         &                   & 2.5$\pm$0.5e11 & 3.4$\pm$1.1 &    23.9$\pm$0.7 \\
         & both         & 5.1$\pm$0.4e11 & 6.2$\pm$0.6 & $-$6.1$\pm$0.4 \\
         &                   & 1.9$\pm$0.3e11 & 4.4$\pm$1.1 & 20.4$\pm$0.7 \\
 & \nl
S III  & 1190.208   & 3.0$\pm$0.4e13 & 16$\pm$4 & 5$\pm$2 \\
 & \nl
C IV  & 1548.195   &9.8$\pm$0.4e12 & 24$\pm$1 & 3$\pm$1 \\
        & 1550.770   &  10.7$\pm$0.5e12 & 22$\pm$2 & 9$\pm$1 \\
        & average        & 10.0$\pm$0.3e12 & 23$\pm$1 & 6$\pm$3 \\
 & \nl
Si IV & 1393.755  &   1.2$\pm$0.1e12 & 18$\pm$2 & $-$6$\pm$1 \\
        &  1402.770  &   1.0$\pm$0.1e12 & 13$\pm$3 & $-3$$\pm$2 \\
        & average        &   1.2$\pm$0.1e12 & 17$\pm$2 & $-$5$\pm$1 \\
 & \nl
N V   & 1238.821  &   3.5$\pm$0.2e12 & 23$\pm$2 & 2$\pm$1 \\
        & 1242.804  &    5.1$\pm$0.5e12 & 27$\pm$3 & $-$2$\pm$2 \\
        & average       &    3.8$\pm$0.2e12 & 24$\pm$2 & 0$\pm$2 \\
\enddata
\tablenotetext{a}{Uncertainties are 1-$\sigma$.}
\tablenotetext{b}{May arise in H I gas ($\S$~3.3.3).}
\end{deluxetable}

\clearpage

\begin{deluxetable}{ccccccc}
\scriptsize
\tablecolumns{7}
\tablenum{13}
\tablecaption{23 Ori --- H II Region (HV, IV) Component Properties\tablenotemark{a} \label{tab:h2prop}}
\tablewidth{0pt}
\tablehead{
\multicolumn{1}{c}{Comp}&
\multicolumn{1}{c}{$N$(C II*)/$N$(C II)}&
\multicolumn{1}{c}{$T_{\rm max}$\tablenotemark{b}}&
\multicolumn{1}{c}{$n_e$ = $n_{\rm H}$\tablenotemark{c}}&
\multicolumn{1}{c}{log(2$n_{\rm H}T$)\tablenotemark{c}}&
\multicolumn{1}{c}{$N$(H II)\tablenotemark{d}}&
\multicolumn{1}{c}{r}\\
\multicolumn{1}{c}{ }&
\multicolumn{1}{c}{ }&
\multicolumn{1}{c}{(K)}&
\multicolumn{1}{c}{(cm$^{-3}$)}&
\multicolumn{1}{c}{(cm$^{-3}$ K)}&
\multicolumn{1}{c}{(cm$^{-2}$)}&
\multicolumn{1}{c}{(pc)}\\}
\startdata
 1H & 0.085$\pm$0.031 & 8700/8900/9150 & 1.4$\pm$0.5 & 4.4 & 2.0e16 & 0.005 \\
 & \nl
 2H & 0.028$\pm$0.004 & 12550/11950/13950 & 0.47$\pm$0.09 & 3.9 & 1.8e17 & 0.12 \\
 & \nl
 3H & 0.031$\pm$0.006 & 8200/11150/12550 & 0.52$\pm$0.12 & 3.9 & 9.0e16 & 0.06 \\
 & \nl
 4H & 0.025$\pm$0.004 & 10250/6850/8400 & 0.42$\pm$0.08 & 3.8 & 1.6e17 & 0.12 \\
 & \nl
 5I &  $<$0.125 & 9200/\nodata/\nodata & $<$2.1 & $<$4.5 & 9.0e15 & $>$0.001 \\
 & \nl
 6I &  0.23$\pm$0.07 & 14400/\nodata/\nodata & 3.9$\pm$1.2 & 4.8 & 8.7e15 & 0.0007 \\
 & \nl
 7I &  0.30$\pm$0.10 & \nodata & 5.0$\pm$1.7 & 4.9 & 9.3e15 & 0.0006 \\
 & \nl
 8I &  $<$0.118 & \nodata & $<$1.9 & $<$4.5 & 9.6e16 & $>$0.02 \\
 & \nl
 9I &  0.091$\pm$0.039 & \nodata & 1.5$\pm$0.7 & 4.4 & 1.9e17 & 0.04 \\
 & \\
\enddata
\tablenotetext{a}{Components (1--4) are HV; components (5--9) are IV.}
\tablenotetext{b}{From $b$-values for C II, Mg II, and Si II.}
\tablenotetext{c}{Assuming $T$ = 8000$\pm$2000 K.}
\tablenotetext{d}{From $N$(C II), assuming solar photospheric abundance, no C III, and no depletion.}
\end{deluxetable}

\end{document}